\documentclass[acmtog,authorversion]{acmartarxiv} 

\makeatletter
\def\set@curr@file#1{%
  \begingroup
    \escapechar\m@ne
    \xdef\@curr@file{\expandafter\string\csname #1\endcsname}%
  \endgroup
}
\def\quote@name#1{"\quote@@name#1\@gobble""}
\def\quote@@name#1"{#1\quote@@name}
\def\unquote@name#1{\quote@@name#1\@gobble"}
\makeatother

\usepackage{algorithm}
\usepackage{algorithmic}
\usepackage{balance}
\usepackage{graphicx}
\usepackage[squaren]{SIunits}
\usepackage{units}

\definecolor{lightgray}{rgb}{0.8,0.8,0.8}
\definecolor{darkblue}{rgb}{0,0,0.7}
\definecolor{darkgreen}{rgb}{0,0.5,0}
\definecolor{darkpurple}{rgb}{0.5,0.0,0.4}
\definecolor{midred}{rgb}{0.8,0,0}
\definecolor{midgreen}{rgb}{0,0.8,0}
\definecolor{midorange}{rgb}{0.8,0.6,0}
\definecolor{shadecolor}{RGB}{255,255,255}

\newcommand{\changed}[1]{\textcolor{darkgreen}{#1}}
\renewcommand{\changed}[1]{#1}
\newcommand{\deleted}[1]{\textcolor{midred}{\sout{#1}}}
\renewcommand{\deleted}[1]{}

\newcommand{\whitebox}[1]{\hspace{-0.5em}
\noindent\colorbox{shadecolor}
{\mbox{#1}}
}

\newcommand{\Fig}[1]{Figure~\ref{#1}}

\newcommand{\Eq}[1]{Equation~\ref{#1}}

\newcommand{\Alg}[1]{Algorithm~\ref{#1}}
\newcommand{\Algs}[1]{Algorithms~\ref{#1}}
\newcommand{\Sec}[1]{Section~\ref{#1}}
\newcommand{\Tab}[1]{Table~\ref{#1}}
\newcommand{\hide}[1]{}

\newcommand{\R}{\mathbb{R}}
\DeclareMathOperator*{\argmin}{arg\,min}

\renewcommand{\vec}[1]{\mathbf{#1}}
\newcommand{\Iref}{\vec{I}_\text{ref}}
\renewcommand{\P}{\vec{P}}
\newcommand{\p}{\vec{p}}
\newcommand{\x}{\vec{x}}
\newcommand{\Popt}{\vec{P}_{\text{opt}}}

\newcommand{\norm}[1]{\left\Vert #1\right\Vert _{2}}

\acmJournal{TOG}
\setcitestyle{square}
\setcopyright{rightsretained}
\title{Non-Line-of-Sight Reconstruction using Efficient Transient Rendering}

\author{Julian Iseringhausen}
\email{iseringhausen@cs.uni-bonn.de}
\orcid{0000-0002-6240-5105}
\author{Matthias B. Hullin}
\email{hullin@cs.uni-bonn.de}
\orcid{0000-0002-8041-5665}
\affiliation{
  \institution{University of Bonn}
  \department{Institute of Computer Science II}
  \city{Bonn}
  \postcode{53115}
  \country{Germany}
}
\renewcommand\shortauthors{Iseringhausen, J. and Hullin, M. B.}

\ccsdesc[500]{Computing methodologies~Computational photography}
\ccsdesc[500]{Computing methodologies~3D imaging}
\ccsdesc[500]{Computing methodologies~Reconstruction}
\ccsdesc[500]{Computing methodologies~Physical simulation}

\makeatletter
\begin{document}
\begin{anonsuppress}
\thanks{This work was supported by the German Research Foundation (DFG) under grant (HU-2273/2-1), the
X-Rite Chair for Digital Material Appearance, and the European Research Council under ERC Starting Grant ECHO. We thank Hendrik Lensch, Michael Wand, Johannes Hanika, Felix Heide, Wolfgang Heidrich, Ivo Ihrke and Ramesh Raskar for valuable discussion over the course of this project. We also gratefully acknowledge the support of NVIDIA Corporation with the donation a Titan Xp GPU that was used in this research.}
\end{anonsuppress}

\begin{abstract}
Being able to see beyond the direct line of sight is an intriguing prospective and could benefit a wide variety of important applications. Recent work has demonstrated that time-resolved measurements of indirect diffuse light contain valuable information for reconstructing shape and reflectance properties of objects located around a corner. In this paper, we introduce a novel reconstruction scheme that, by design, produces solutions that are consistent with state-of-the-art physically-based rendering. Our method combines an efficient forward model (a custom renderer for time-resolved three-bounce indirect light transport) with an optimization framework to reconstruct object geometry in an analysis-by-synthesis sense. We evaluate our algorithm on a variety of synthetic and experimental input data, and show that it gracefully handles uncooperative scenes with high levels of noise or non-diffuse material reflectance.
\end{abstract}

\begin{teaserfigure}\centering
\begin{tabular}{ccc}
\includegraphics[width=0.35\textwidth]{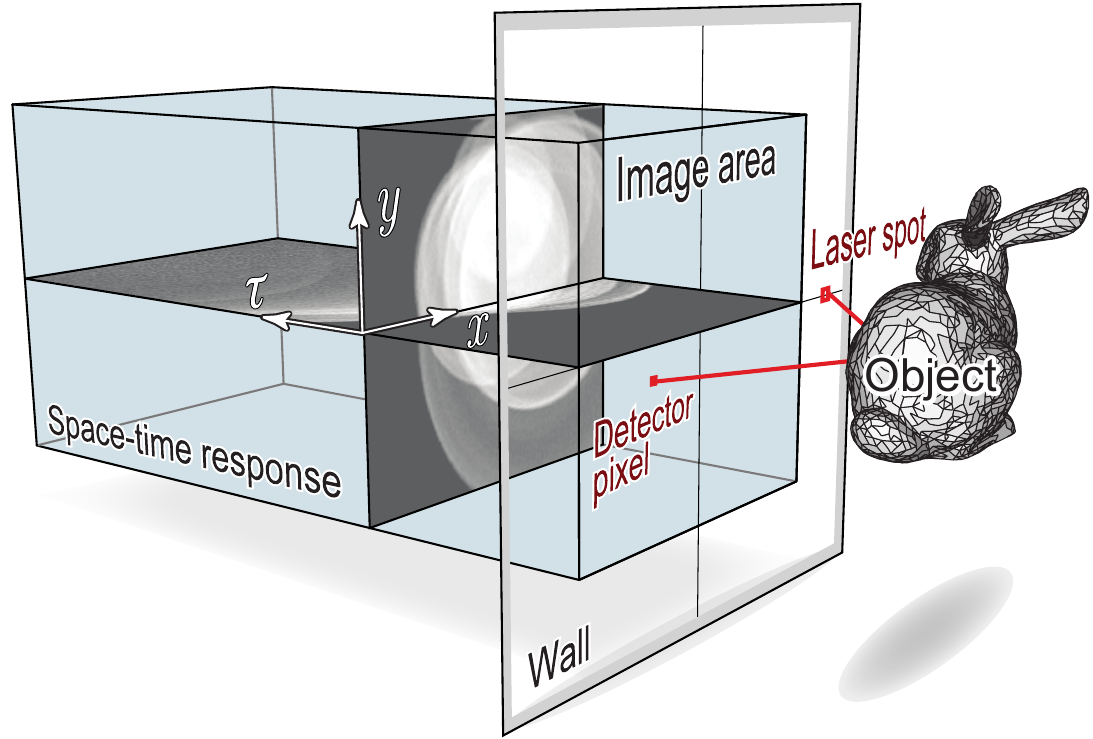}%
&\raisebox{0em}{\includegraphics[height=0.24\textwidth]{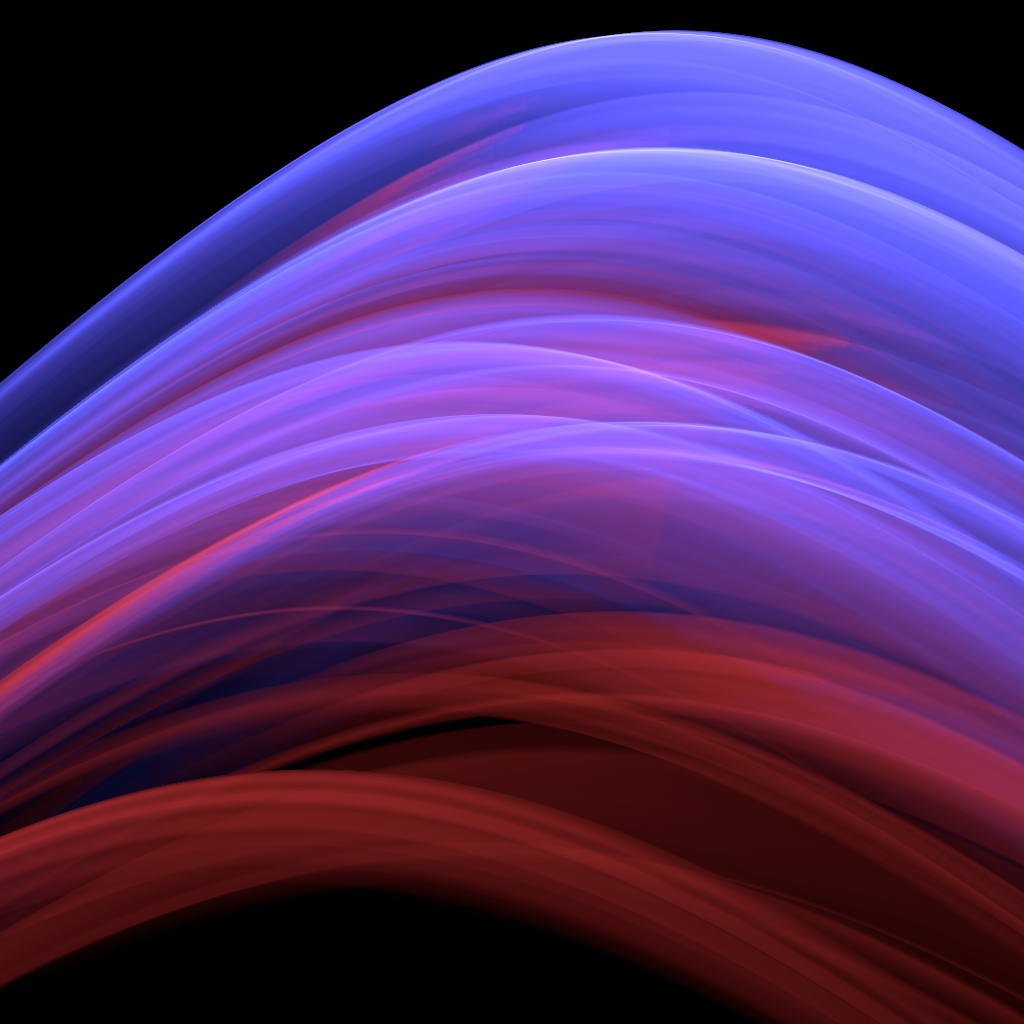}}&\raisebox{0em}{\includegraphics[height=0.24\textwidth]{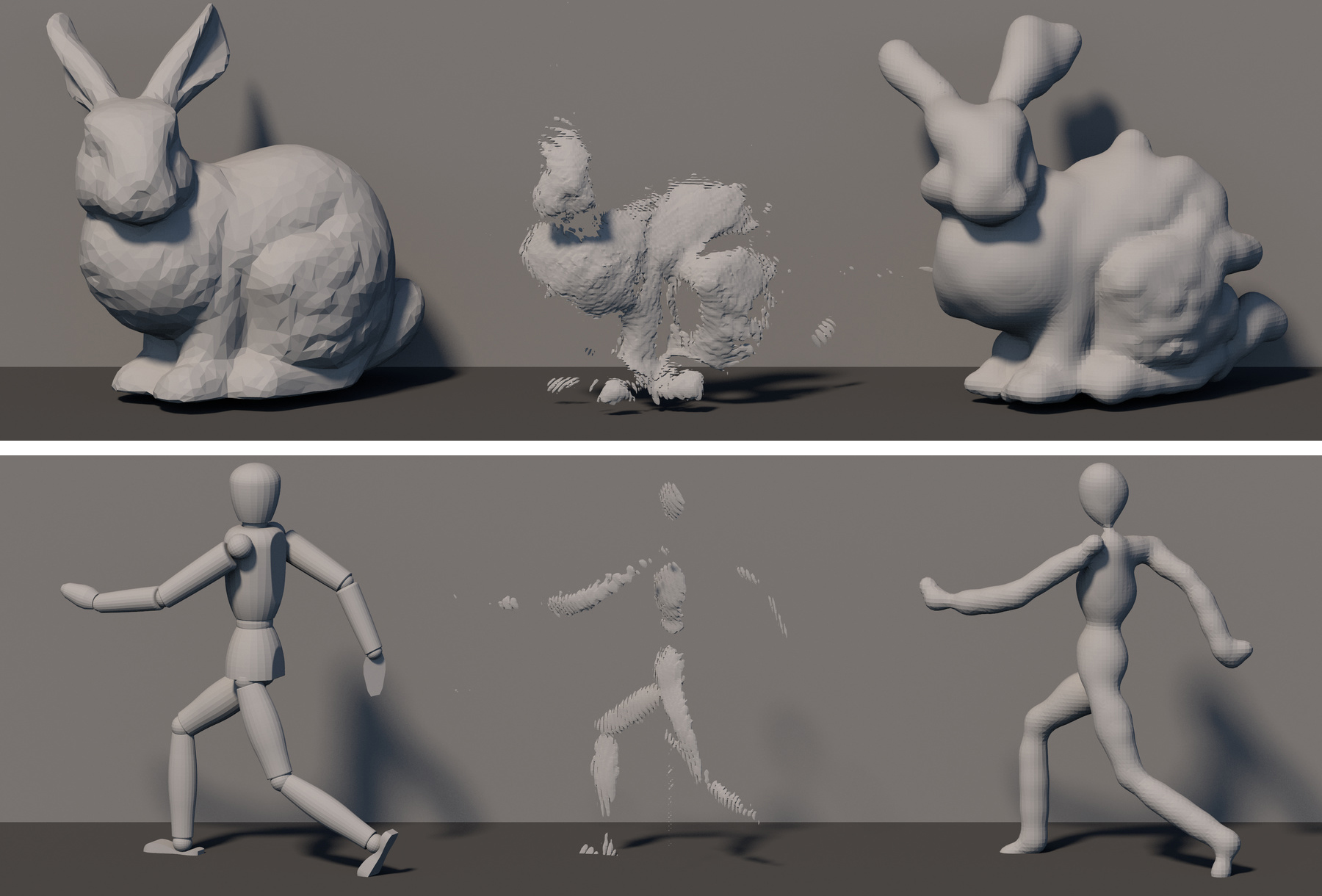}}\\
\textsf{(a)}&\textsf{(b)}&\textsf{(c)}
\end{tabular}
\caption{(a) The challenge of looking around the corner deals with the recovery of information about objects beyond the direct line of sight. In this illustration of a setting proposed by Velten et al.~\shortcite{Velten:2012:Recovering}, an unknown object is located in front of a wall, but additional obstacles occlude the object from any optical devices like light sources or cameras. Our only source of information are therefore indirect reflections off other surfaces (here, a planar ``wall''). A point on the wall that is illuminated by an ultrashort laser pulse turns into an omnidirectional source of indirect light (``laser spot''). After scattering off the unknown object, some of that light arrives back at the wall, where it forms an optical ``echo'' or space-time response (shown are 2D slices) that can be picked up by a suitable camera. Locations on the wall can be interpreted as omnidirectional detector pixels that receive different mixtures of backscattered light contributions at different times. We assume that neither camera nor laser can directly illuminate or observe the object, leaving us with the indirect optical space-time response as the only source of information. Note that for the sake of clarity, laser source, camera, and occluder are not shown here. The complete setup is illustrated in \Fig{fig:scenario2d}. (b) We propose a novel transient renderer to simulate such indirectly scattered light transport efficiently enough for use as a forward model in inverse problems. In this artistic visualization, light contributions removed by the shadow test are marked in red, and the net intensity in blue. Together with an optimization algorithm, the renderer can be used to reconstruct the geometry of objects outside the line of sight.  (c) Left to right: ground-truth object geometry; reconstruction using a state-of-the-art method (ellipsoidal backprojection); reconstruction using the technique presented in this paper. Top row: \texttt{BunnyGI} dataset; bottom row: \texttt{Mannequin1Laser} dataset. Our method relies on highly efficient and near-physical forward simulation, and it exemplifies the use of computer graphics as a technical tool to solve inverse problems in other fields. 
}
\label{fig:teaser}
\end{teaserfigure}

\maketitle
\thispagestyle{empty}

\makeatletter

\section{Motivation}
Every imaging modality from ultrasound to x-ray knows situations where the target is partially or entirely occluded by other objects and therefore cannot be directly observed.
In a recent strand of work, researchers have aimed to overcome this limitation, developing a variety of approaches to extend the line of sight of imaging systems, for instance using wave optics \cite{Katz:2014,boger2018non} or by using the occluder itself as an accidental imager \cite{bouman2017turning}.
Among all the techniques proposed, a class of methods has received particular attention within the computer vision and imaging communities.
The main source of information for these methods are indirect reflections of light within the scene, represented by time-resolved impulse responses.
From such responses, it has been shown that the presence and position of objects ``around a corner'' \cite{Kirmani:2009}, or even their shape \cite{Velten:2012:Recovering} and/or reflectance \cite{Naik:2011} can be reconstructed.
In this paper, we focus on the archetypal challenge of reconstructing the shape of an unknown object from 3-bounce indirect and (more or less) diffuse reflections off a planar wall (\Fig{fig:teaser}(a)) \cite{Kirmani:2009}.
The overwhelming majority of approaches to this class of problem rely on ellipsoidal \emph{backprojection}, where intensity measurements are smeared out over the loci in space (ellipsoidal shells) that correspond to plausible scattering locations under the given geometric constraints \cite{Velten:2012:Recovering,Buttafava:2015,Gariepy:2016,Kadambi:2016:OIT:2882845.2836164,ArellanoOpEx2017}.
Ellipsoidal backprojection implicitly assumes that the object is a volumetric scatterer, and it does not take into account surface orientation and self-occlusion of the object.
More importantly, unlike linear backprojection used in standard emission or absorption tomography, ellipsoidal backprojection is not the adjoint of a physically plausible forward light transport operator.
Where such operators have been identified \cite{lamanna}, they are typically constrained to rudimentary volumetric, non-opaque, isotropic scattering models.
This necessitates \deleted{heavy heuristic} filtering, and the reconstructed shapes are typically flat and low in detail. On the other hand, algorithms based on ellipsoidal backprojection generally have much shorter runtimes than our approach, since they do not require a global optimization scheme.

Here, we propose an alternative approach that mitigates some of the problems of backprojection by formulating the non-line-of-sight sensing problem in an analysis-by-synthesis sense. In other words, we develop a physically plausible and efficient forward simulation of light transport (transient renderer) and combine it with a nonlinear optimizer to determine the scene hypothesis that best agrees with the observed data. The method is enabled by a number of technical innovations, which we consider the key contributions of this work:
\begin{itemize}
\item a scene representation based on level sets and a surface-oriented scattering model for time-resolved light transport around a corner (wall to object to wall) based on time-resolved radiative transfer,
\item an extremely efficient GPU-based custom renderer for three-bounce backscatter that features near-physical handling of occlusion effects and a novel temporal filtering scheme for triangular surfaces, and
\item a global, self-refining optimization strategy to minimize the reconstruction error.
\end{itemize}
We evaluate our method on a number of synthetic and experimental datasets and find that it is capable of achieving significantly higher object coverage and detail than ellipsoidal backprojection, even on greatly reduced and degraded input data. Our renderer not only naturally accommodates surface BRDFs, but is also open to extensions like higher-order light bounces or advanced background models that will be needed in order to tackle future non-line-of-sight sensing problems. The method, as proposed here, is not capable of delivering high reconstruction rates in this first implementation. However, we believe that being able to generate transient renderings for the around-the-corner setting very efficiently will enable novel approaches to the problem, for instance based on machine learning.
 \section{Related work}
The research areas of transient imaging and non-line-of-sight reconstruction have recently received tremendous attention from the computer vision, graphics, imaging and optics communities. For a structured overview on the state of the art, we refer the interested reader to a recent survey \cite{jarabo2017recent}.

\subsection{Transient imaging}
Imaging light itself as it propagates through space and time poses the ultimate challenge to any imaging system. To obtain an idea of the frame rate required, consider that in vacuum, light only takes about 3 picoseconds ($3\cdot 10^{-12}\,s$) per millimeter of distance traversed.  
The typical transient imaging system consists of an ultrashort (typically, sub-picosecond) light source and an ultrafast detector. Oddly, three of the highest-performing detection technologies are over 40 years old: streak tubes \cite{Velten:2011} wherein a single image scanline is ``smeared out'' over time on a phosphor screen; holography using ultrashort pulses \cite{Abramson:78}, and gated image intensifiers \cite{Laurenzis:2014}. More common nowadays, however, are semiconductor devices that achieve comparable temporal resolution without the need for extreme light intensities or voltages. Among the technologies reported in literature are regular reverse-biased photodiodes \cite{Kirmani:2009}, as well as time-correlated single-photon counters 
which conveniently map to standard CMOS technology \cite{gariepy2015single}.
On the low end, it has also been shown that transient images can be computationally reconstructed from multi-frequency correlation time-of-flight measurements \cite{Heide:2013:LBT}, although data thus obtained typically suffers from the low temporal bandwidth of these devices, which necessitates heavy regularization.

\subsection{Transient rendering}
The simulation of transient light transport, when done na\"ively, is no different from regular physically-based rendering, except that for each light path that contributes to the image, its optical length must be calculated and its contribution stored in a time-of-flight histogram \cite{smith2008transient}. A number of offline transient renderers have been made available to the public \cite{toftracer,jarabo2014framework}. Even with advanced temporal sampling \cite{jarabo2014framework} and efficiency-increasing filtering strategies such as photon beams \cite{marco17transient}, such  renderers still take on the order of hours to days to produce converged results. In contrast, the special-purpose renderer introduced in this paper is capable of producing close-to-physical renderings of around-the-corner settings in a matter of milliseconds. 
Finally, there have been efforts to simulate the particular characteristics of single-photon counters \cite{hernandez2017computational}, an emerging type of sensor that can be expected to assume a major role in transient imaging.

\subsection{Analysis of transient light transport and looking around corners}
The information carried by transient images has been the subject of several investigations. Wu et al.~laid out the geometry of space-time streak images for lensless imaging \shortcite{wu2012frequency}, and discussed the influence of light transport phenomena such as subsurface scattering on the  shape of the temporal response \shortcite{Wu2014}. Economically, the most important use of transient light transport analysis today is likely in multi-path backscatter removal for correlation-based time-of-flight ranging \cite[and many others]{fuchs2010multipath}. 

\begin{figure}%
\centering
\includegraphics[width=0.8\columnwidth]{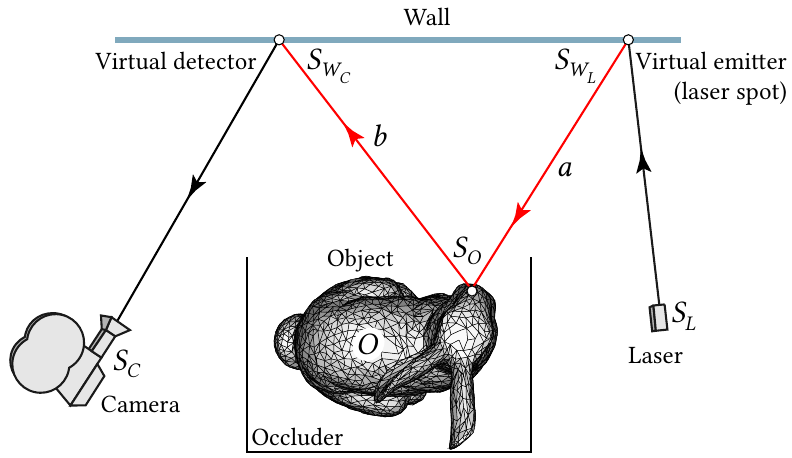}%
\caption{Schematic top view of the scene arrangement, where the unknown object is occluded from direct observation. We assume that the temporal response has been ``unwarped'' (e.g., \cite{Kadambi:2016:OIT:2882845.2836164}), so only the occluded segments $a$ and $b$ contribute to the total time of flight and to the shading in \Eq{eq:rendering}.}%
\label{fig:scenario2d}%
\end{figure}

In this paper, we direct our main attention to the idea of exploiting time-resolved measurements of indirect reflections for the purpose of extending the direct line of sight and, in effect, looking around corners \cite{Kirmani:2009,Velten:2012:Recovering}.
While a variety of geometric settings have been investigated, the bulk of work in this area relies on the arrangement illustrated in \Fig{fig:teaser}(c) and \Fig{fig:scenario2d} and further introduced in the following \Sec{sec:problem}.

The reconstruction strategies can be roughly grouped in two classes.
One major group is formed by backprojection approaches where each input measurement casts votes on those locations in the scene where the light could have been scattered \cite{Velten:2012:Recovering,Laurenzis:2014,Buttafava:2015,Gariepy:2016,Kadambi:2016:OIT:2882845.2836164,ArellanoOpEx2017}.
A smaller but more diverse group of work relies on the use of forward models to arrive at a scene hypothesis that best agrees with the measured data.
Here, reported approaches fall into several categories.
A combinatorial labeling scheme was developed by Kirmani et al.\@ \shortcite{Kirmani:2009}.
If the capture geometry is sufficiently constrained, frequency-domain inverse filtering \cite{otoole2018} can be employed.
Variational methods using simple linearized light transport tensors \cite{Naik:2011,Heide:2014} and simplistic models based on radiative transfer \cite{klein2016tracking,pediredla2017reconstructing} are (in principle) capable of expressing opacity effects like shadowing and occlusion, and physically plausible shading.
These approaches are closest to our proposed method.
In concurrent work, Heide et al.~\shortcite{heideNLOS2017} added such extra factors as additional weights into their least-squares data term, achieving non-line-of-sight reconstructions of significantly improved robustness.
Thrampoulidis et al.~\shortcite{thrampoulidis2017exploiting} applied a similar idea on the reconstruction of 2D albedo maps on known geometry that are further obscured by known occluders between object and wall.
For homogeneous volumetric media in direct sight, Gkioulekas et al.~\shortcite{gkioulekas2013inverse} extensively relied on physically-based rendering to recover their scattering parameters and phase function.
With the proposed method, we demonstrate what we believe is the first reconstruction scheme for non-line-of-sight object geometry that is based on a near-physical yet extremely efficient special-purpose renderer and, by design, produces solutions that are self-consistent.
We believe that our work can serve as an example for other uses of computer graphics methodology as a technical tool for solving inverse problems in imaging and vision.\vspace{0cm} \section{Problem statement}
\label{sec:problem}
Here we introduce the geometry of the non-line-of-sight reconstruction problem as used in the remainder of the paper. For simplicity, we neglect the constant factor $c$ (the speed of light) connecting \emph{time} and \emph{(optical) path length}. Thus, time and distance can be used synonymously and all discussions become independent of the absolute scale. 

\subsection{Problem geometry and transient images}
We model our setting after the most common scenario from literature (\Fig{fig:scenario2d}), where the unknown object is observed indirectly by illuminating a wall with a laser beam and measuring light reflected back to the wall. Following Kadambi et al.~\shortcite{Kadambi:2016:OIT:2882845.2836164}, the laser spot on the wall acts as an area light source, and observed locations on the wall are equivalent to omnidirectional detectors that produce an ``unwarped'' transient image \cite{Velten:2013:FCV:2461912.2461928} (\Fig{fig:teaser}(a)). The extent of the observed wall, the size of the object and its distance to the wall are usually on the same order of magnitude.
The \emph{transient image} or \emph{space-time response} $\vec I\!\in\!\R^{n_x\times n_\tau}$ is the entirety of measurements taken using this setup, $n_x$ being the number of combinations of detector pixels and illuminated spots and $n_\tau$ the number of bins in a time-of-flight histogram recorded per location. For a two-dimensional array of observed locations (for instance, when using a time-gated imager), the space-time response can be interpreted as a three-dimensional data cube similar to a video. 

\subsection{Problem formulation}
The idea underlying ellipsoidal backprojection is that any entry in the transient image, or the response of a pair of emitter and detector positions for a given travel time, corresponds to an ellipsoidal locus of possible candidate scattering locations. If no further information is available, any measured quantity of light therefore ``votes'' for all locations on its ellipsoid. Finally, the sum or product of all such votes is interpreted as occupancy measure, or probability of there being an object at any point in space. We refer to a recent study \cite{lamanna} that discusses the design options for such algorithms in great detail. 

\begin{figure}[t]%
\includegraphics[width=\columnwidth]{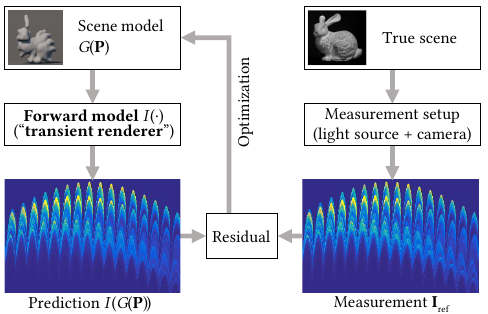}%
\caption{Overview of our analysis-by-synthesis scheme for looking around a corner. Our pipeline heavily relies on custom-made components (scene representation, renderer, residual function, optimizer) to make this approach viable.}%
\label{fig:abys}%
\end{figure}

\hide{
\begin{figure}%
\centering
\includegraphics[width=0.7\columnwidth]{figures/surface-vs-volume.pdf}%
\caption{Our light transport model relies on a surface-based object representation (left) with well-defined, opaque, oriented primitives (with normal vector $\vec n$) that allow for shading and visibility tests including self-occlusion. Most other reconstruction techniques such as backprojection or O'Toole's method \shortcite{otoole2018}, on the other hand, use} a volumetric representation (right) where each voxel carries a scattering density but no information on surface localization and orientation.%
\label{fig:surface-volume}%
\end{figure}
}
In contrast, we formulate the reconstruction task as a non-linear least-squares minimization problem
\begin{equation}
\underset{\P}{\min} \| \Iref - I\left(G\left(\P\right)\right)\|_2^2,
\label{eq:objective}
\end{equation}
where $\vec P$ is a parameter vector describing the scene geometry, $G(\cdot)$ is a function that generates explicit scene geometry (a triangle mesh), $\Iref$ is the measured space-time scene response, and $I(\cdot)$ is a forward model (renderer) that predicts the response under the scene hypothesis passed as argument. The purpose of the optimization is to find the scene geometry $G(\P)$ that minimizes the sum of squared pixel differences between the predicted and the observed responses. \Fig{fig:abys} illustrates this principle.

A key feature of this formulation is that the solution by its very definition is optimally consistent with the chosen physical model of light transport, and that ongoing improvements in forward modeling will also benefit the reconstruction. Furthermore, our approach naturally handles opaque, oriented surfaces, whereas in backprojection, surface geometry is implicitly defined and needs to be derived using additional filtering steps \hide{(\Fig{fig:surface-volume})}. Furthermore, our method is able to handle arbitrary surface BRDFs, where current  backprojection methods implicitly assume diffuse cloud-like scattering \cite{lamanna}. A downside of our approach is that it requires a full model of the scene, and that any unknowns (such as background or noise) can distort the solution in ways that are hard to predict. On the other hand, we believe that our approach lends itself for future extensions like higher-order light bounces.
 \section{Method}
\label{sec:method}
In the following, we introduce the components of our reconstruction algorithm in detail.

\subsection{Geometry representation}
\label{sec:geometry}
\begin{figure}[t]
\includegraphics[width=\columnwidth]{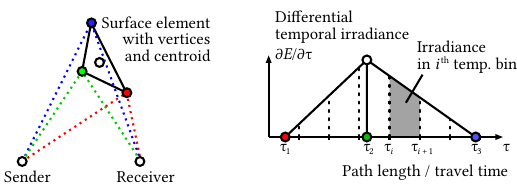}%
\caption{To compute the total irradiance $\alpha_t$ contributed by a surface triangle to a given detector pixel, we evaluate the radiative transfer using the element's centroid. We then use a first-order filter to distribute this irradiance over the temporal bins that are affected by the triangle. To this end, we compute the three optical path lengths, or travel times, $\tau_{1\dots 3}$ belonging to the triangle's three vertices. The irradiance ending up in any temporal bin is then obtained by constructing a triangular function of total area $\alpha_t$ using the three arrival times as illustrated, then geometrically integrating over the  time interval that corresponds to the bin. The true temporal distribution depends on the position and orientation of the triangle. However, the effectiveness of the temporal filter can be seen in \Tab{tab:render_error} and Figures \ref{fig:filtercheck} and \ref{fig:renderererror}.}
\label{fig:filtering}
\end{figure}

We seek to parameterize the scene geometry in terms of a vector $\P$ that has a small number of degrees of freedom to make the optimization in \Eq{eq:objective} tractable. Rather than using $\P$ to directly store a mesh representation with vertices and faces, we express the geometry implicitly as an isosurface of a scalar field $M_\P(\vec x)$ composed of globally supported basis functions. This approach is also common in surface reconstruction from point clouds \cite{carr2001reconstruction}. In our case, the vector $\vec P,$
\begin{equation}
\begin{aligned}
\vec{P} &= (\vec{p}_1, \ldots, \vec{p}_m)\\
&=((\vec{x}_1, \sigma_1), \ldots, (\vec{x}_m, \sigma_m)),
\end{aligned}
\end{equation}
lists the centers $\vec x_i$ and standard deviations $\sigma_i$ of $m$ isotropic 
Gaussian blobs. From the scalar field
\begin{equation}
M_\vec P(\vec{x}) = \sum_{i=1}^m e^{-\|\vec x - \vec{x}_i\|_2^2 / (2 \sigma_i^2)}
\end{equation}
we extract the triangle mesh $G(\vec P)$ using a GPU implementation of Marching Cubes \cite{Lorensen:1987:MCH:37401.37422}. For all our reconstructions, we used a fixed resolution of $128^3$ voxels for the reconstruction volume, and  a fixed threshold of $\frac{3}{4}$ for the isosurface. The extension to other implicit functions, such as anisotropic Gaussians or general radial basis functions, is trivial.

\subsection{Rendering (synthesis)}
\label{sec:rendering}
We propose a custom renderer that is suitable for use as forward model $I(\cdot)$ inside the objective function, \Eq{eq:objective}. In order to be suited for this purpose, the renderer must be sufficiently close to physical reality. At the same time, it has to be very efficient because hundreds of thousands of renderings may be required over the course of the optimization run. We achieve this efficiency by restricting the renderer to a single type of light path and rendering only light bounces from the wall to the object and back to the wall.
Following the notation of \cite{Pharr:2010:PBR:1854996} and by dropping any constant terms, we can write the incoming radiance for each camera pixel as
\begin{equation}
\label{eq:rendering}
L = {\int_{O}} {f \left( S_{W_L} {\rightarrow} S_O {\rightarrow} S_{W_C} \right)} {\eta\left( S_O {\leftrightarrow} S_{W_C} \right)} {\eta \left( S_{W_L} {\leftrightarrow} S_O \right)} \text{d}S_O,
\end{equation}
where $O = G(\vec{P})$ denotes the object, $f$ the object's BRDF and $S_{\_}$ surface points as shown in \Fig{fig:scenario2d}. The geometric coupling term $\eta$ is defined as
\begin{equation}
\eta \left( S_1 {\leftrightarrow} S_2 \right) = V \left( S_1 {\leftrightarrow} S_2 \right) \frac{\left|\cos(\theta_1)\right| \left|\cos(\theta_2)\right|}{\norm{S_1 - S_2}^2},
\end{equation}
with $V$ being the binary visibility function and $\theta_i$ the angle of the ray connecting $S_1$ and $S_2$ to the respective surface normal. Since our object is already represented as a triangle mesh, we are able to approximate \Eq{eq:rendering} by assuming a constant radiance over each triangles' surface,
\begin{equation}
\begin{aligned}
L &\approx \sum_{t \in T} {f \left( S_{W_L} {\rightarrow} S_t {\rightarrow} S_{W_C} \right)} {\eta\left( S_t {\leftrightarrow} S_{W_C} \right)} {\eta \left( S_{W_L} {\leftrightarrow} S_t \right)} A_t \\
&\eqqcolon \sum_{t \in T}\alpha_t. 
\end{aligned}
\end{equation}
Here, $T$ is the set of all triangles of our object, $P_t$ is the centroid, and $A_t$ the area. We denote the total irradiance contributed by triangle $t$ as $\alpha_t$. In our experiments, we use Lambertian and metal BRDFs, but other reflectance functions can be used as well. This approximation can be seen as an extension of the one found in \cite{klein2016tracking}. We further add two important features to increase physical realism and generate a smooth transient image.

Our first addition are visibility tests ($V$) for both segments of the light path, which is necessary for handling non-convex objects. We first connect the laser point and the triangle centroid by a straight line, and test whether this segment intersects with any of the other triangles of the object mesh. For all visible triangles for which no intersection is found, we test the visibility of the second path segment (return of scattered light to the wall) in the same way.
This shadow test avoids overestimation of backscatter from self-occluding object surfaces. We note, however, that our way of performing the test only for the triangle centroid leads to a binary decision (triangle entirely visible or entirely shadowed) and therefore potentially makes the objective non-continuous. This can be reduced by using a triangle grid of sufficiently high resolution.

\begin{figure}[t]
\begingroup
\setlength{\tabcolsep}{1pt}
\begin{tabular}{lll}
\includegraphics[width=0.32\columnwidth]{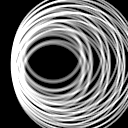}&%
\includegraphics[width=0.32\columnwidth]{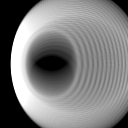}&%
\includegraphics[width=0.32\columnwidth]{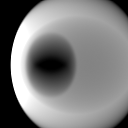}\\[-5mm]
\,\whitebox{\unit{5.44}{ms}}&\,\whitebox{\unit{19.02}{ms}}&\,\whitebox{\unit{1032.2}{ms}}\\[0.3mm]%
\includegraphics[width=0.32\columnwidth]{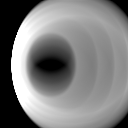}&%
\includegraphics[width=0.32\columnwidth]{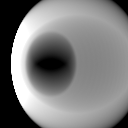}&%
\includegraphics[width=0.32\columnwidth]{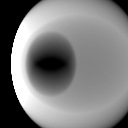}\\[-5mm]
\,\whitebox{\unit{6.36}{ms}}&\,\whitebox{\unit{21.99}{ms}}&\,\whitebox{\unit{1181.5}{ms}}
\end{tabular}%
\endgroup
\caption{The temporal filter also results in overall smoother spatial slices of the space-time response. Here we verify the performance of the filter by rendering the response generated by a planar square using different levels of detail. Shown is a single time slice without (top row) and with temporal filtering (bottom row). From left to right: coarse tesselation (4$\times$4 quads), medium tesselation (16$\times$16 quads), fine tesselation (128$\times$128 quads). Numbers indicate the rendering time for the entire transient data cube (128$\times$128 pixels, 192 time bins) on an NVIDIA GTX 980. Note the significant quality improvement at only 14--17\% increased computational cost.}
\label{fig:filtercheck}%
\end{figure}

To render a transient image, we extend the pixels of the steady-state renderer to record time-of-flight histograms. The light contribution $\alpha_t$ enters into this histogram according to the geometric length of the corresponding light path; this length is simply the sum of the two Euclidean distances from laser point to point on triangle and back to the receiving point on the wall (see \Fig{fig:scenario2d}). We found that the temporal response is prone to artifacts if only the centroid of the triangle is taken into account for the path length. Instead, we use the path lengths for the triangle's three corner vertices to determine the temporal footprint of the surface element. Using a linear filter, we then distribute the contribution $\alpha_t$ over the temporal domain (\Fig{fig:filtering}). This procedure ensures that the rendered outcome is smooth in the temporal and spatial domains even when a single surface element covers dozens of temporal bins (\Fig{fig:filtercheck}).

\begin{figure*}[t]
\begingroup
\setlength{\tabcolsep}{1pt}
\begin{tabular}{cccccccc}
\includegraphics[width=0.12\textwidth]{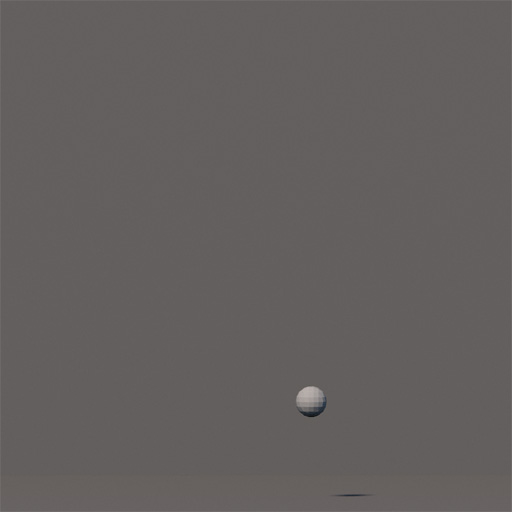} &
\includegraphics[width=0.12\textwidth]{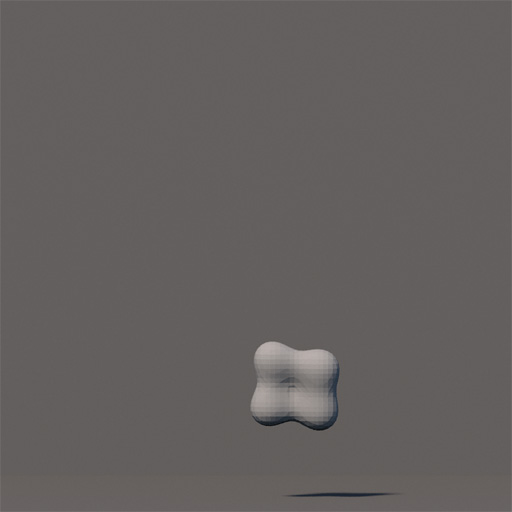} &
\includegraphics[width=0.12\textwidth]{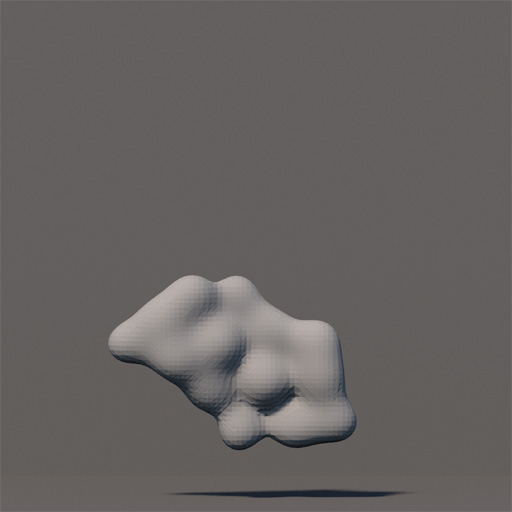} &
\includegraphics[width=0.12\textwidth]{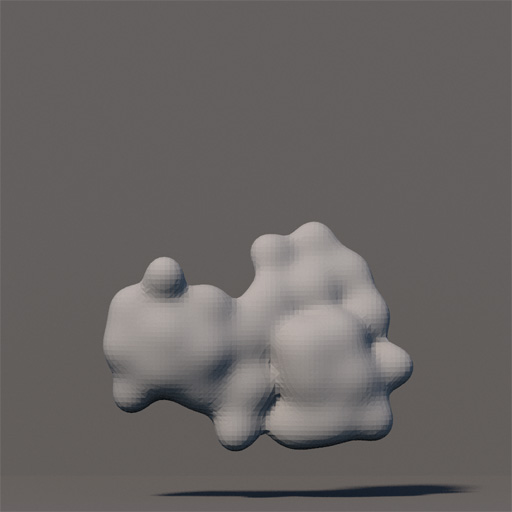} &
\includegraphics[width=0.12\textwidth]{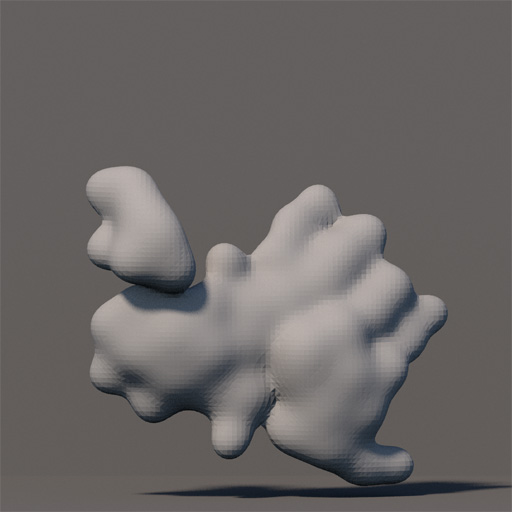} &
\includegraphics[width=0.12\textwidth]{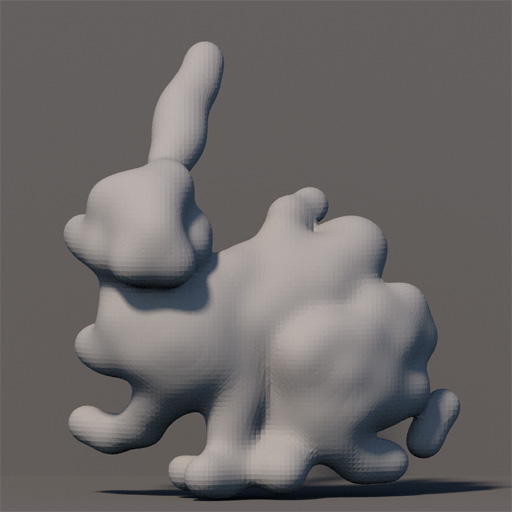} &
\includegraphics[width=0.12\textwidth]{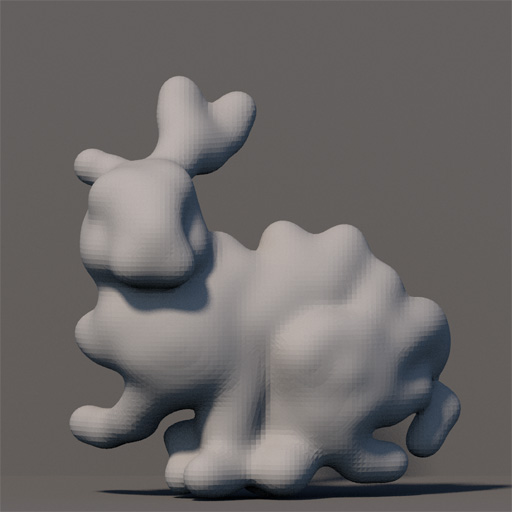} &
\includegraphics[width=0.12\textwidth]{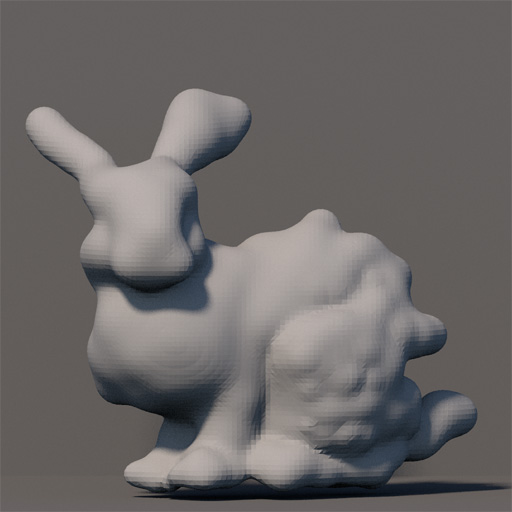} \\
\small 001: 9.88e$-$01 &
\small 004: 8.66e$-$01 &
\small 014: 3.90e$-$01 &
\small 026: 1.88e$-$01 &
\small 044: 5.93e$-$02 &
\small 098: 9.93e$-$03 &
\small 146: 8.43e$-$03 &
\small 534: 3.98e$-$03 
\end{tabular}
\endgroup
\caption{Convergence of reconstructed geometry for the \texttt{Bunny} dataset over the course of the optimization. Number pairs denote  iteration number and value of cost function (relative to start value).}%
\label{fig:convergence}%
\end{figure*}

\subsection{Optimization (analysis)}
The optimization problem in \Eq{eq:objective} is non-convex and non-linear, so special care has to be taken to find a solution (a set of blobs) that, when rendered, minimizes the cost function globally. While it would be desirable to optimize over the whole parameter vector $\P$ simultaneously, this is computationally prohibitive. To address this problem, we developed the iterative optimization scheme summarized in \Alg{alg:global}, with subroutines provided in \Algs{alg:inner} and \ref{alg:subroutines}. \Fig{fig:convergence} shows several intermediate results during execution of the optimization scheme.

\begin{algorithm}
\caption{Global optimization scheme}
\label{alg:global}
\begin{algorithmic}[1]
\REQUIRE Reference image $\Iref$, Threshold $c_\text{thresh}$
\ENSURE Parameter vector $\P$, Cost $c$
\STATE $\vec{x} \gets \textsc{sample}(\varnothing)$
\STATE $\P,c  \gets  \textsc{add\_blob}(\varnothing, \x)$
\WHILE {$c > c_\text{thresh}$}
\STATE $\vec{x} \gets \textsc{sample}(\P)$
\STATE $\P_1, c_1 \gets \textsc{add\_blob}(\P, \x)$
\STATE $\P_2, c_2 \gets \textsc{duplicate\_blob}(\P, \x)$
\STATE $\P_3, c_3 \gets \textsc{delete\_blob}(\P, \x)$
\STATE $i \gets \argmin_x c_x$
\IF {$c_i < c$}
\STATE $\P, c \gets \P_i, c_i$
\ENDIF
\STATE $\P_r,c_r \gets \textsc{reiterate}(\P)$
\IF {$c_r < c$}
\STATE $\P, c \gets \P_r, c_r$
\ENDIF
\STATE $\P,c \gets \textsc{check\_delete}(\P,c)$
\ENDWHILE
\end{algorithmic}
\end{algorithm}

\begin{algorithm}
\caption{Inner optimization scheme}
\label{alg:inner}
\begin{algorithmic}[1]
\STATE $\textbf{function } \textsc{iterate}(\p, \P)$
\begin{ALC@g}
\STATE $\Popt \gets \textsc{find\_neighbors}(\p, \P, 10)$
\STATE $\textsc{set\_fixed}(\P)$
\FORALL {$(\tilde{\p}, \tilde{\sigma}) \in \Popt$}
\STATE $\textsc{set\_variable}(\tilde{\p})$
\ENDFOR
\STATE $\P \gets \textsc{levenberg\_marquardt}(\P)$
\FORALL {$(\tilde{\p}, \tilde{\sigma}) \in \Popt$}
\STATE $\textsc{set\_variable}(\tilde{\p})$
\STATE $\textsc{set\_variable}(\tilde{\sigma})$
\ENDFOR
\STATE $\P \gets \textsc{levenberg\_marquardt}(\P)$
\STATE $c \gets \textsc{compute\_cost}(\P)$
\STATE $\textbf{return } \P, c$
\end{ALC@g}
\end{algorithmic}
\end{algorithm}

\begin{algorithm}
\caption{Subroutines to \Alg{alg:global}.}\label{alg:subroutines}
\begin{algorithmic}[1]
\STATE $\textbf{function } \textsc{add\_blob}(\P, \vec{x})$
\begin{ALC@g}
\STATE $\vec{p} \gets (\vec{x}, \sigma_0)$
\STATE $\textbf{return } \textsc{iterate}(\vec{p}, \P \cup \p)$
\end{ALC@g}
\end{algorithmic}
\begin{algorithmic}[1]
\STATE $\textbf{function } \textsc{check\_delete}(\P)$
\begin{ALC@g}
\FORALL{$\vec{p} \in \P$}
\IF {$\textsc{compute\_cost}(\P \setminus \vec{p}) < \eta \cdot c$}
\STATE $\P \gets \P \setminus \vec{p}$
\ENDIF
\ENDFOR
\STATE $c \gets \textsc{compute\_cost}(\P)$
\STATE $\textbf{return } \P, c$
\end{ALC@g}
\end{algorithmic}
\begin{algorithmic}[1]
\STATE $\textbf{function } \textsc{duplicate\_blob}(\P, \vec{x})$
\begin{ALC@g}
\STATE $\p \gets \textsc{find\_nearest}(\P, \x)$
\STATE $\p_1, \p_2 \gets \textsc{split}(\p)$
\STATE $\textbf{return } \textsc{iterate}(\p, \P \setminus \p \cup \p_1 \cup \p_2)$
\end{ALC@g}
\end{algorithmic}
\begin{algorithmic}[1]
\STATE $\textbf{function } \textsc{reiterate}(\P)$
\begin{ALC@g}
\STATE $\p \gets \textsc{choose\_random}(\P)$
\STATE $\textbf{return } \textsc{iterate}(\p, \P)$
\end{ALC@g}
\end{algorithmic}
\begin{algorithmic}[1]
\STATE $\textbf{function } \textsc{remove\_blob}(\P, \x)$
\begin{ALC@g}
\STATE $\p \gets \textsc{find\_nearest}(\P, \x)$
\STATE $\textbf{return } \textsc{iterate}(\p, \P \setminus \p)$
\end{ALC@g}
\end{algorithmic}
\end{algorithm}

The heart of our optimization algorithm is the inner optimization loop \textsc{iterate}($\p$, $\P$), which determines the $k=10$ nearest neighbors of a given pivot blob $\p$ using the routine \textsc{find\_neighbors}($\p$, $\P$). It then optimizes the \emph{positions} of those blobs using the Levenberg-Marquardt algorithm, \textsc{levenberg\_marquardt}($\P$) \cite{levenberg,marquardt}. The function \textsc{set\_variable}($\vec{x}$) is used to label these parameters as variable to the solver, while all other blobs are kept fixed during the optimization run using \textsc{set\_fixed}($\vec{x}$). Derivatives for the Jacobian matrix are computed numerically using finite differences (by repeatedly executing our forward renderer with the perturbed parameter vector). In a subsequent step, the sizes of the selected blobs are also included in a second optimization run, with a parameter $\sigma_{\text{max}}$ defining an upper limit for the blob size. We found that this two-stage approach is necessitated by the strong non-convexity of the objective function. {By optimizing over multiple blobs simultaneously, we allow the optimizer to recover complex geometry features that are influenced by more than a single blob.} 

The algorithm starts with a single blob as initial solution, then performs an outer loop over four phases: sampling, mutation, reiteration, and regularization. In the following, we provide a full description of the individual phases and explain our design choices. The parameters used in our reconstructions are shown in \Tab{tbl:parameters}.

\paragraph{Sampling. }
Our algorithm pivots around locations in the reconstruction volume that are randomly chosen according to a distribution (PDF) that aims to give problematic regions a higher probability of being sampled. We obtain the PDF by backprojecting the absolute value of the current residual image into the working volume. For locations $\x$ that are sampled by the function \textsc{sample}(), our working hypothesis is that \emph{something} about the solution should change there; we address this by selecting the nearest blob to this location (\textsc{find\_nearest}($\P$, $\x$)) and applying and testing our three mutation strategies on it. Since each mutation probably increases the cost function, it is followed by a relaxation of the neighborhood of the pivot blob.

\paragraph{Mutation. }
We employ three mutation strategies to generate variations of the current solution. \textsc{add\_blob}($\P$, $\x$) adds a new blob $(\x, \sigma_0)$ to $\P$.
\textsc{delete\_blob}($\P$, $\x$) deletes the blob $\p \in \P$ that is closest to $\x$.
\textsc{duplicate\_blob}($\P$, $\x$)  replaces the blob $\p \in \P$ by two new blobs that are displaced by a vector $\pm\vec d$ from the original position so they can be separated by the optimizer. 
Out of the three solutions (each one after performing an inner optimization \textsc{iterate}($\p$, $\P$) on the neighborhood), the one with the lowest cost $c_i$ is chosen to be the new solution.
A call to \textsc{iterate} consists of two non-linear optimizations, one solely over the blob positions, followed by an optimization over both blob positions and sizes. This procedure is essential due to the non-convexity of the cost function, initial experiments have shown that skipping the first optimization generally results in unwanted, strong local minima, where a single blob spans large parts of the reconstruction volume.

\paragraph{Reiteration. }
As the next step, another call to \textsc{iterate} is performed on a random group of neighboring blobs. This re-evaluation of previously relaxed blobs is necessary to avoid being stuck in local minima during early iterations, when the hypothesis does not yet contain enough blobs to properly describe the transient response.

\paragraph{Regularization. }
Finally, the algorithm first checks each blob for its significance to the solution (\textsc{check\_delete}), and deletes it if doing so does not worsen the total cost by more than a small factor $\eta$. This regularizing step prevents the build-up of excess geometry in hidden regions that is not supported by the data.
It is the only step that can lead to an increase in the cost $c$; all other heuristics ensure that the cost falls monotonically. 

\subsection{Implementation details} 
Our reconstruction software is written in C\texttt{++}. Geometry generation and rendering are implemented on the GPU, using NVIDIA CUDA and the Thrust parallel template library for the bulk of the tasks and the NVIDIA OptiX prime ray-tracing engine for the shadow tests. The optimization algorithm is implemented using the Ceres solver \cite{ceres-solver}. Intermediate results are visualized on-the-fly using the VTK library \cite{VTK4}. {We used various workstations in our experiments, with Intel Core i7 CPUs and NVIDIA GeForce GPUs ranging from GTX 780 to Titan Xp.}
 \section{Evaluation}\label{sec:eval}
In this section, we verify the correctness of our renderer, and use it to reconstruct geometry from simulations and experimental measurements of around-the-corner scattered light. Input data, \changed{relay wall parameters, scene geometry,} as well as output volumes and meshes of our proposed method and the state-of-the-art ellipsoidal backprojection method of \cite{ArellanoOpEx2017} can be found in the supplemental material.

\subsection{\changed{Scene geometry}}
\label{sec:scene_geometry}
\changed{All our synthetic experiments were conducted in a consistent scene setup and our} models use the same arbitrary unit for length and time.
The standard temporal resolution (size of histogram bin) of our virtual detectors is 0.4 units.
Typical time resolutions of real-world devices are \unit{10}{ps} for streak cameras or \unit{100}{ps} for SPAD detectors.
Equating the bin size with these time constants results in a conversion factor to real-world distances of \changed{\unit{7.5}{mm}} and \changed{\unit{75}{mm}} per world unit, respectively.
We arranged the scene such that the \changed{relay} wall is a diffuse plane at $z=0$ with normal in positive $z$ direction.
The object, with a typical size of 50 units, was located on the $z$ axis at $z=45$.
The laser spot was modeled as a cosine-lobe light source pointing in positive $z$ direction at one of four wall locations $(45,0,0)$, $(-45,0,0)$, $(0,45,0)$ and $(0,-45,0)$.
The range of observed points on the wall was represented by an area of $80\times80$\,units$^2$ which was observed by an orthographic camera centered at $(x,y)=(0,0)$.
\changed{The synthetic scene parameters can also be found in \Tab{tbl:parameters} and the capture geometry of the measured scenes can be found in the respective publications \cite{Buttafava:2015,otoole2018}.}

\subsection{Correctness of renderer}
\label{sec:rendereval}
Before we evaluate the performance of our overall reconstruction system, we test correctness and performance of the forward model that is at its heart, our custom renderer. To this end, we prepare test scenes and render reference images using Microsoft's Time of Flight Tracer \cite{toftracer}, a transient renderer based on \texttt{pbrt} version 2 \cite{Pharr:2010:PBR:1854996}.

\begin{figure}[t]
\begin{center}
\includegraphics[height=0.33\linewidth]{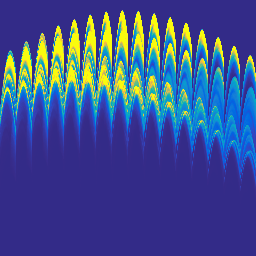}\hfill
\includegraphics[height=0.33\linewidth]{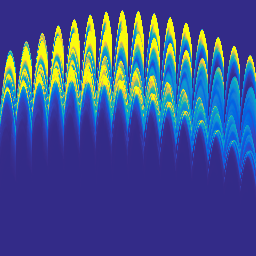}\hfill\includegraphics[height=0.33\linewidth]{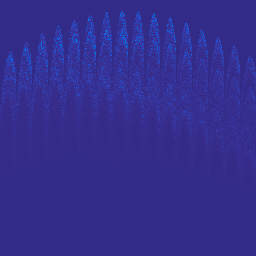}
\end{center}
\caption{The physically-based renderings with and without global illumination are virtually indistingishuable. From left to right: Rendering with global illumination; rendering without global illumination; difference of the two renderings.}
\label{fig:2bouncevsfull}
\end{figure}

Using a $30 \%$ reflective triangle mesh model of the Stanford Bunny, we generated two reference renderings of $16\times 16\times 256$ spatio-temporal resolution using the physically-based renderer, one with full global illumination and one with a maximum path length of 2 reflections. With the cosine light source representing the spot lit by the laser, a path length of 2 includes light scattering from the wall to the object and back to the wall, but not light that has been interreflected at the object or that has bounced between object and wall multiple times. 
In \Fig{fig:2bouncevsfull}, both versions are shown along with the difference. At least for our around-the-corner setting, we found that the error caused by truncating the path length to 2 is not very significant, with \unit{69.809}{\deci\bel} peak signal-to-noise ratio (PSNR) or a relative $L2$ difference of $0.486 \%$.

\begin{figure}[ht]
\includegraphics[width=0.24\columnwidth,trim=0 1.5cm 0 2mm,clip]{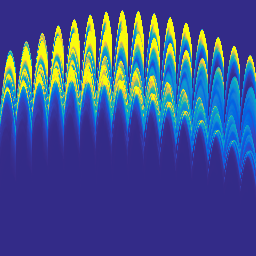}\hfill%
\includegraphics[width=0.24\columnwidth,trim=0 1.5cm 0 2mm,clip]{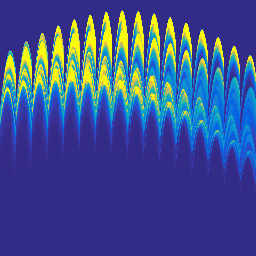}\hfill%
\includegraphics[width=0.24\columnwidth,trim=0 1.5cm 0 2mm,clip]{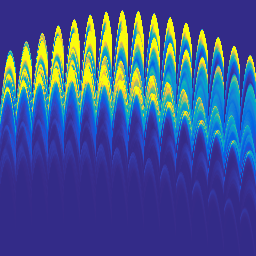}\hfill%
\includegraphics[width=0.24\columnwidth,trim=0 1.5cm 0 2mm,clip]{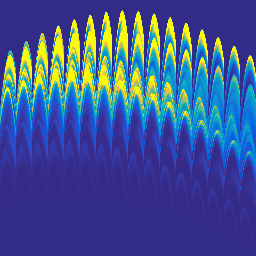}
\\[1mm]
\includegraphics[width=0.24\columnwidth,trim=0 1.5cm 0 2mm,clip]{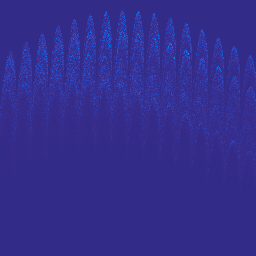}\hfill%
\includegraphics[width=0.24\columnwidth,trim=0 1.5cm 0 2mm,clip]{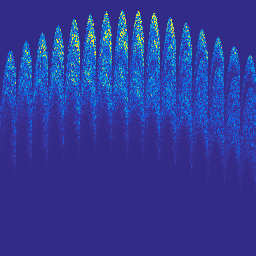}\hfill%
\includegraphics[width=0.24\columnwidth,trim=0 1.5cm 0 2mm,clip]{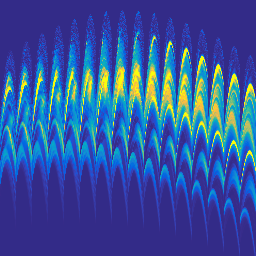}\hfill%
\includegraphics[width=0.24\columnwidth,trim=0 1.5cm 0 2mm,clip]{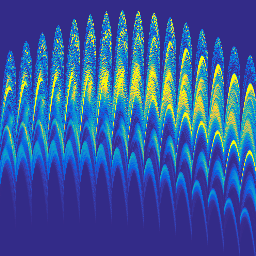}
\caption{The effect of our augmentations on the rendering error. The top row shows transient renderings made with our renderer, the bottom row shows the respective difference to the ground truth \texttt{toftracer} rendering (range scaled for print). From left to right: Our renderer with all features turned on; temporal filtering turned off; shadow tests turned off; temporal filtering and shadow tests turned off. Error metrics for these renderings are provided in \Tab{tab:render_error}.}
\label{fig:renderererror}
\end{figure}

\begin{figure}[ht]
\includegraphics{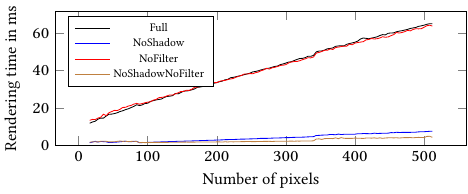}
\caption{Rendering performance of four versions of our algorithm (with/without filtering, with/without shadow test) as a function of output pixel count.}
\label{fig:rendererTimings}%
\end{figure}

\begin{figure}[ht]
\includegraphics{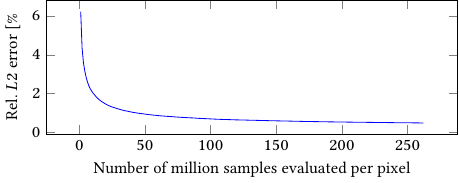}
\caption{Difference between our fast renderer and the ray-traced reference solution with a varying number of samples per pixels.}
\label{fig:toftracerConvergence}%
\end{figure}

\begin{table}[ht]%
\small\centering
\begin{tabular}{ll|r|r}
\multicolumn{2}{l|}{\bf Comparison} & {PSNR [dB]} & {Rel. $L2$ error [$\%$]} \\
\hline
RTFull &/ RTTrunc & $69.809$ & $0.486$ \\
RTTrunc &/ OursFull & $69.796$ & $0.489$ \\
RTTrunc &/ OursNoFilter & $53.379$ & $3.237$ \\
RTTrunc &/ OursNoShadow & $45.638$ & $7.892$ \\
RTTrunc &/ OursNoShadowNoFilter & $44.942$ & $8.550$ 
\end{tabular}
\caption{Using the Stanford Bunny as test object, we compare our renderer to ray-traced renderings with maximum path lengths of 2 (RTTrunc) and $\infty$ (RTFull). With all the features enabled (OursFull), our renderer matches the ray-traced solution for the 3-bounce setting (wall-object-wall) to $0.49 \%$, which is on the same order as the influence of global illumination (RTFull) on this scene. Omission of shadow tests and temporal filtering result in significantly higher error values. }
\label{tab:render_error}
\end{table}

\begin{figure*}[ht]
\begin{center}
\newcommand{\etsize}{0.13}
\begin{tabular}{cc|cc|ccc}
\multicolumn{2}{c}{\texttt{Bunny}} &
\multicolumn{2}{c}{\texttt{Mannequin1Laser}} & 
\multicolumn{2}{c}{\texttt{Mannequin}} \\
\multicolumn{2}{c}{\includegraphics[width=0.22\linewidth]{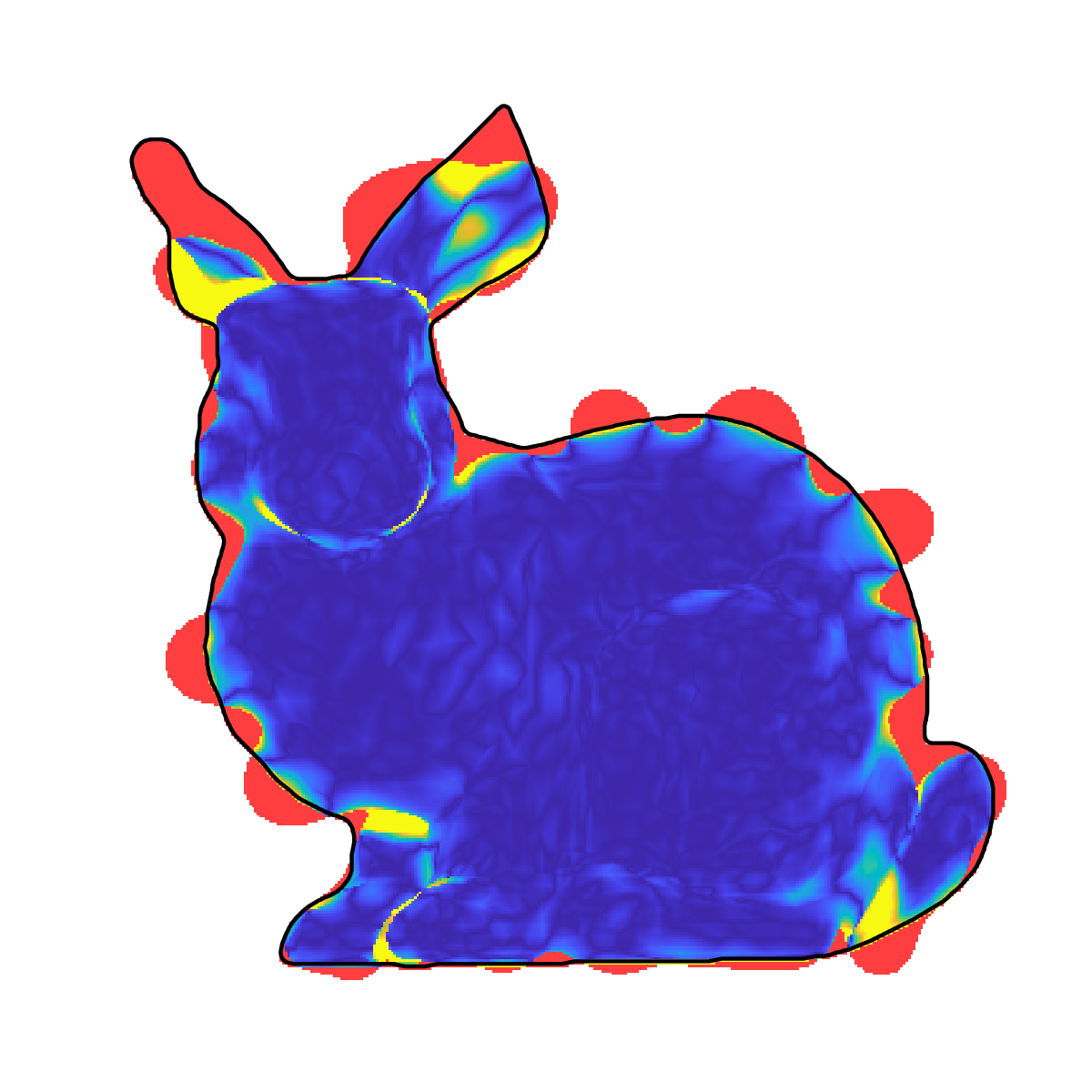}} &
\multicolumn{2}{c}{\includegraphics[width=0.22\linewidth]{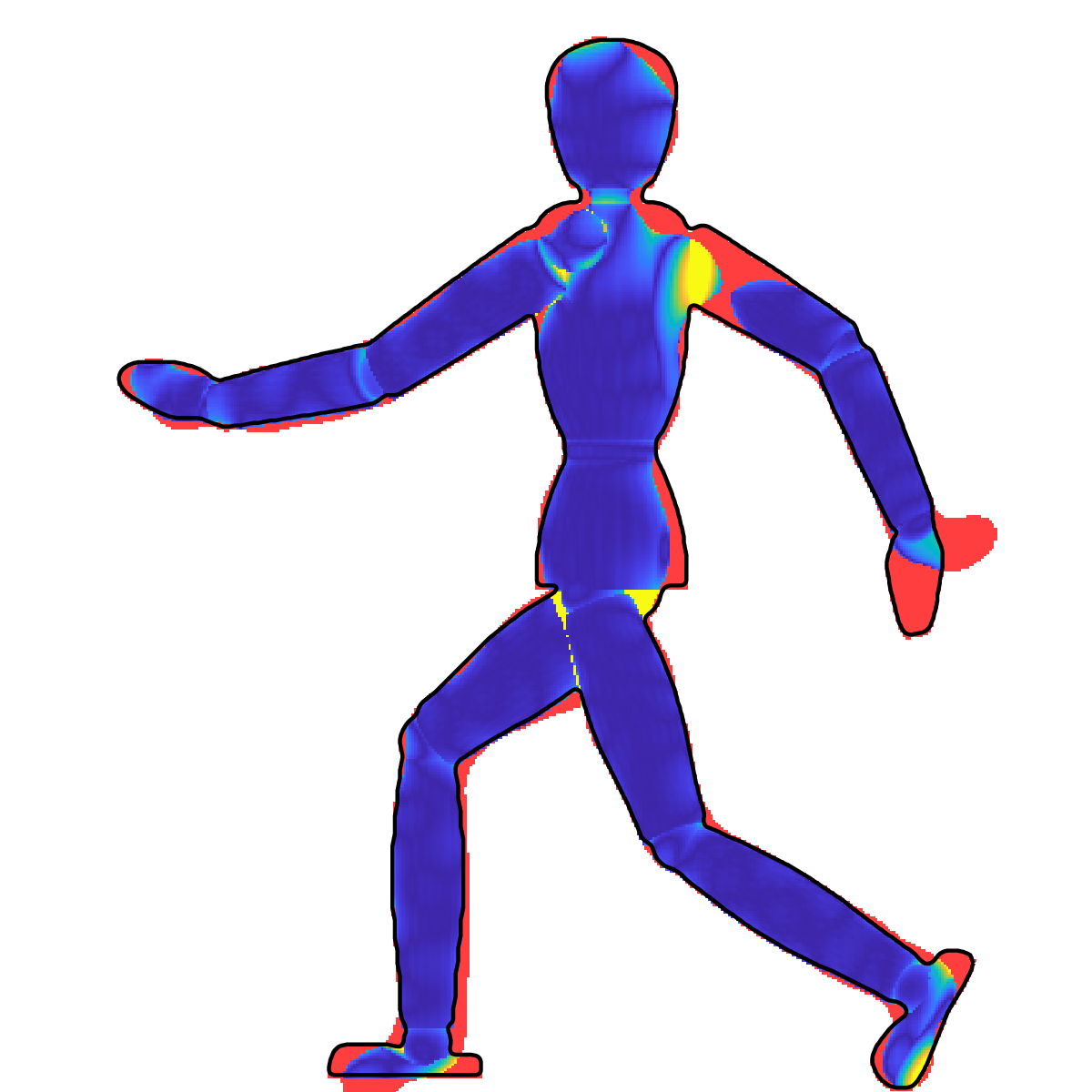}} &
\multicolumn{2}{c}{\includegraphics[width=0.22\linewidth]{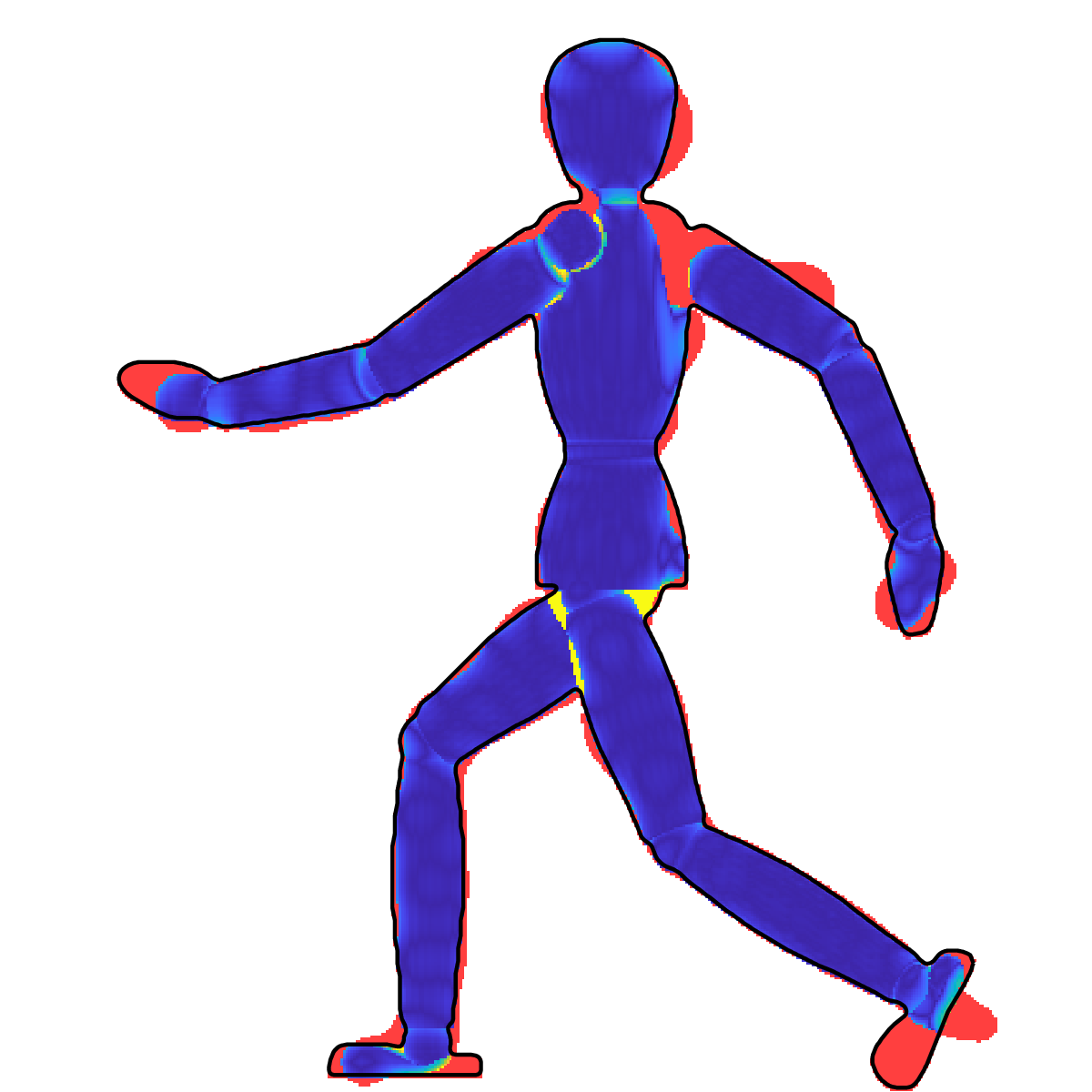}}\\
\multicolumn{2}{c}{ours} &
\multicolumn{2}{c}{ours} &
\multicolumn{2}{c}{ours}\\
\cline{1-6} \\
\includegraphics[width=\etsize\linewidth]{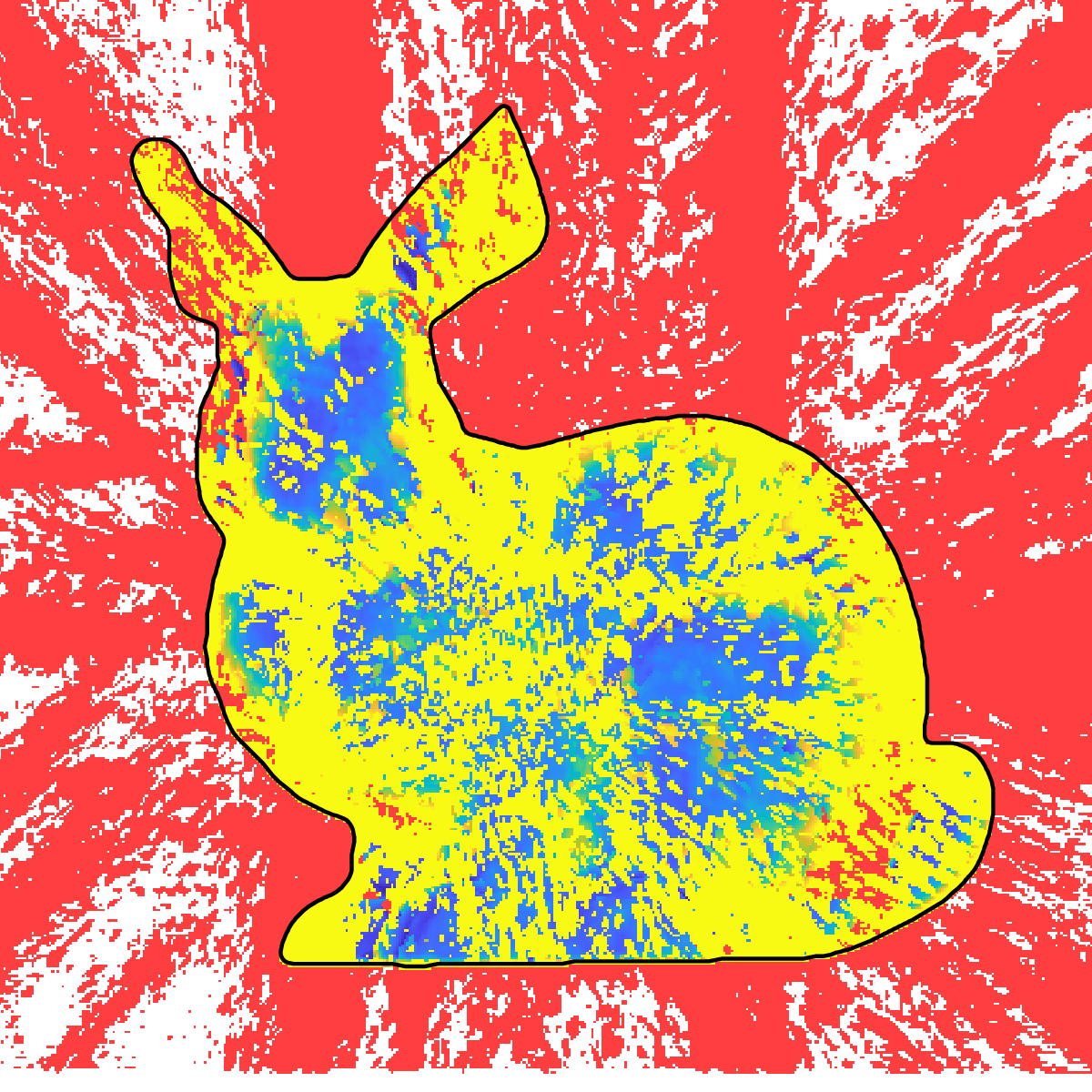} &
\includegraphics[width=\etsize\linewidth]{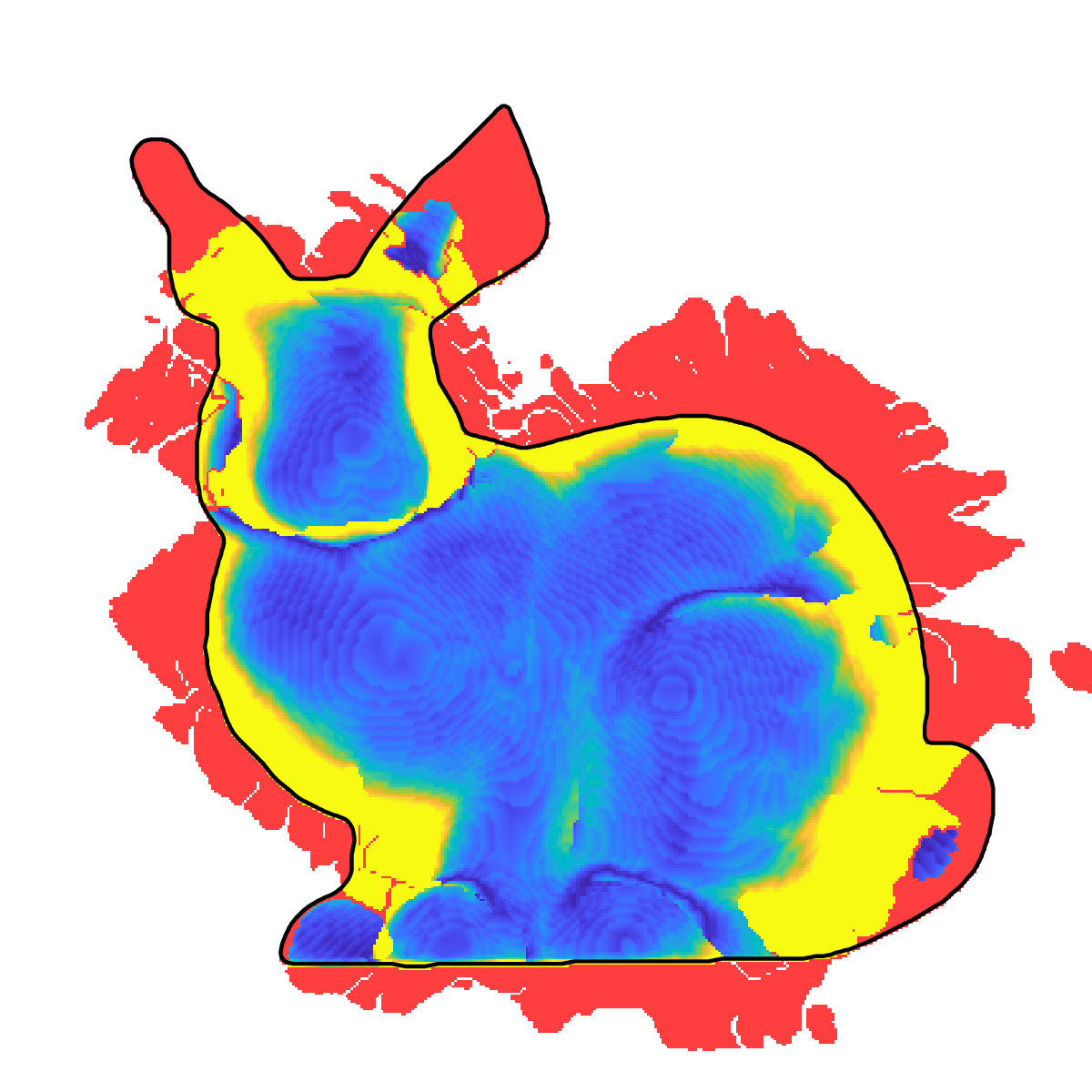} &
\includegraphics[width=\etsize\linewidth]{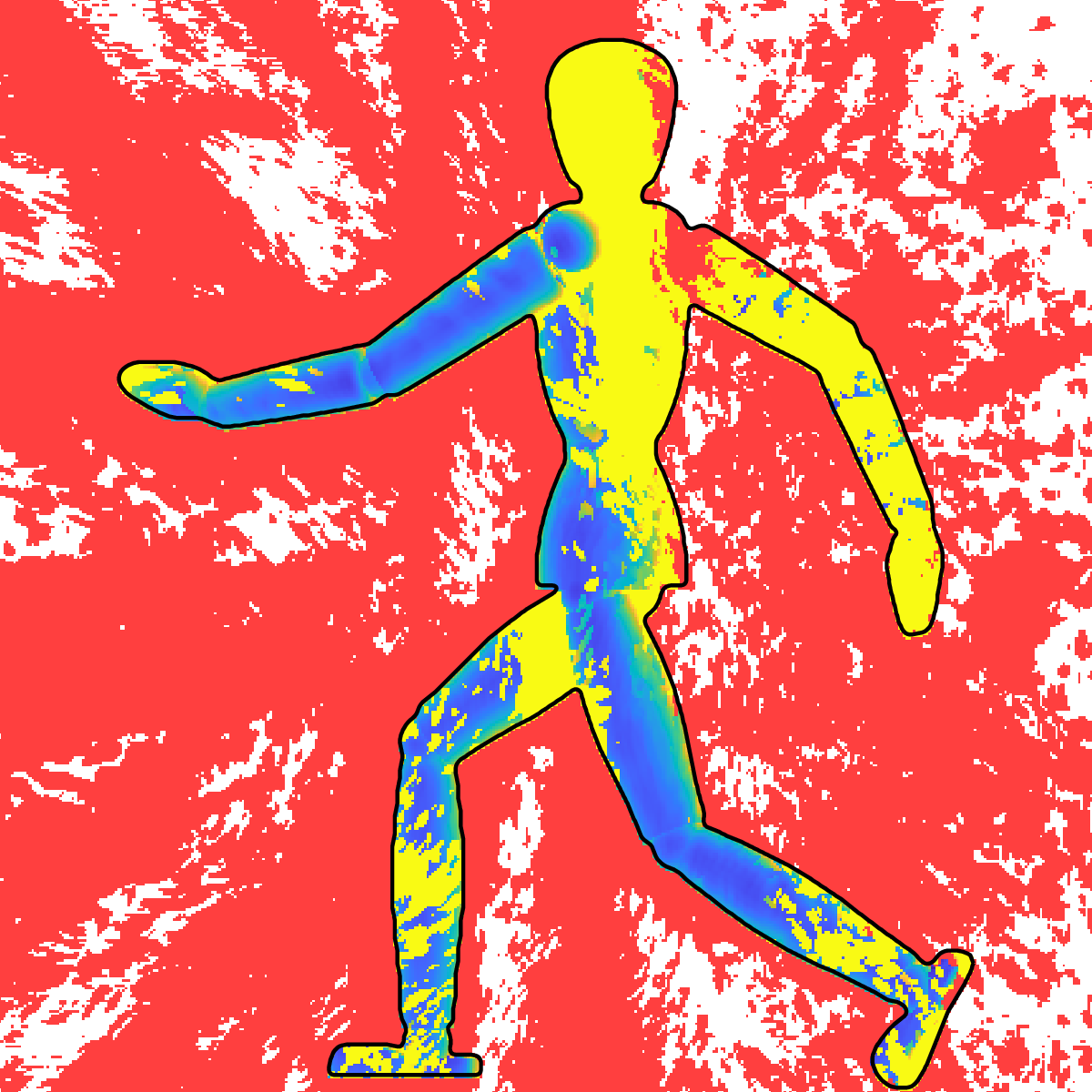} &
\includegraphics[width=\etsize\linewidth]{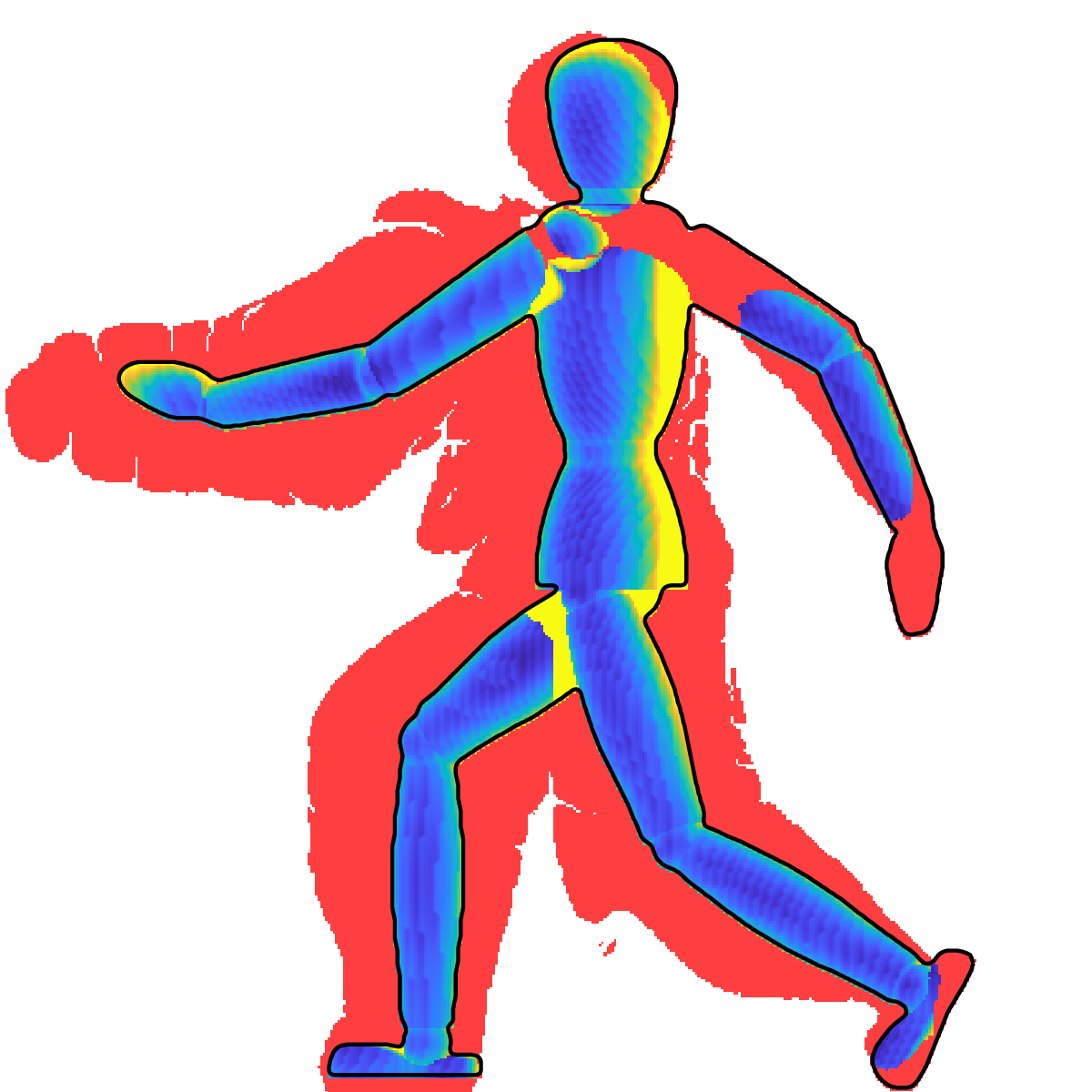} &
\includegraphics[width=\etsize\linewidth]{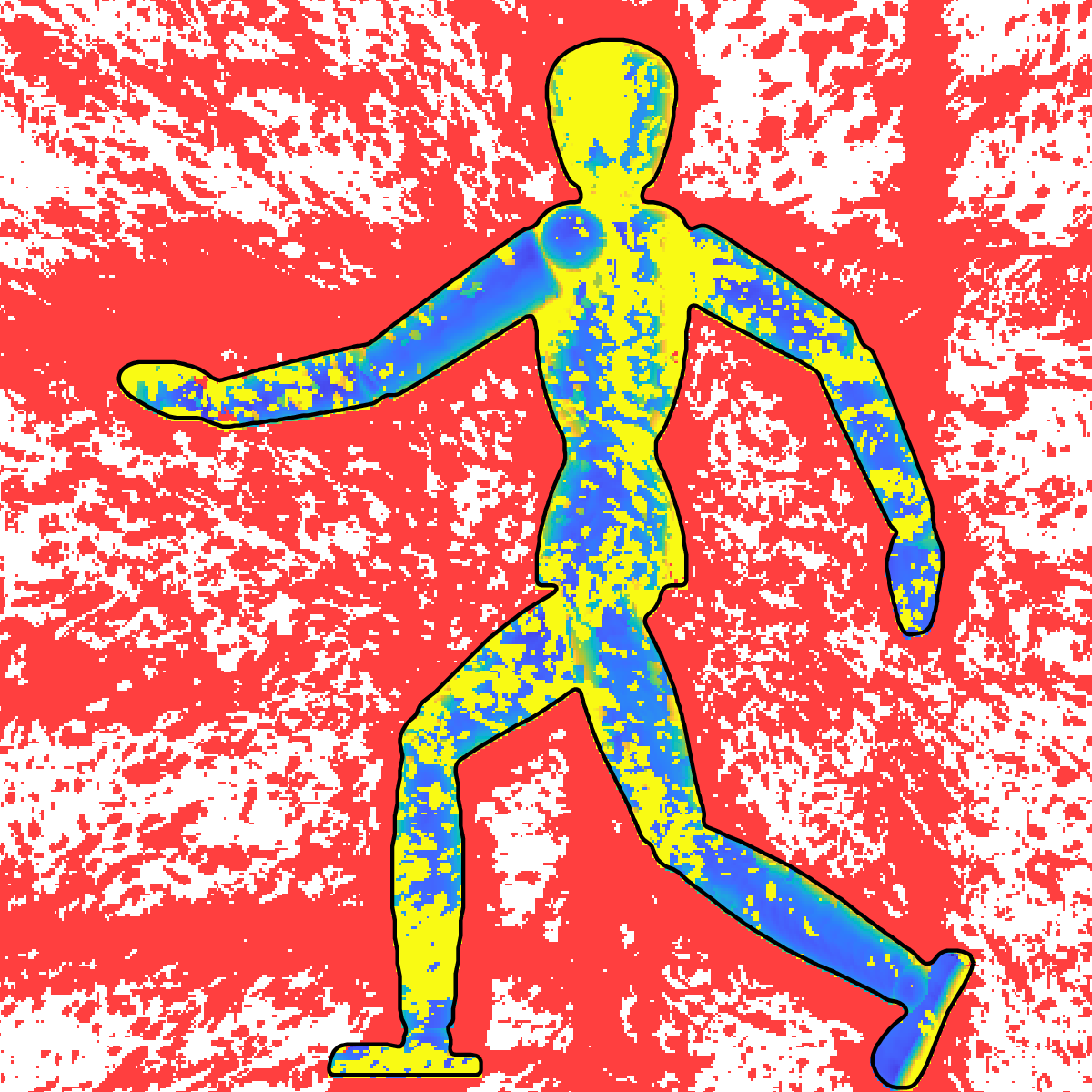} &
\includegraphics[width=\etsize\linewidth]{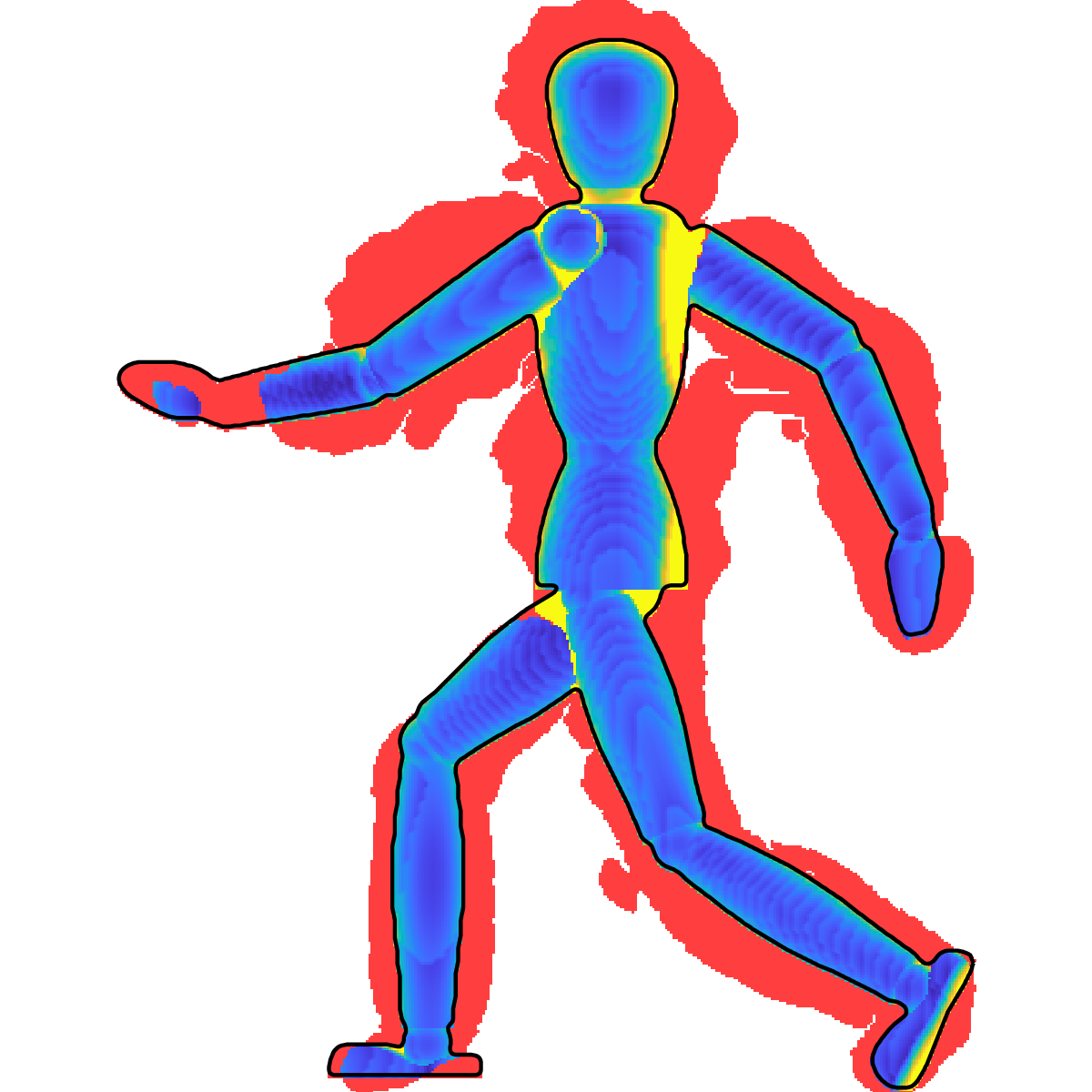} \\
\includegraphics[width=\etsize\linewidth]{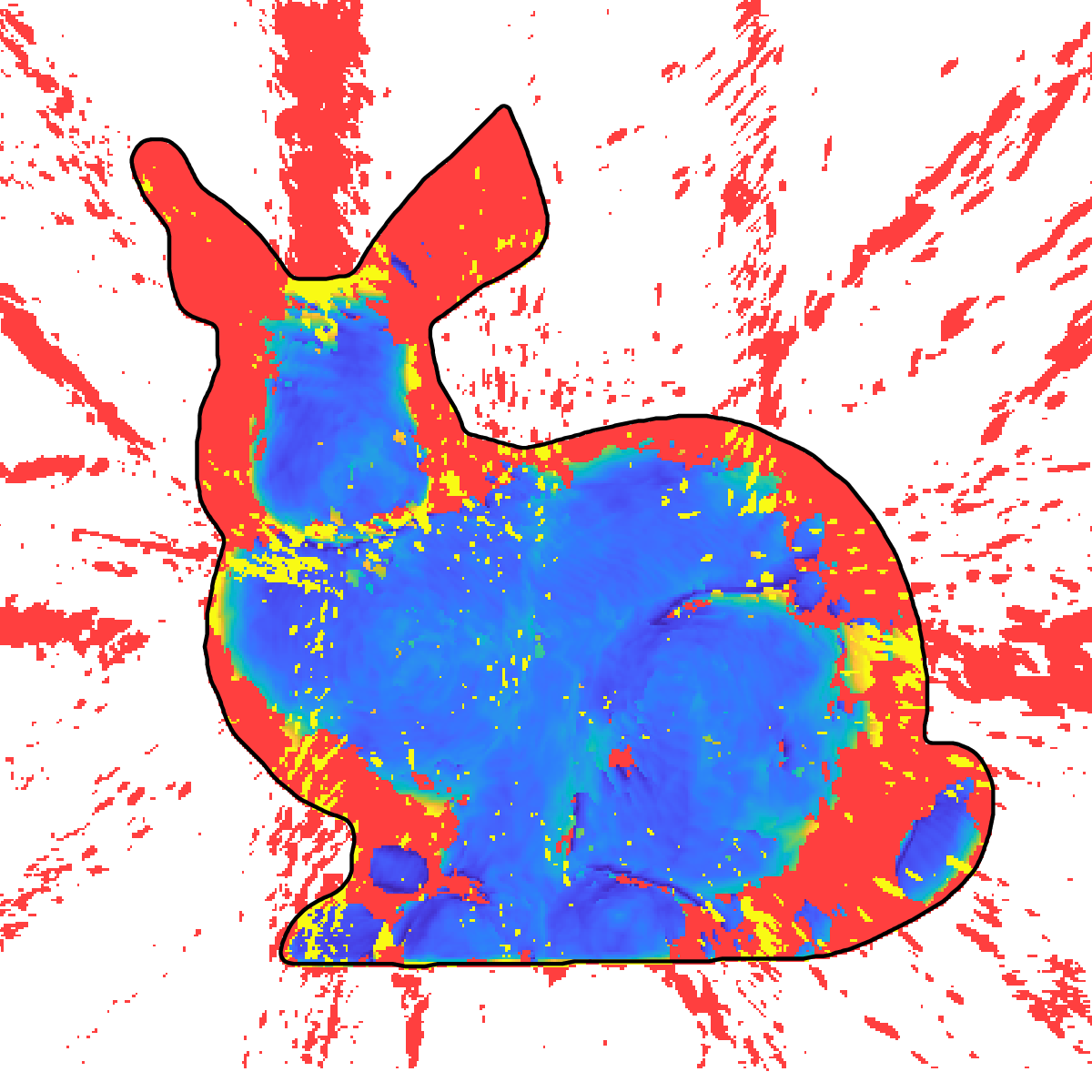} &
\includegraphics[width=\etsize\linewidth]{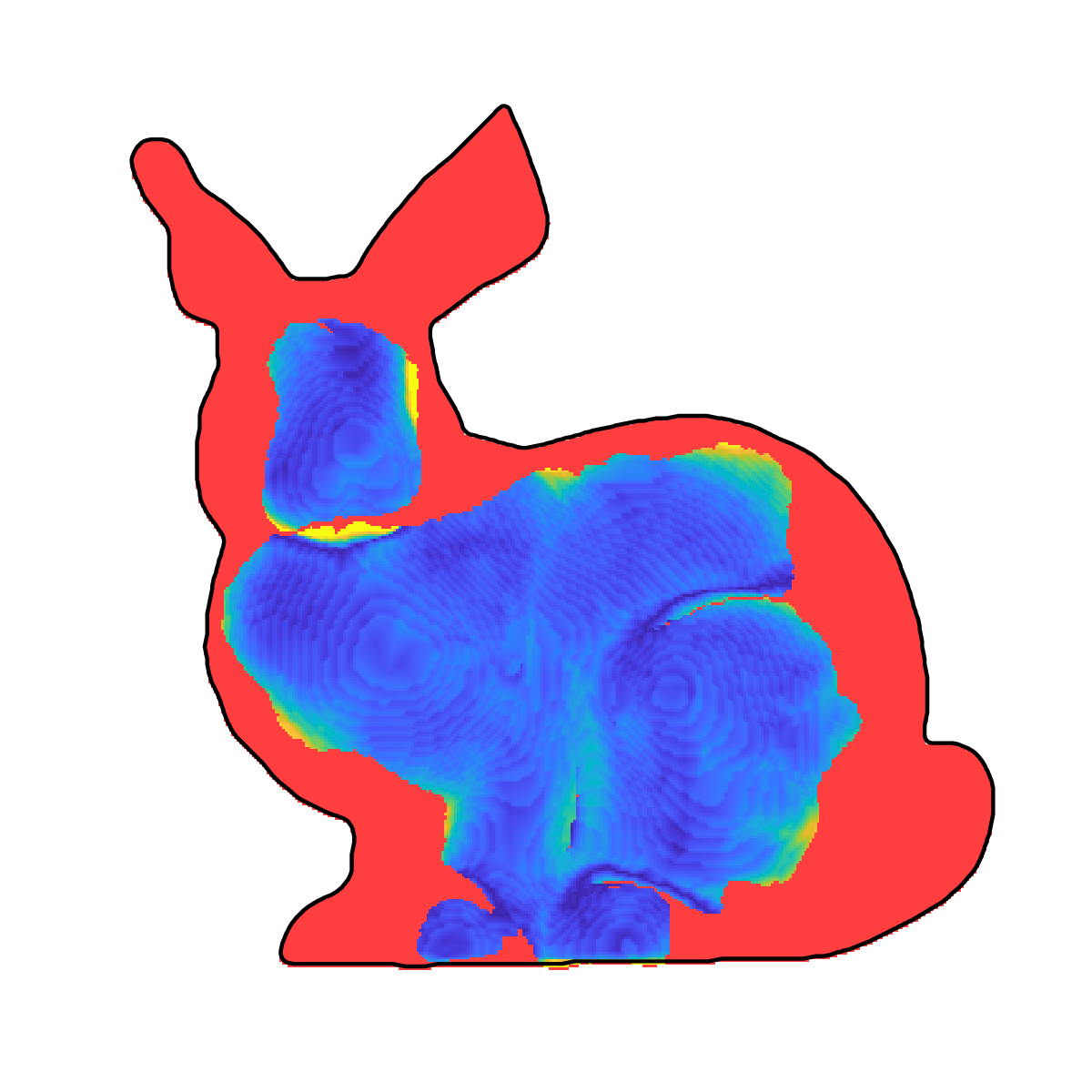} &
\includegraphics[width=\etsize\linewidth]{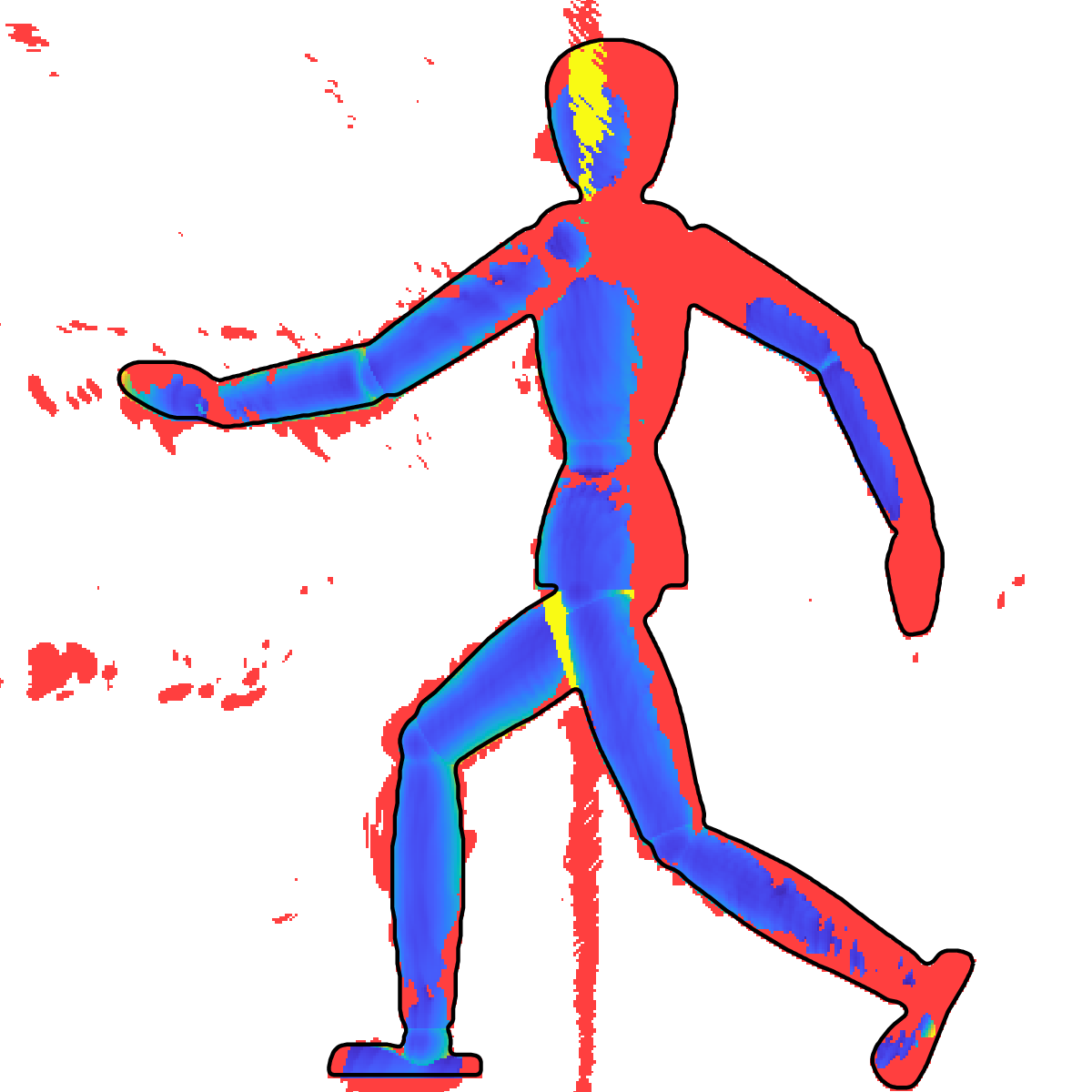} &
\includegraphics[width=\etsize\linewidth]{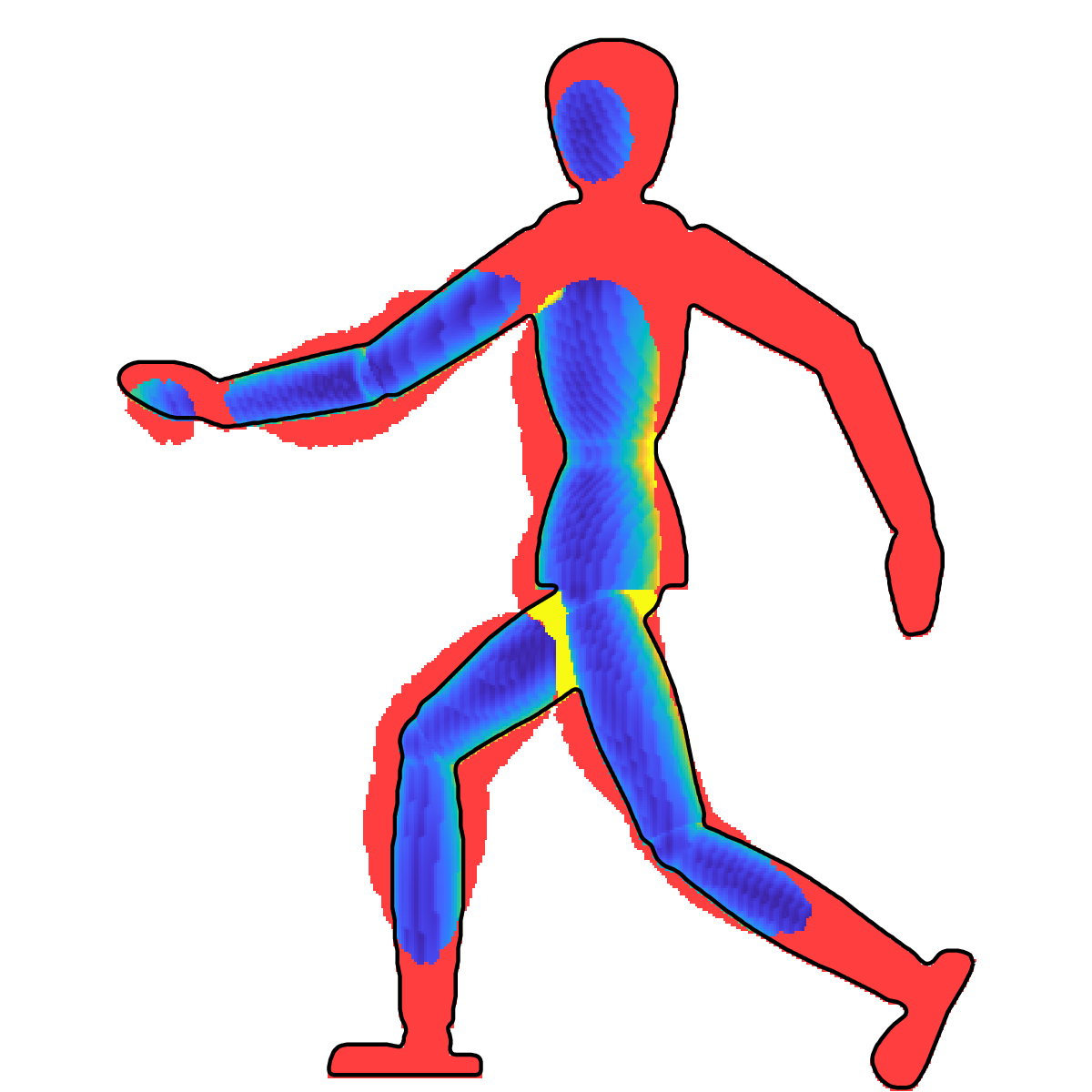} &
\includegraphics[width=\etsize\linewidth]{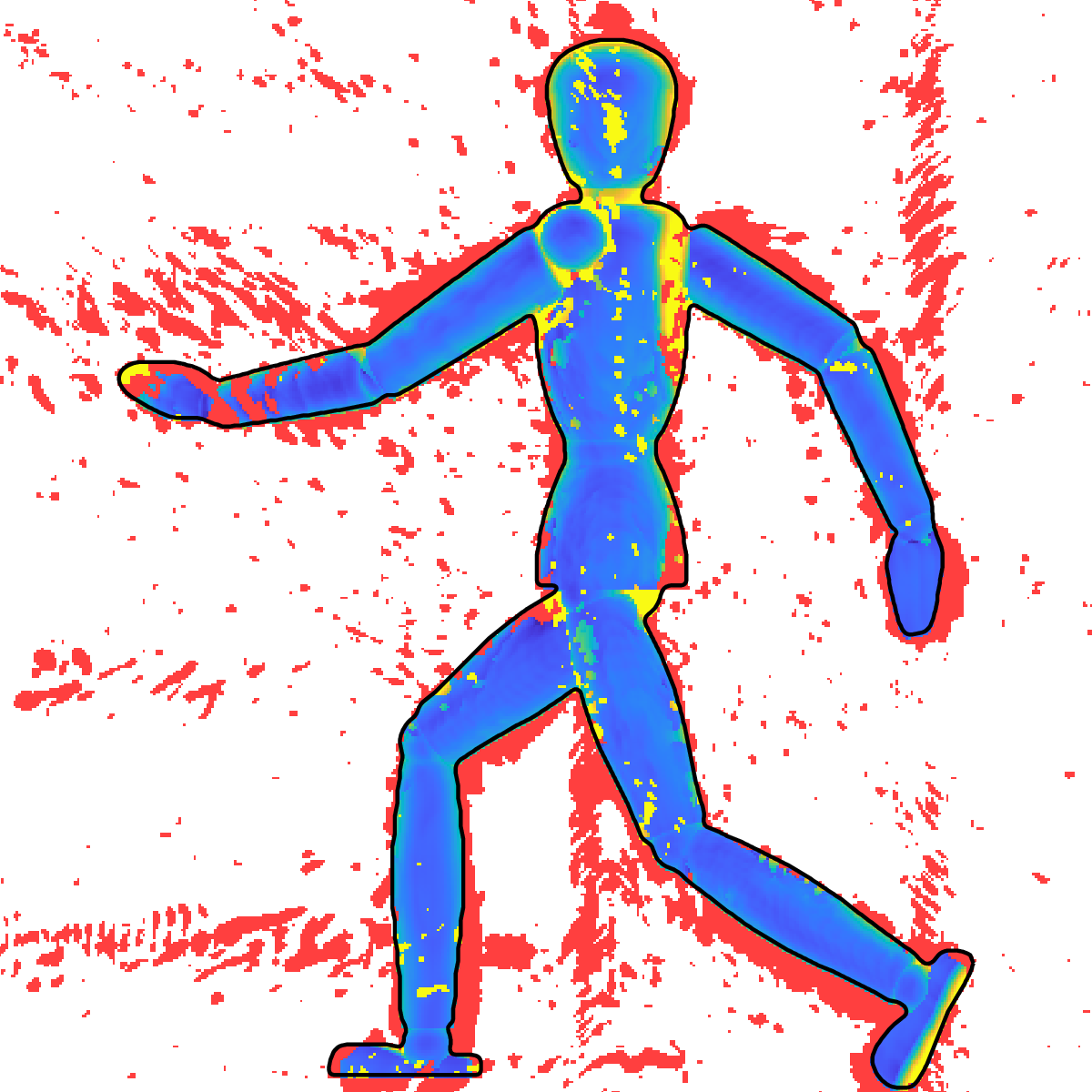} &
\includegraphics[width=\etsize\linewidth]{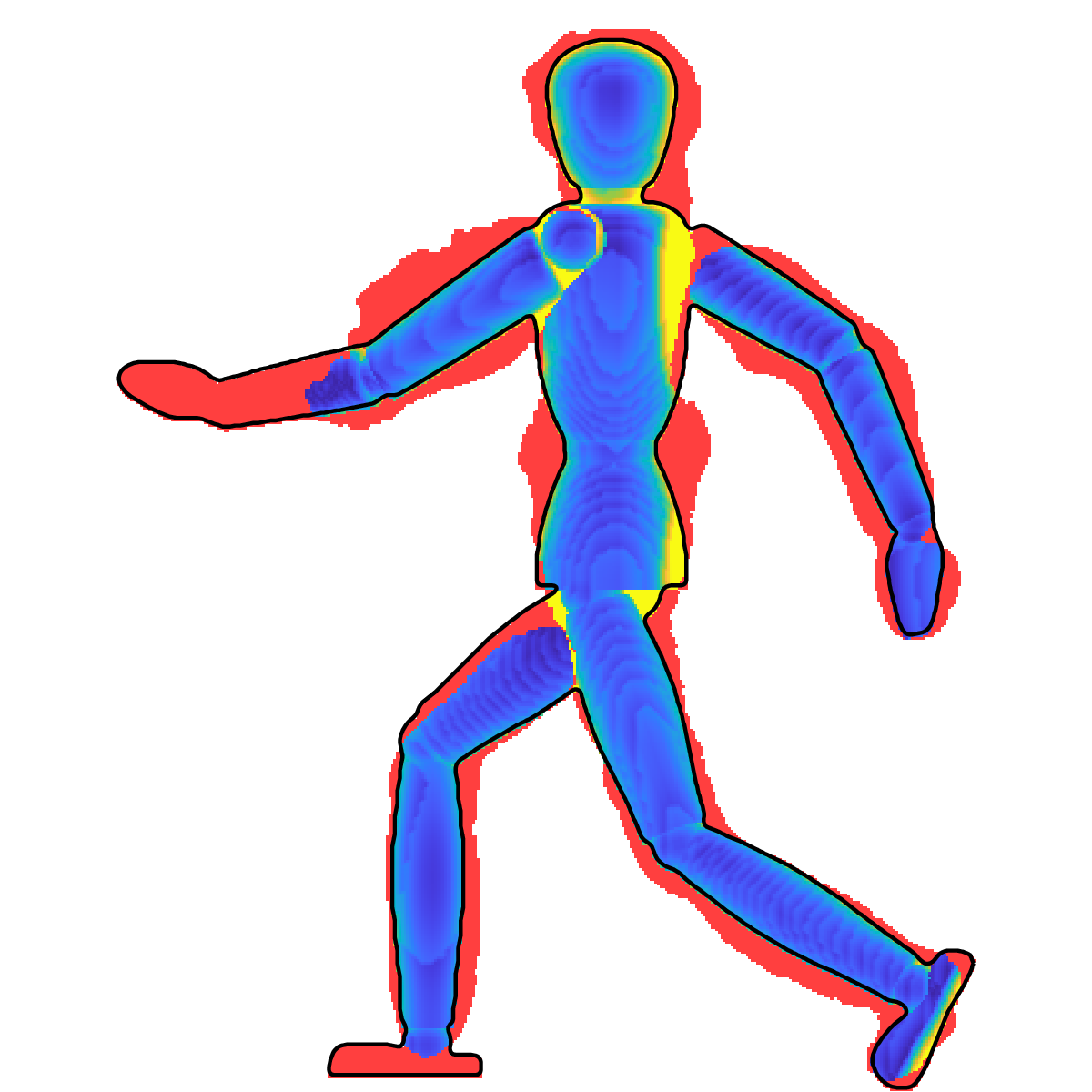} \\
\includegraphics[width=\etsize\linewidth]{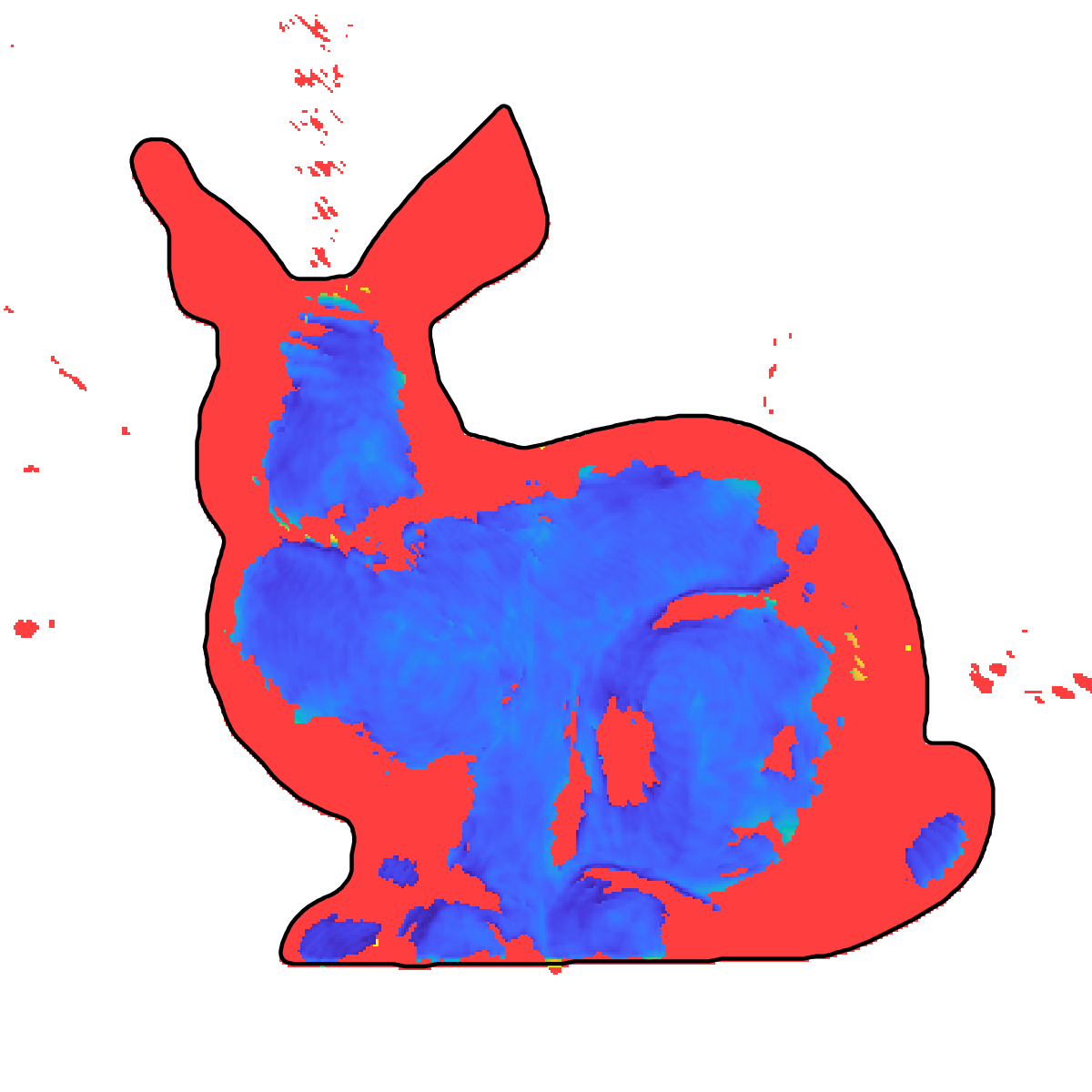} &
\includegraphics[width=\etsize\linewidth]{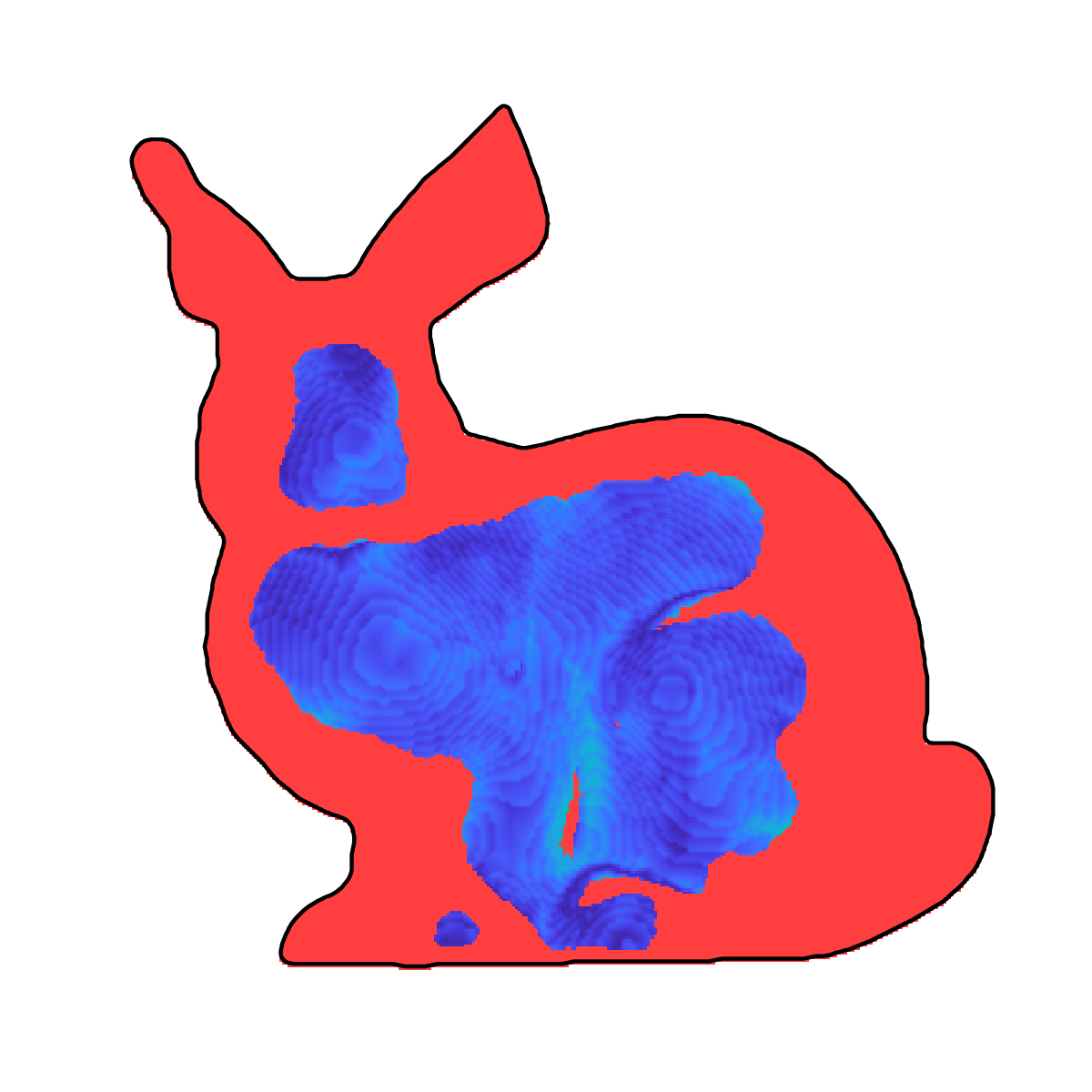} &
\includegraphics[width=\etsize\linewidth]{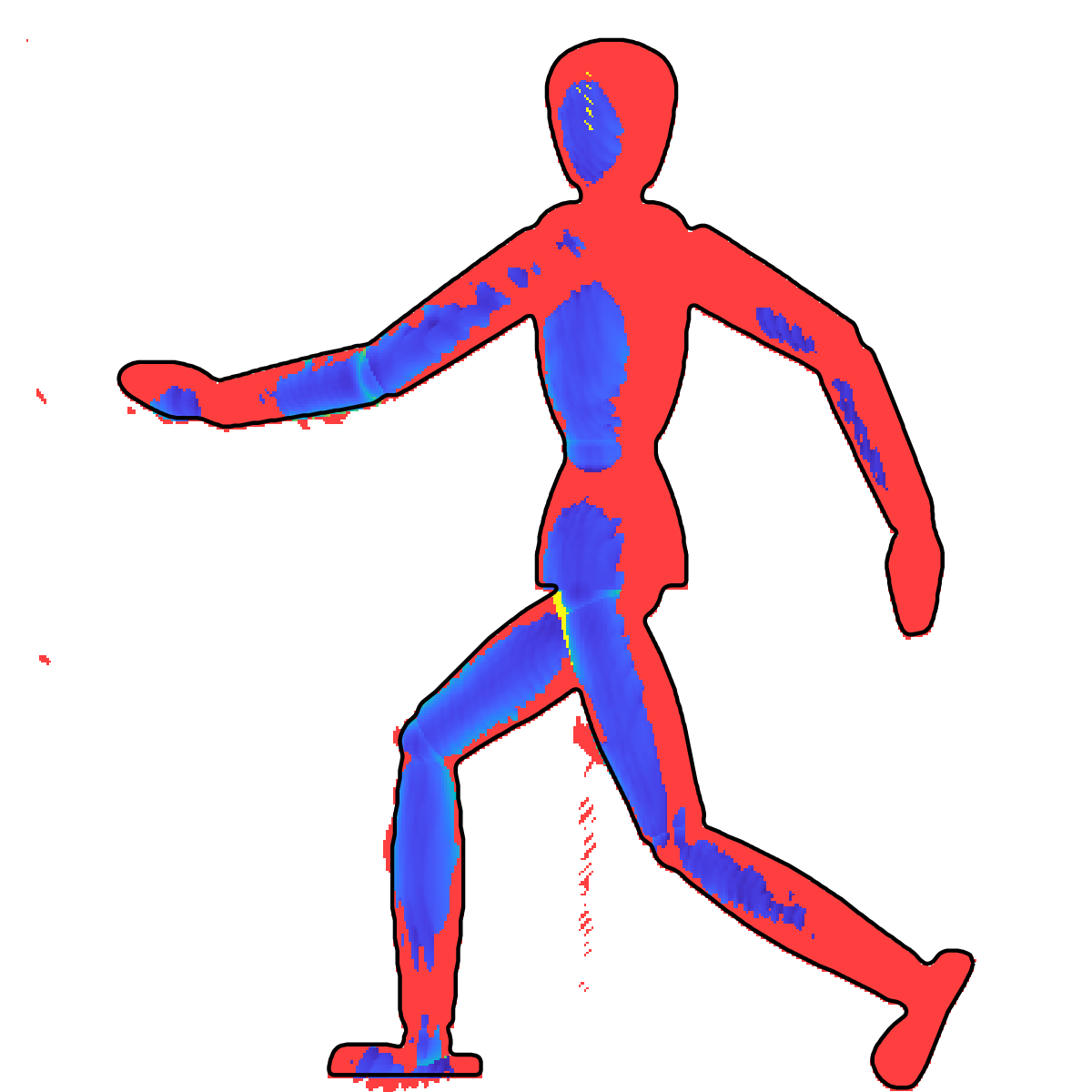} &
\includegraphics[width=\etsize\linewidth]{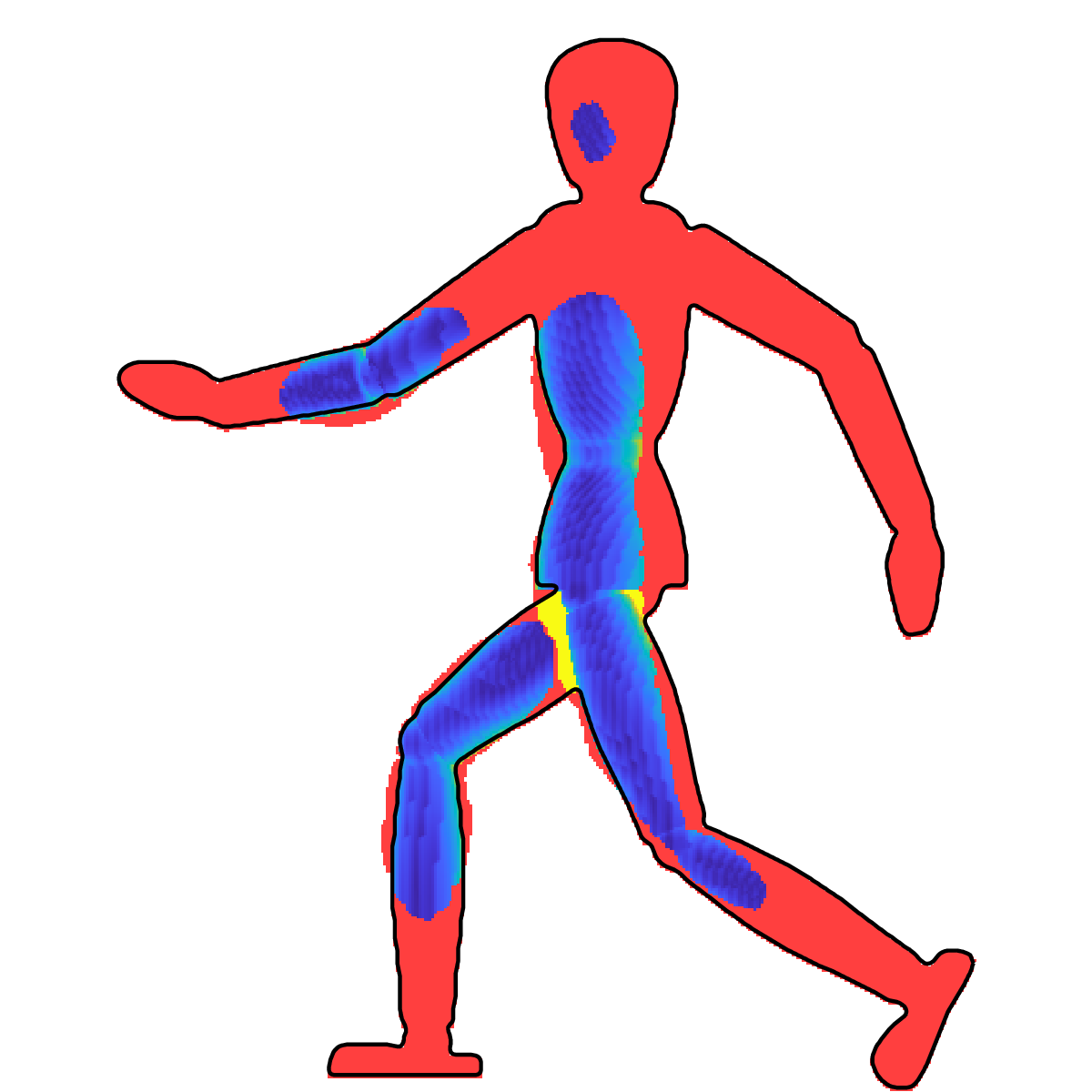} &
\includegraphics[width=\etsize\linewidth]{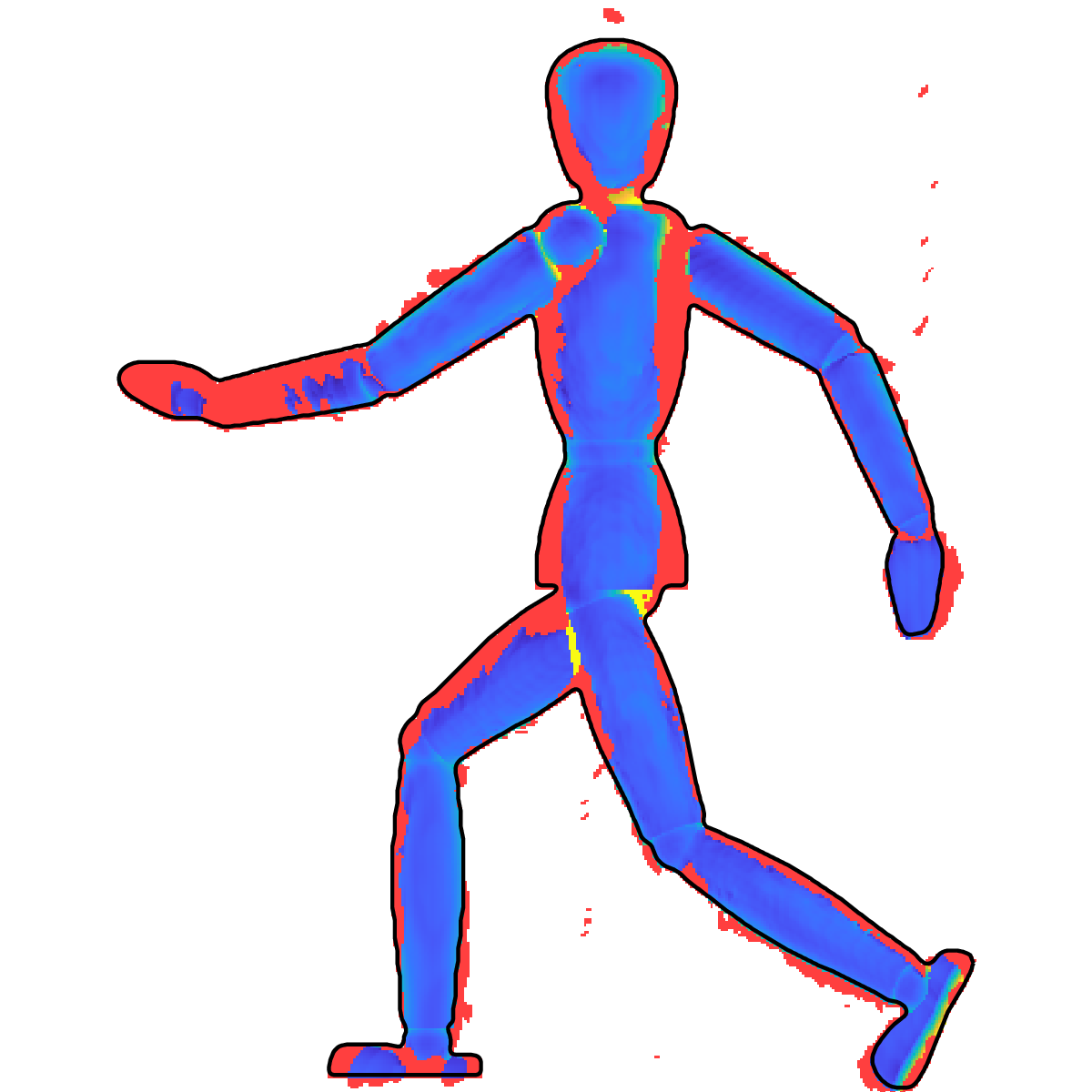} &
\includegraphics[width=\etsize\linewidth]{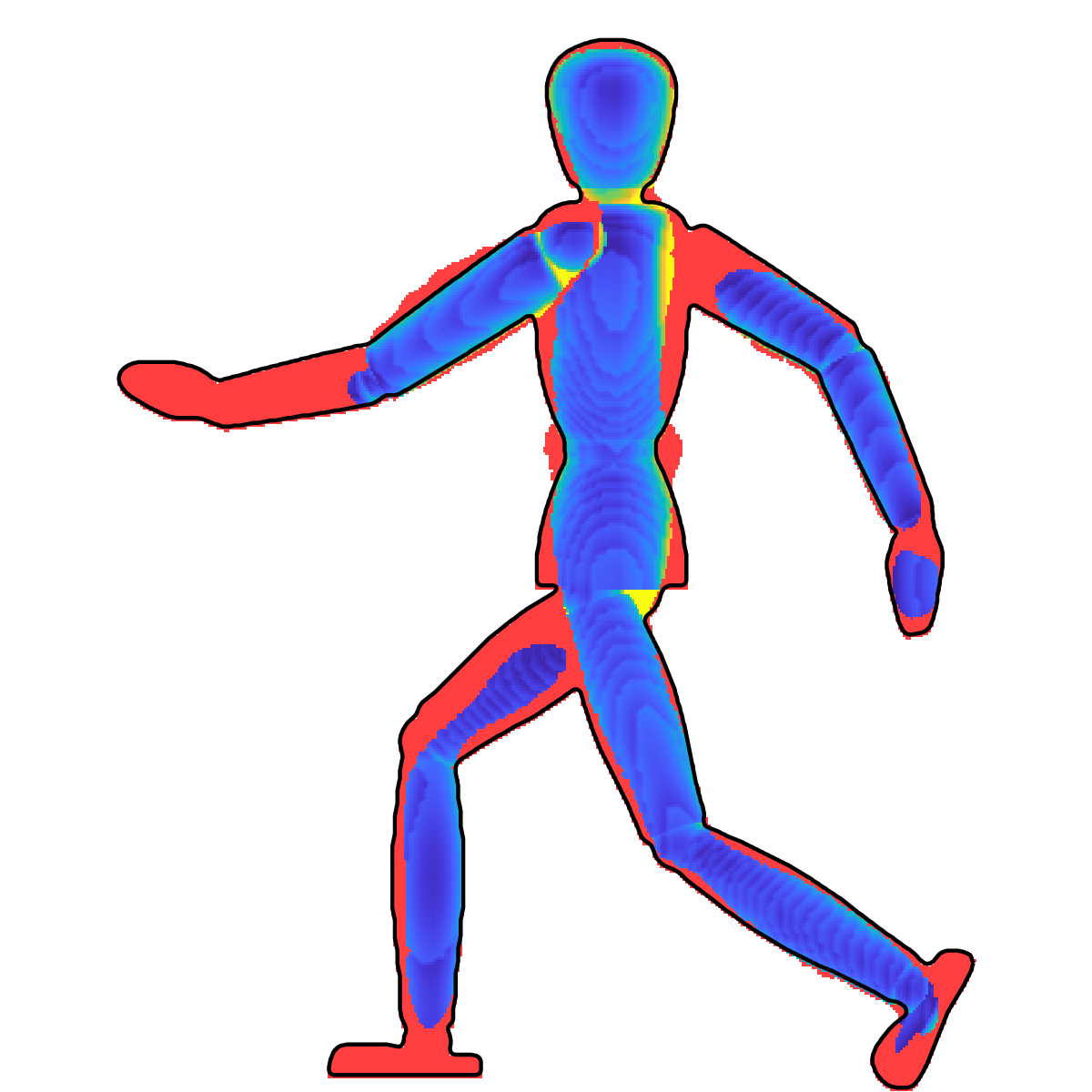} \\
\includegraphics[width=\etsize\linewidth]{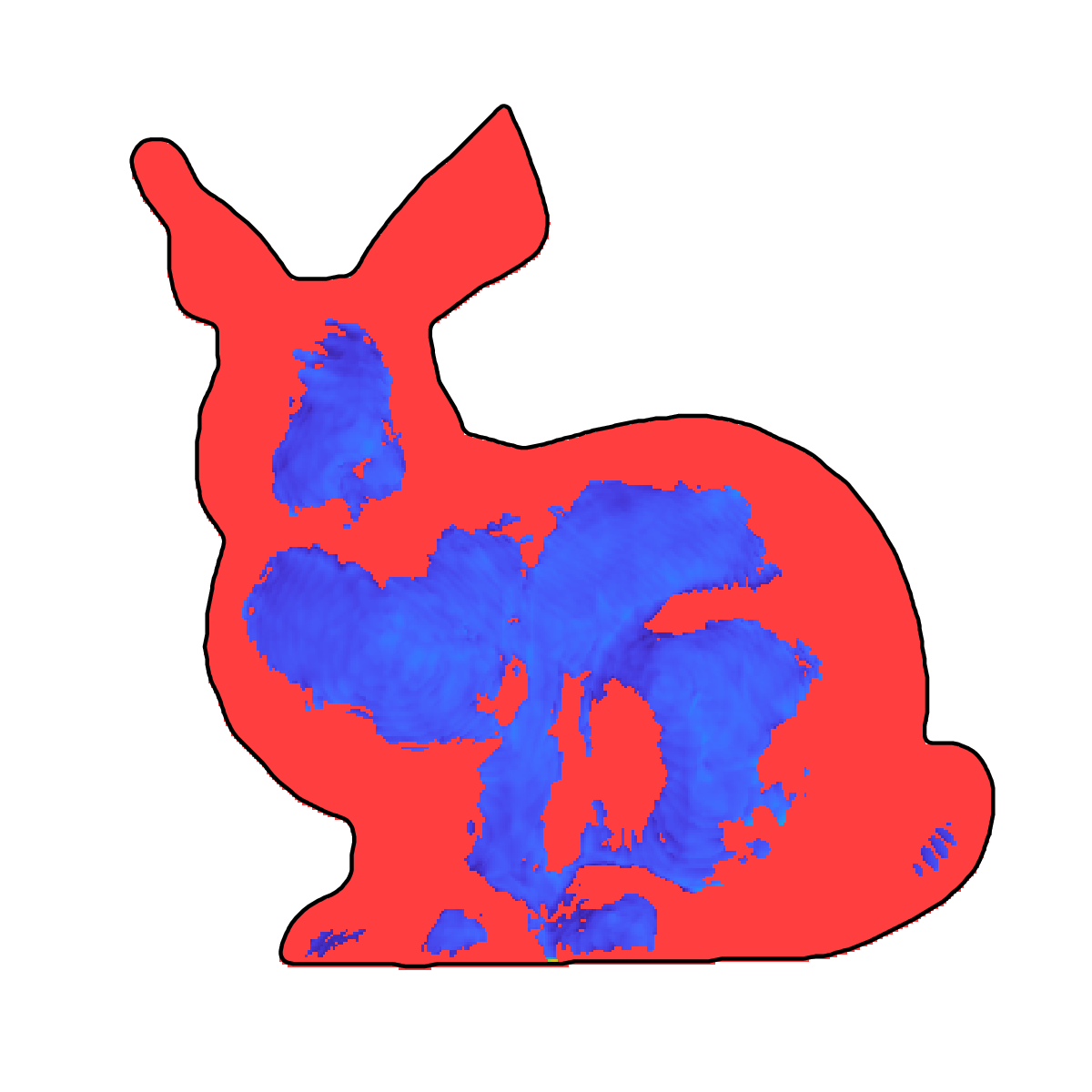} &
\includegraphics[width=\etsize\linewidth]{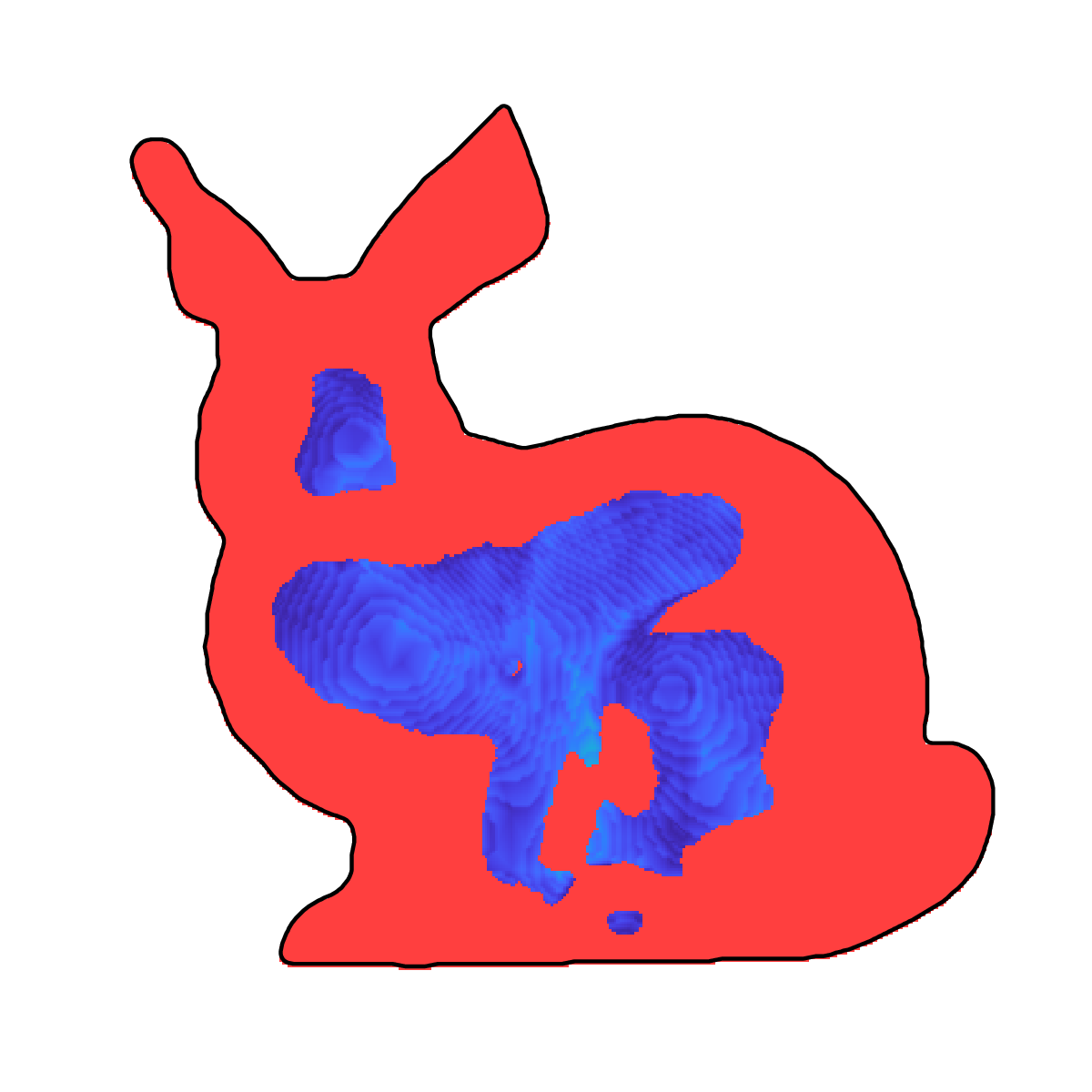} &
\includegraphics[width=\etsize\linewidth]{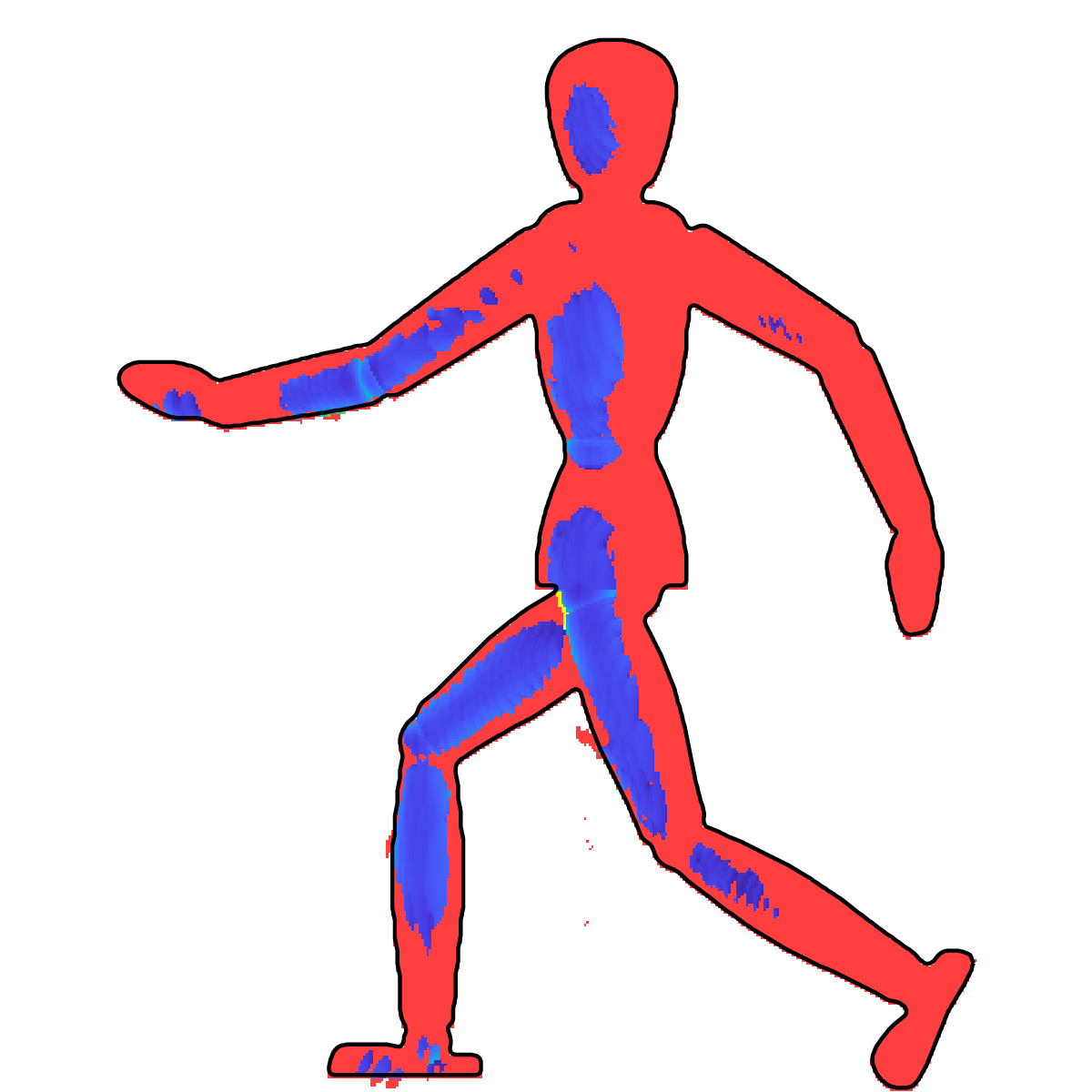} &
\includegraphics[width=\etsize\linewidth]{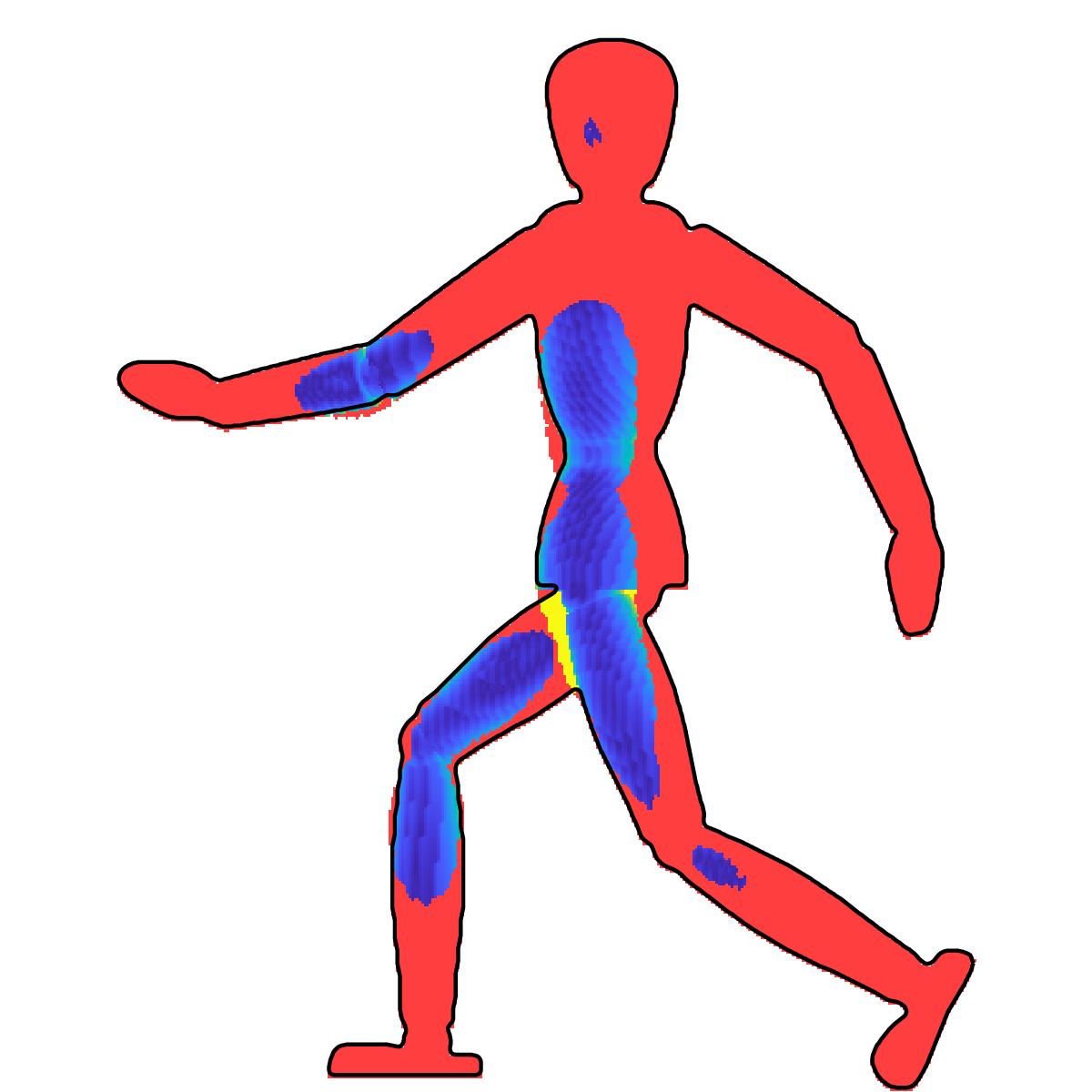} &
\includegraphics[width=\etsize\linewidth]{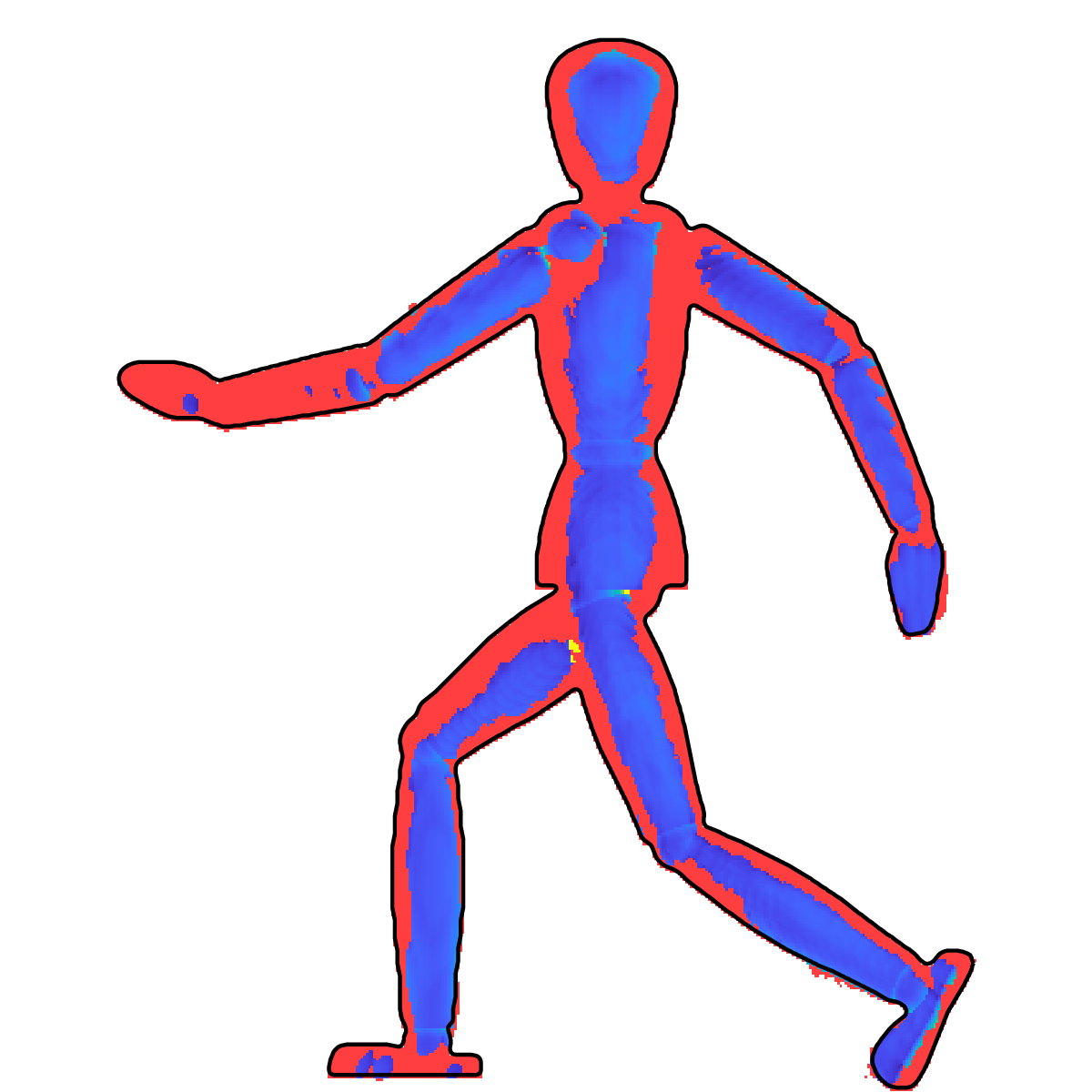} &
\includegraphics[width=\etsize\linewidth]{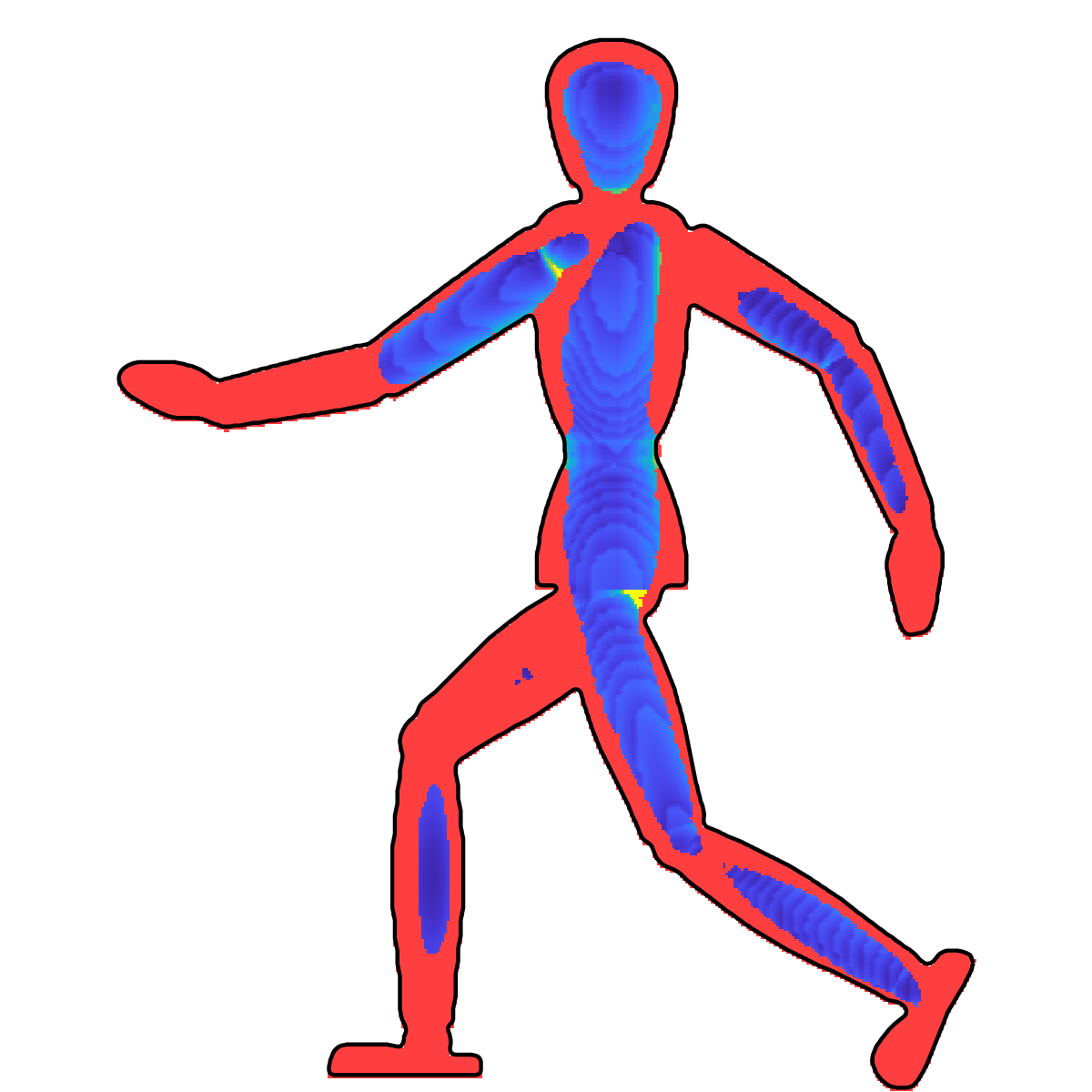}
\\
BP+Laplacian & BP+modified DoG & BP+Laplacian & BP+modified DoG & BP+Laplacian & BP+modified DoG
\end{tabular}%
\begin{minipage}[t]{4.5cm}%
\vspace{-2.5cm}%
\includegraphics[height=9.5cm]{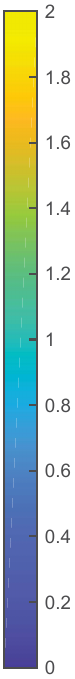}%
\end{minipage}
\end{center}
\caption{\changed{Absolute depth error (in world units) in the reconstructions obtained from the synthetic \texttt{Bunny} (left), \texttt{Mannequin1Laser} (middle), and \texttt{Mannequin} (right) datasets. The top row shows error plots for our method, the bottom part shows depth errors for backprojection (BP) using the Laplacian filter and the modified Difference of Gaussian filter \cite{Laurenzis:2014a} respectively, with increasing isovalues from top to bottom. The black line indicates the ground-truth object silhouette. Red color inside the silhouette indicates a missing (false-negative) surface and outside a silhouette it indicates excess (false-positive) geometry. Note that the range is clamped to $[0,2]$ for visualization; values plotted in yellow can be significantly higher. See \Fig{fig:deptherror_plots} for a quantitative analysis.}}
\label{fig:deptherror}
\end{figure*}

We then used the truncated rendering as reference for our own renderer, and tested the effect of temporal filtering and shadow testing on the difference (\Fig{fig:renderererror}). A na\"ive version of our renderer, with all refinements disabled, reached the reference up to an error of a little under $10 \%$. After activating the temporal filtering and the shadow tests, our fast renderer delivered a close approximation to the ray-traced reference with with \unit{69.796}{\deci\bel} peak signal-to-noise ratio (PSNR) or a relative $L2$ difference of $0.489 \%$. All error values are provided at a glance in \Tab{tab:render_error}. The main result from this investigation is that both features are essential to our renderer. The gain in accuracy comes at the expense of significantly increased runtime when using the shadow test (\Fig{fig:rendererTimings}). For small numbers of pixels, a significant part of that runtime is caused by the construction of acceleration structures---here, about \unit{10}{ms} for an object with approximately 55,000 triangles.
Another noteworthy observation is that the Monte-Carlo rendering used as reference was likely not fully converged (\Fig{fig:toftracerConvergence}) even after evaluating 250 million samples per pixel. We expect that more exhaustive sampling would likely have further reduced the error.

\begin{figure}[ht]
\centering
\includegraphics{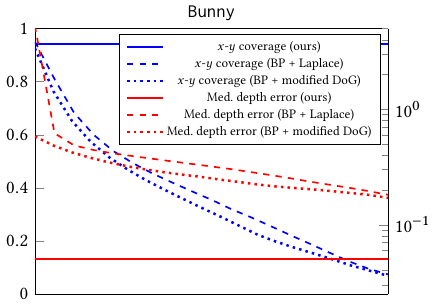}\\[-1mm]
\includegraphics{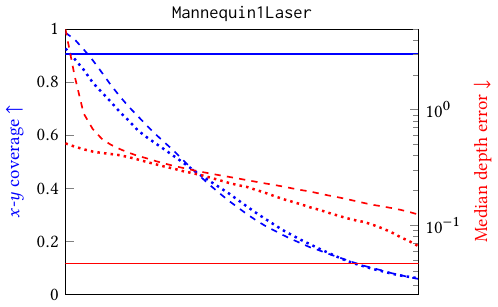}\\[-1mm]
\includegraphics{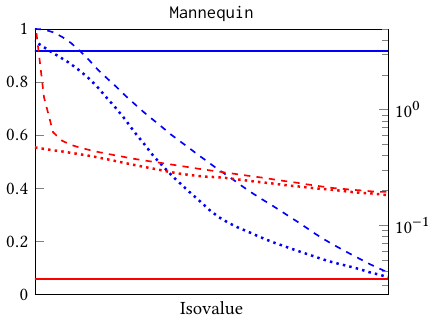}
\caption{Evaluation of the depth map coverage in the $x$-$y$ plane (higher is better) and the median absolute depth error in $z$ direction (lower is better) for the \texttt{Bunny}, \texttt{Mannequin1Laser}, and \texttt{Mannequin} datasets. The proposed method achieves coverage values above $90 \%$ with a median depth error as low as 0.03 to 0.05 world units. For the state-of-the-art method (evaluated using the Laplacian filter and the modified Difference of Gaussian filter \cite{Laurenzis:2014a}), no isovalue is capable of simultaneously achieving high coverage and low depth error. We note that the modified DoG filter generates less noise than the Laplacian filter for low isovalues. A qualitative visualization of this study can be found in \Fig{fig:deptherror}.}
\label{fig:deptherror_plots}
\end{figure}

\subsection{Geometry reconstruction}
 We used various types of input data to test our algorithm: synthetic data generated using a path tracer or our own fast renderer, as well as experimental data obtained from other sources. The results from these reconstructions are scattered throughout the paper, referencing the datasets from \Tab{tbl:parameters} by their respective names. Meshes are rendered in a daylight environment using Mitsuba \cite{Mitsuba}, with a back wall and ground plane added as shadow receivers for better visualization of the 3D shapes. Note that these planes are not part of the experimental setup.

\begin{figure}[ht]
\includegraphics[height=0.33\linewidth]{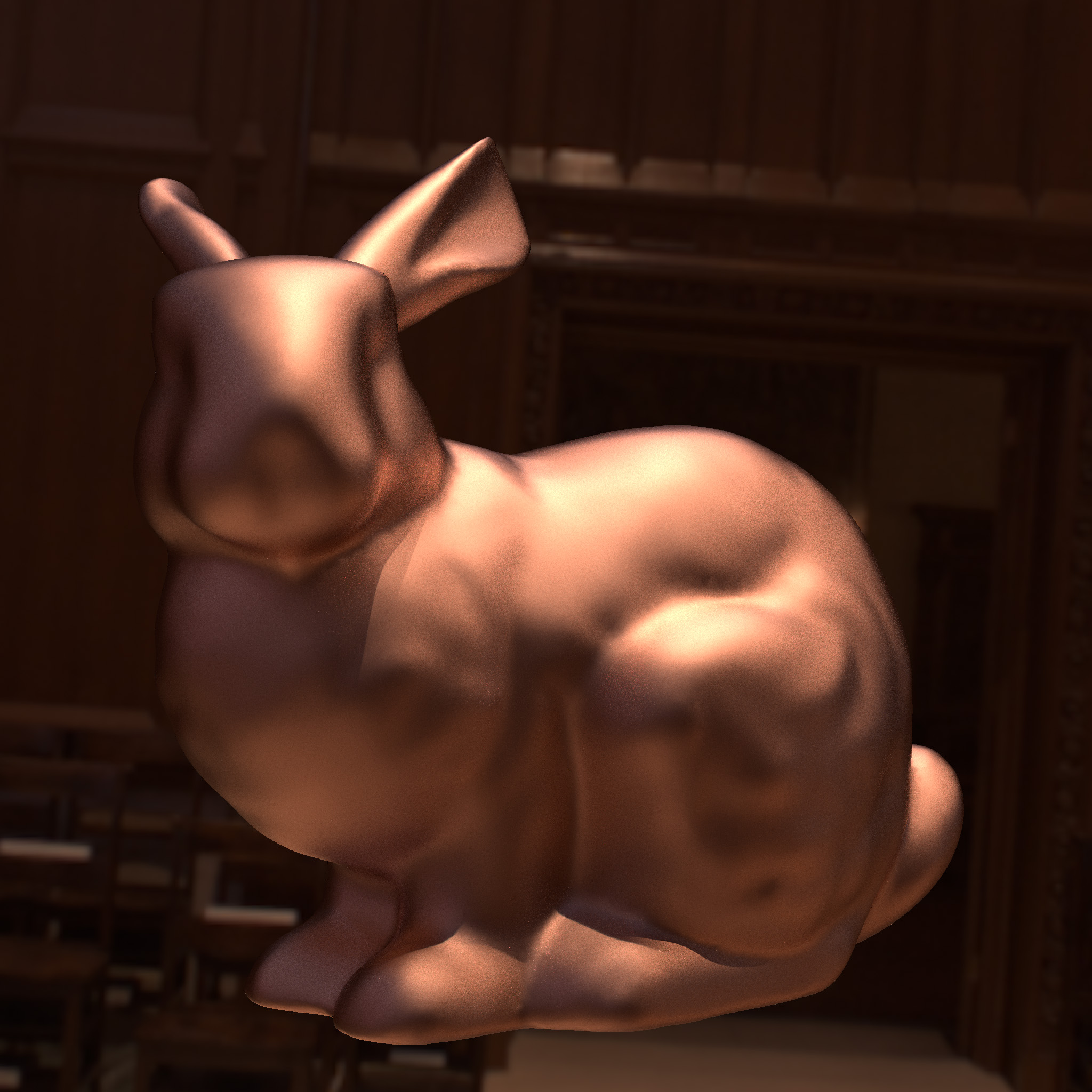}\hfill
\includegraphics[height=0.33\linewidth]{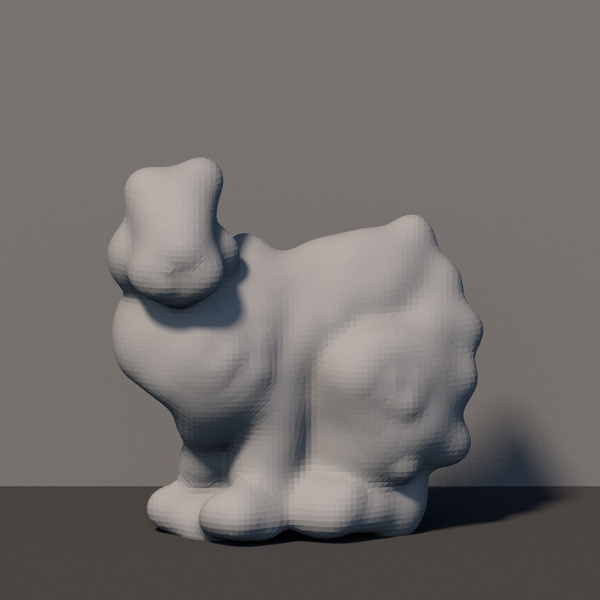}\hfill
\includegraphics[height=0.33\linewidth]{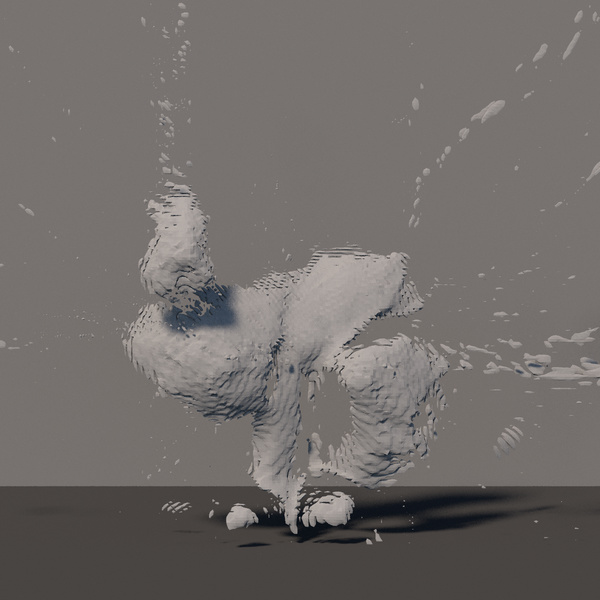}\\
\includegraphics[height=0.33\linewidth]{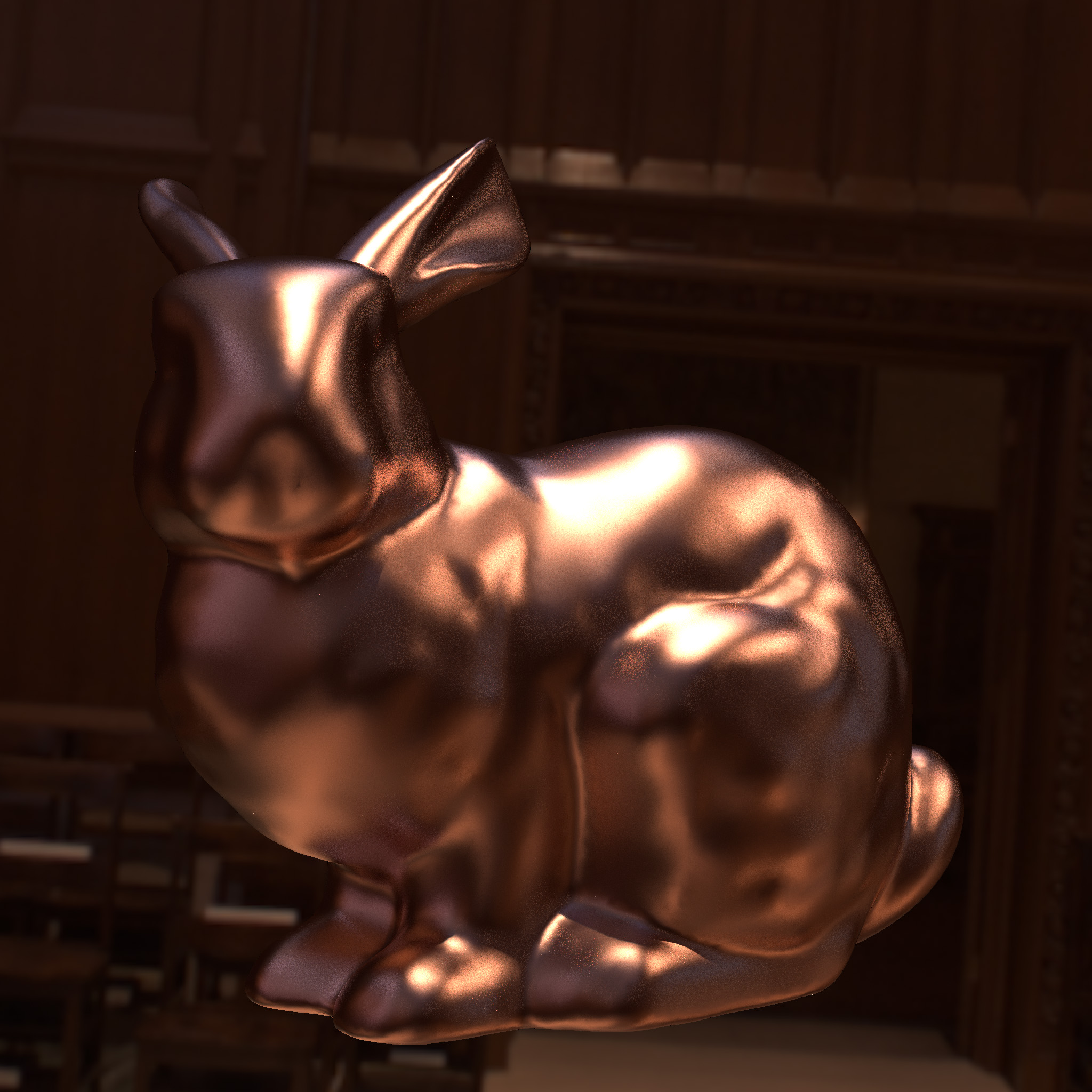}\hfill
\includegraphics[height=0.33\linewidth]{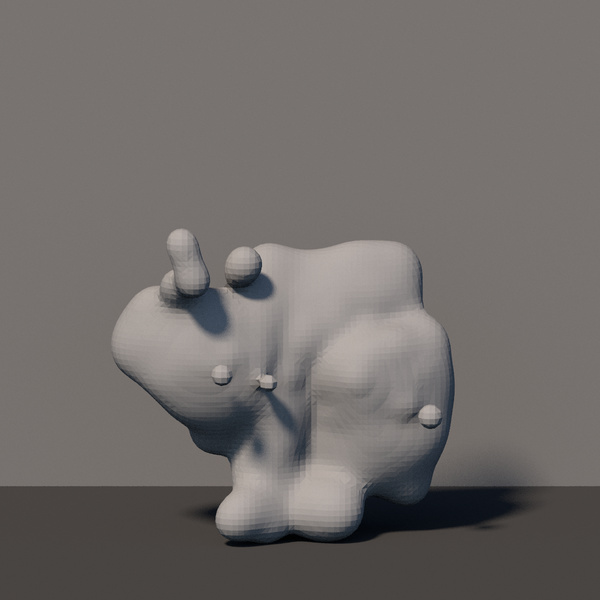}\hfill
\includegraphics[height=0.33\linewidth]{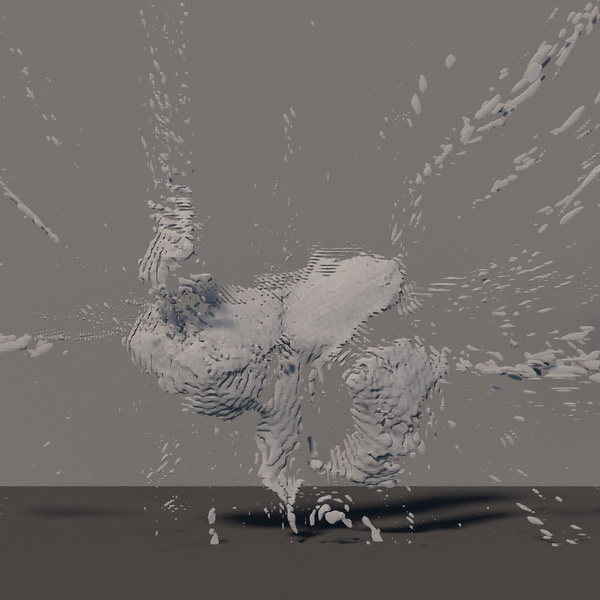}\\
\caption{Reconstruction of the \texttt{BunnyMetal*} scenes with \texttt{pbrt}'s \texttt{metal} BRDF applied to the object (top row: Blinn roughness 0.05; bottom row: Blinn roughness 0.01). From left to right: reference rendering in Grace Cathedral environment \cite{debeveclightprobes}; our proposed method; backprojection.}
\label{fig:metalbunny}
\end{figure}

\begin{figure}[ht]
\includegraphics[height=0.33\linewidth]{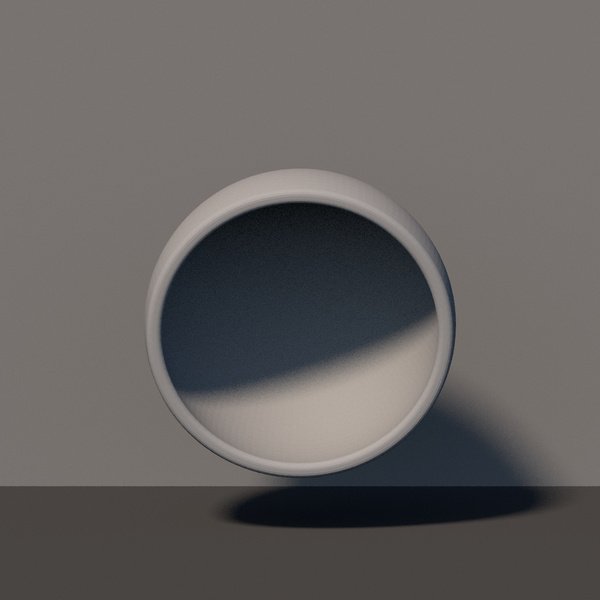}\hfill
\includegraphics[height=0.33\linewidth]{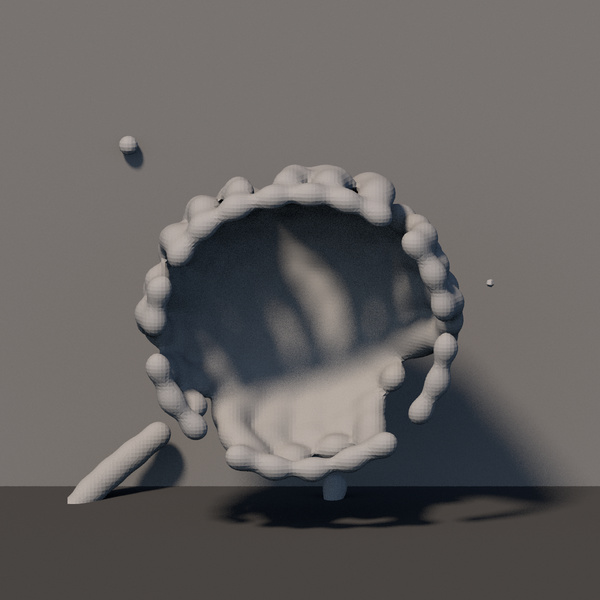}\hfill
\includegraphics[height=0.33\linewidth]{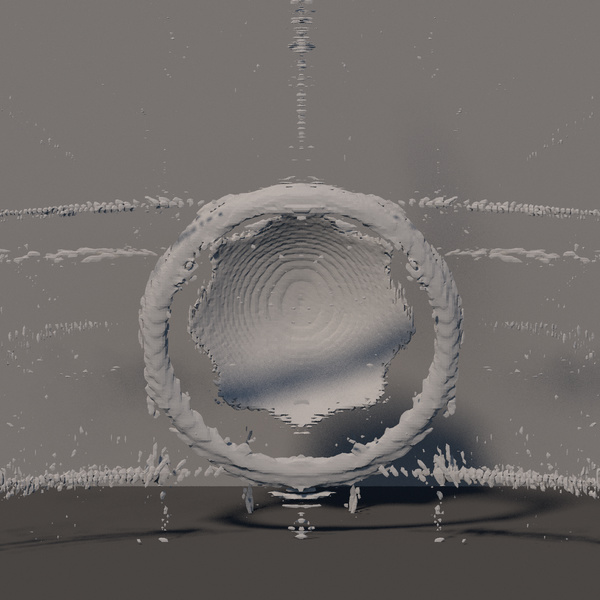}\\
\includegraphics[height=0.33\linewidth]{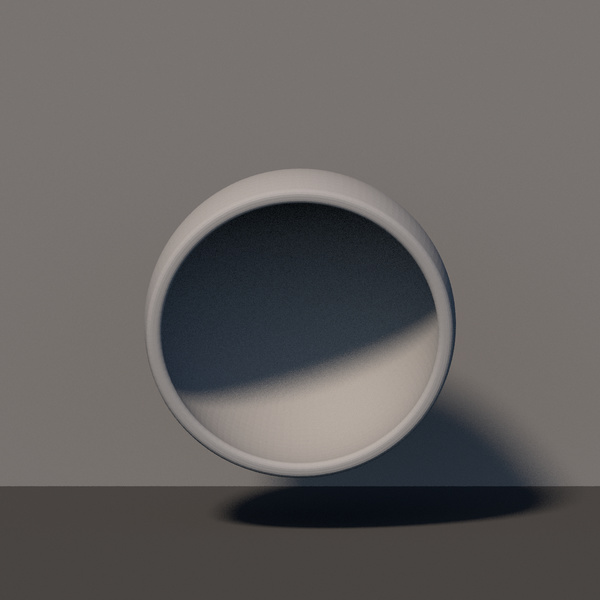}\hfill
\includegraphics[height=0.33\linewidth]{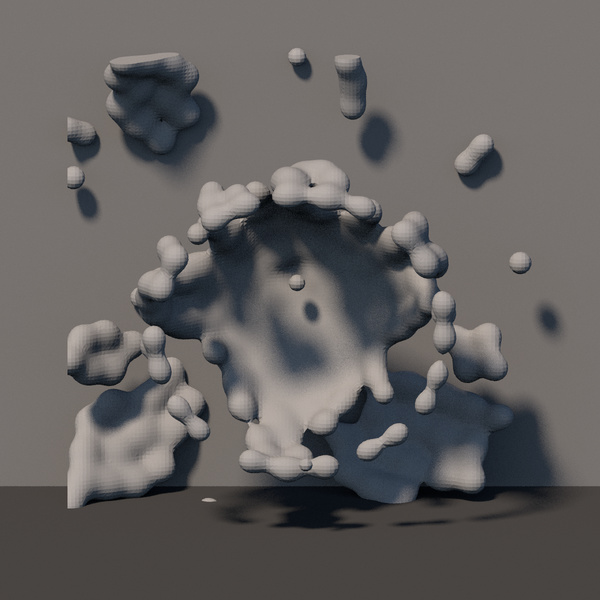}\hfill
\includegraphics[height=0.33\linewidth]{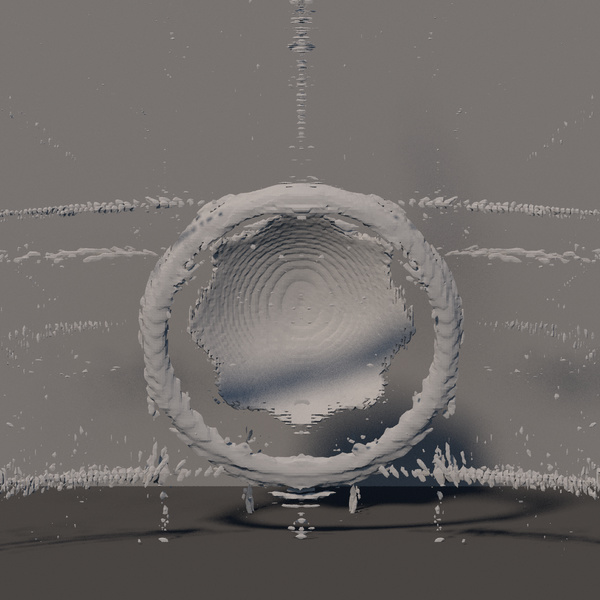}\\
\caption{\texttt{Bowl} scene. A strongly concave shape with high albedo (top row: $30 \%$; bottom row: $100 \%$) features large amounts of interreflected light in the input data, which leads to spurious features in the reconstructed geometry. From left to right: reference geometry; our proposed method; backprojection.}
\label{fig:bowls}
\end{figure}

\begin{figure*}[ht]
\begin{tabular}{ccccc}
\includegraphics[height=0.18\linewidth]{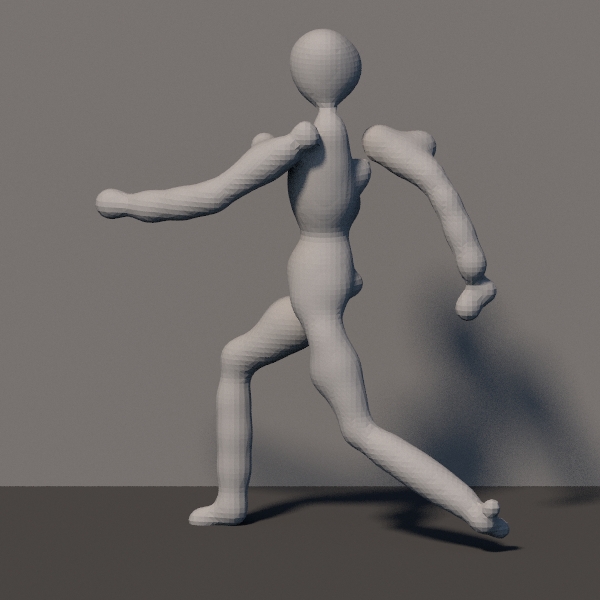} &%
\includegraphics[height=0.18\linewidth]{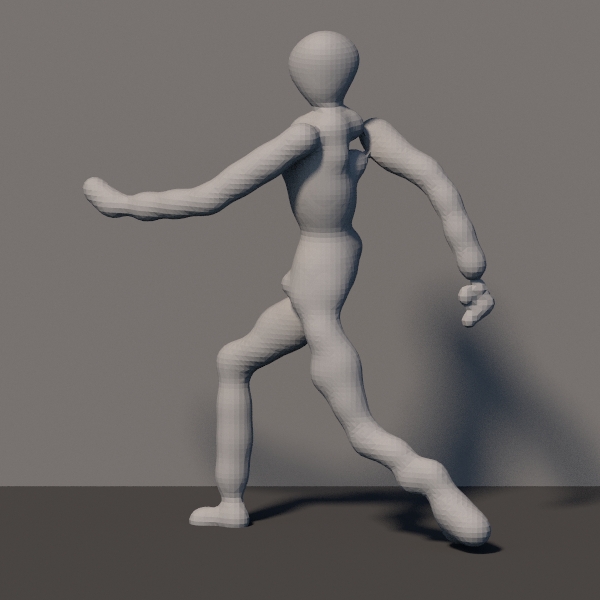} &%
\includegraphics[height=0.18\linewidth]{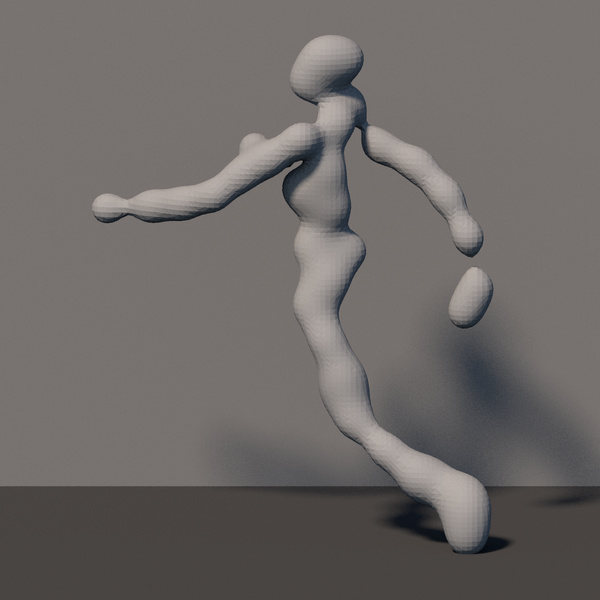} &%
\includegraphics[height=0.18\linewidth]{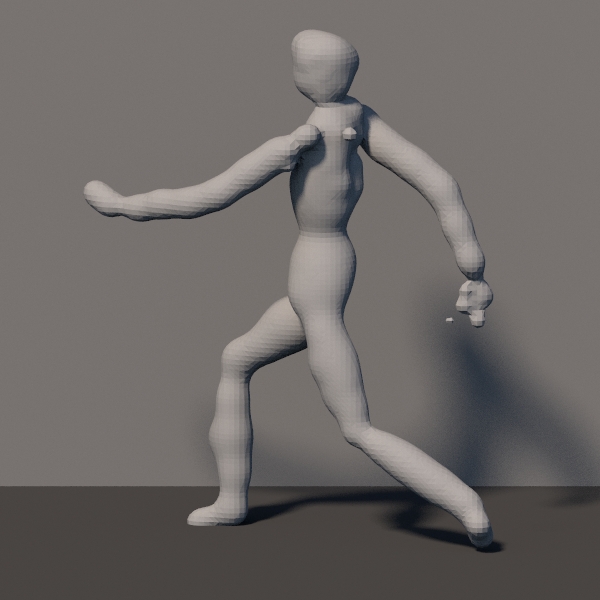} &%
\includegraphics[height=0.18\linewidth]{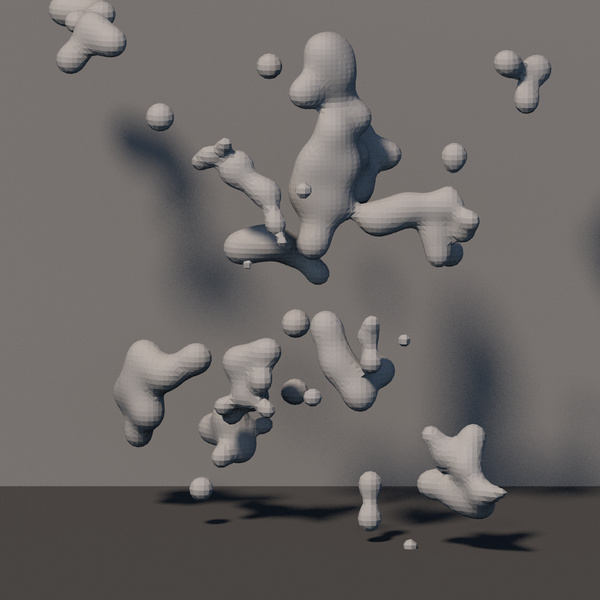} \\
\includegraphics[height=0.18\linewidth]{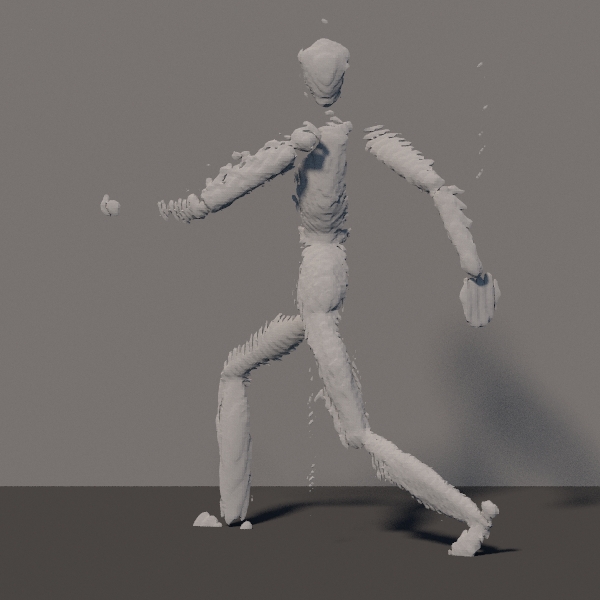} &%
\includegraphics[height=0.18\linewidth]{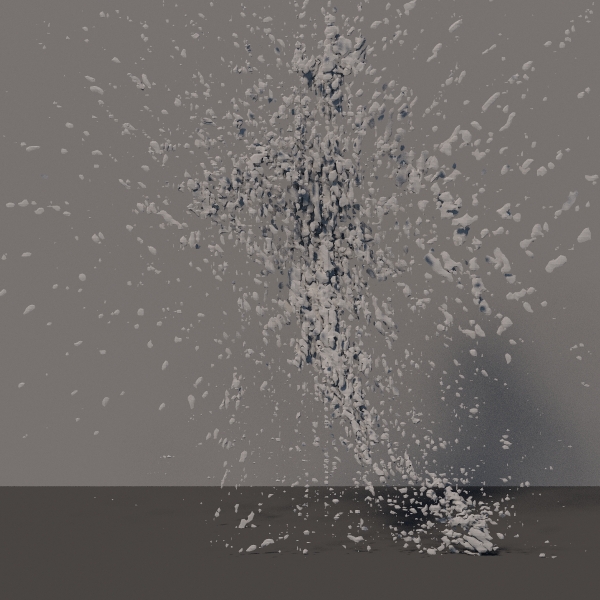} &%
\includegraphics[height=0.18\linewidth]{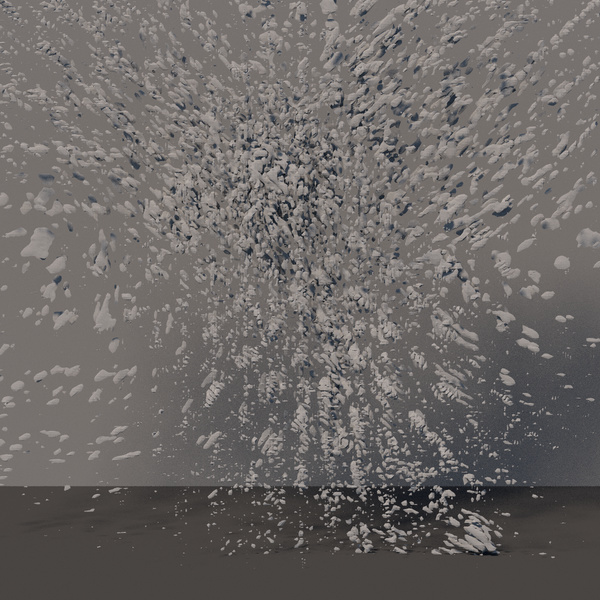} &%
\includegraphics[height=0.18\linewidth]{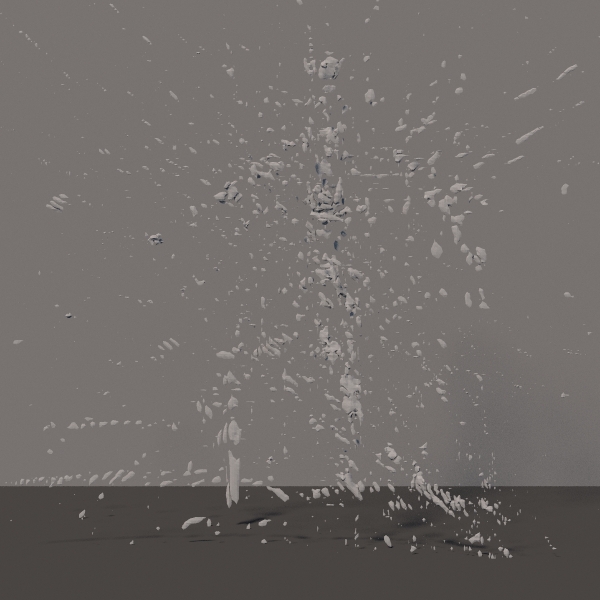} &%
\includegraphics[height=0.18\linewidth]{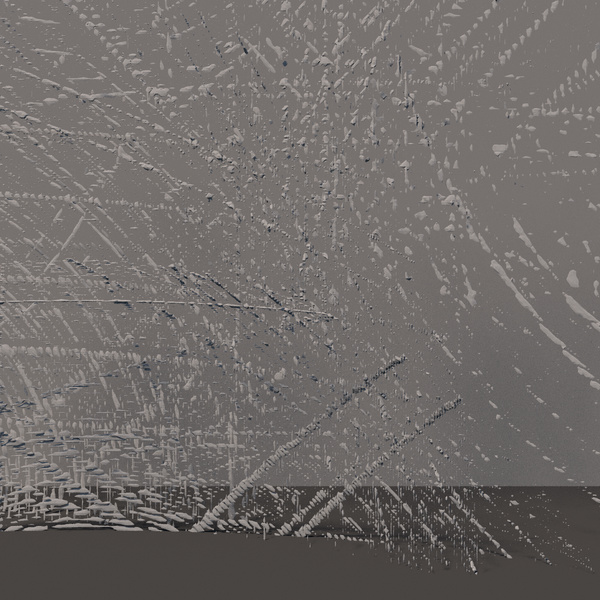} \\
\scriptsize{\texttt{Mannequin} ($16 \times 16 \times 256$)} &%
\scriptsize{\texttt{MannequinLowTemp} ($16 \times 16 \times 32$)} &%
\scriptsize{\texttt{MannequinMinTemp} ($16 \times 16 \times 8$)} &%
\scriptsize{\texttt{MannequinLowRes} ($4 \times 4 \times 256$)} &%
\scriptsize{\texttt{MannequinMinRes} ($2 \times 2 \times 256$)}
\end{tabular}
\caption{Reconstruction of the \texttt{Mannequin*} dataset using different levels of degradation. From left to right: \texttt{Mannequin, MannequinLowTemp, MannequinMinTemp, MannequinLowRes, MannequinMinRes}. Top row: Our reconstruction, bottom row: backprojection. Unlike backprojection, our reconstruction method handles degradations in the input data quite gracefully. Even an extremely low spatial resolution of $2 \times 2$ pixels or a temporal resolution of only 8 bins still produces roughly identifiable results. }
\label{fig:mannequin:degradations}
\end{figure*}
\begin{figure*}[ht]
\begin{tabular}{ccccc}
\includegraphics[height=0.18\linewidth]{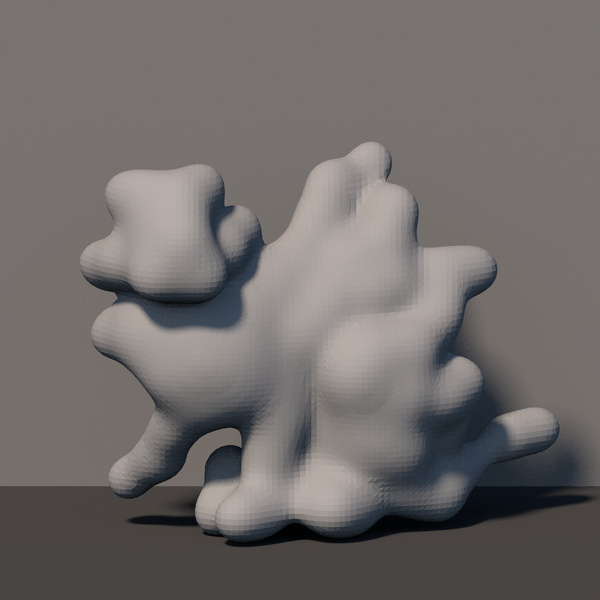} &%
\includegraphics[height=0.18\linewidth]{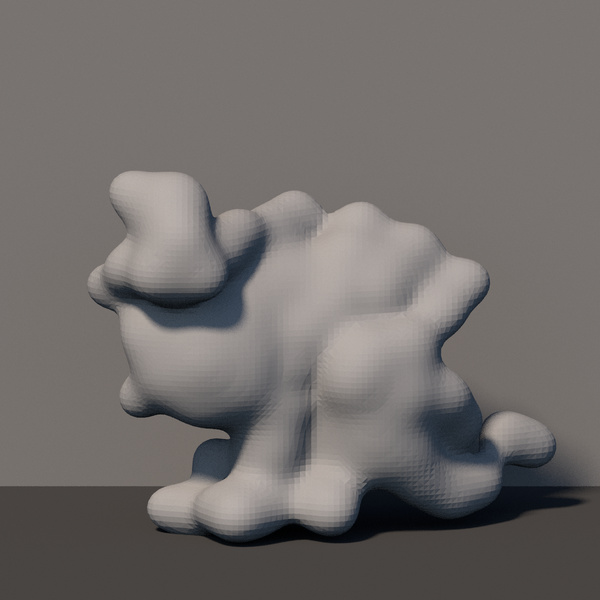} &%
\includegraphics[height=0.18\linewidth]{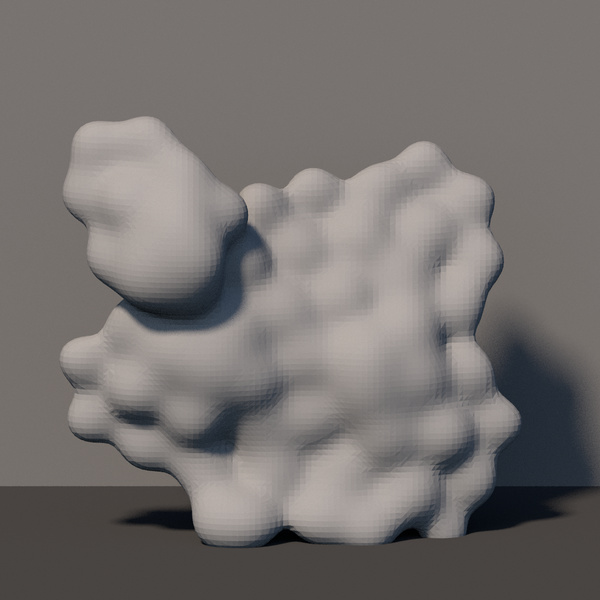} &%
\includegraphics[height=0.18\linewidth]{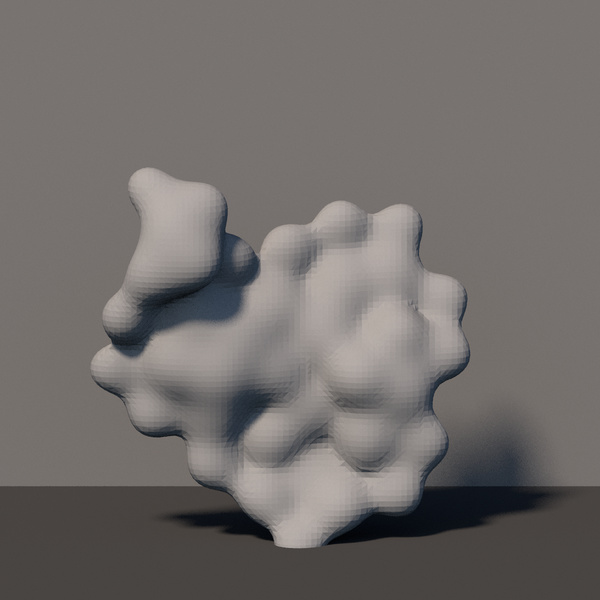} &%
\includegraphics[height=0.18\linewidth]{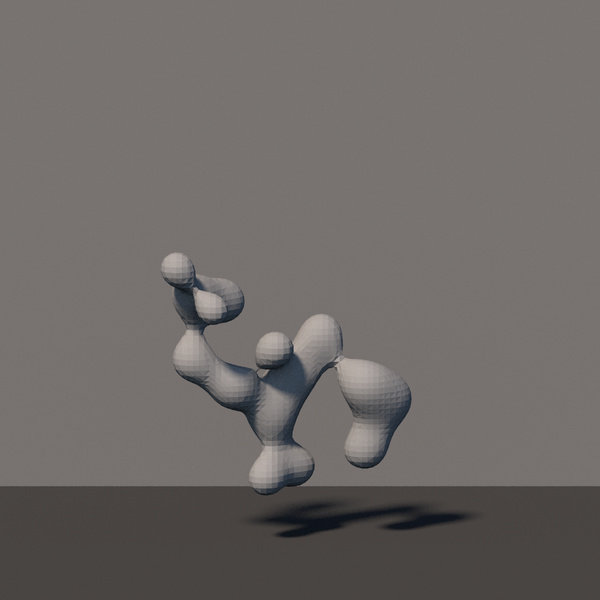} \\
\includegraphics[height=0.18\linewidth]{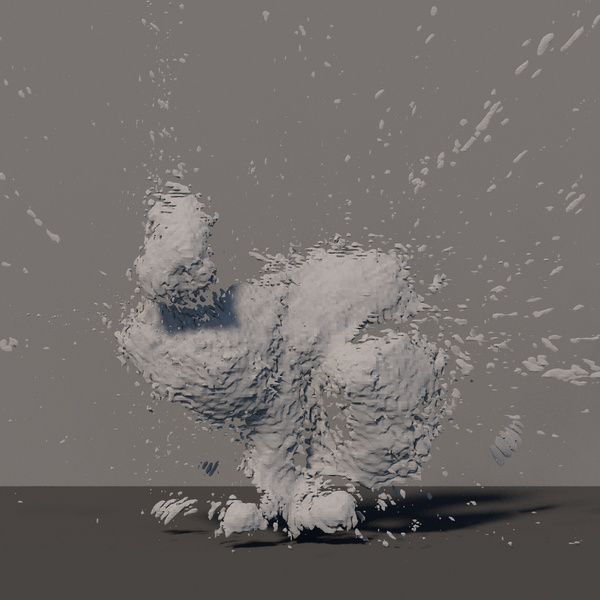} &%
\includegraphics[height=0.18\linewidth]{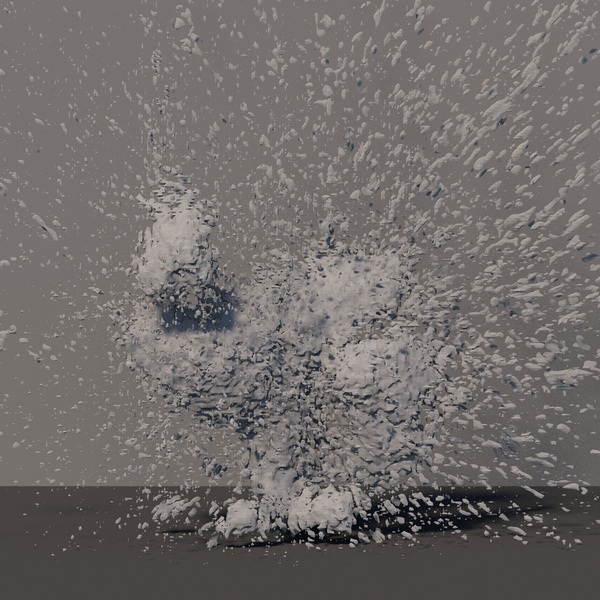} &%
\includegraphics[height=0.18\linewidth]{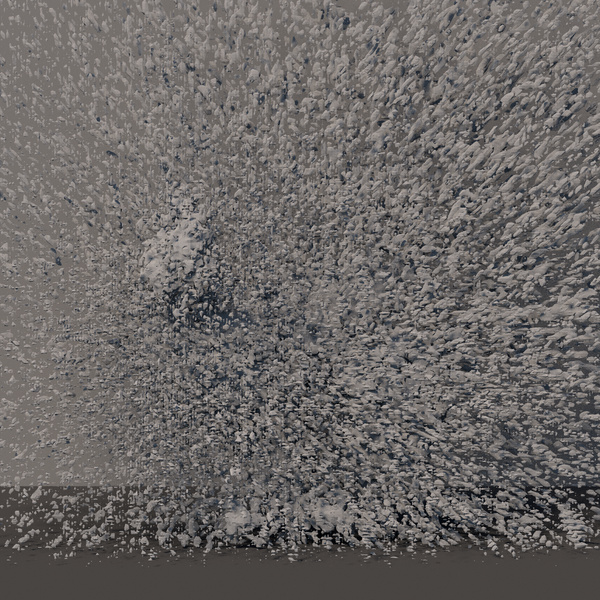} &%
\includegraphics[height=0.18\linewidth]{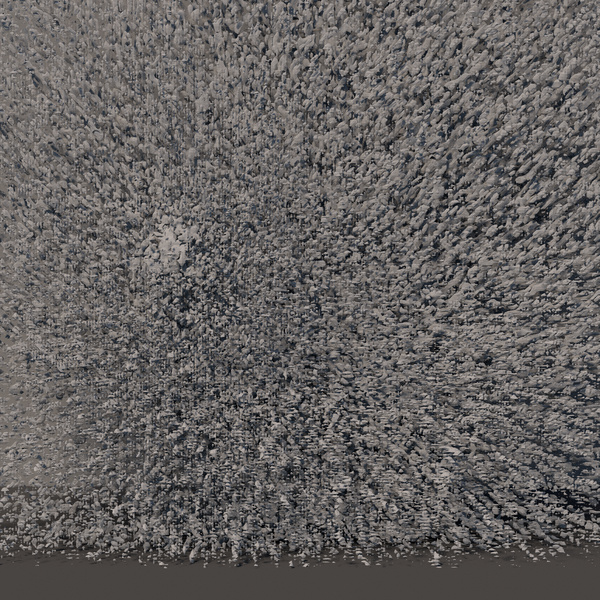} &%
\includegraphics[height=0.18\linewidth]{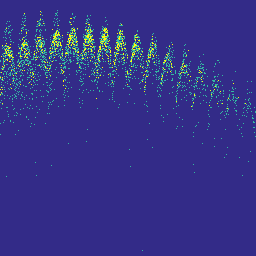} \\
\scriptsize{$\text{noise}_{L2,\text{rel}}= 14.9 \%$} & \scriptsize{$\text{noise}_{L2,\text{rel}}= 25.9 \%$} & \scriptsize{$\text{noise}_{L2,\text{rel}}= 47.1 \%$} & \scriptsize{$\text{noise}_{L2,\text{rel}} = 81.5 \%$} & \scriptsize{$\text{noise}_{L2,\text{rel}}= 149.3 \%$}
\end{tabular}
\caption{Reconstruction of the \texttt{BunnyGI} dataset with different levels of Poisson noise applied to the input data. Relative $L2$ error from left to right: $14.9 \%$, $25.9 \%$, $47.1 \%$, $81.5 \%$, $149.3 \%$. Top row: Our reconstruction, bottom row: backprojection. Our algorithm  is based on a noise-free forward model. It therefore manages to localize the object reliably even under very noisy conditions (albeit at reduced reconstruction quality). In the rightmost example (streak plot), at most two photons have been counted per pixel, resulting in data that contains $50 \%$ more noise than signal.}
\label{fig:bunny:noise}
\end{figure*}

\paragraph{Synthetic datasets} After establishing in \Sec{sec:rendereval} that our fast renderer produces outcomes that are almost identical to the ray-traced reference, we used both the path tracer and our fast renderer to generate a variety of around-the-corner input data. In particular, we prepared several variations of the \texttt{Mannequin}  scene, reducing the number of pixels, the number of laser spots, as well as the temporal resolution. An overview of our datasets, as well as the parameters used for reconstruction, can be found in \Tab{tbl:parameters}. Like the backprojection method, our method has a small number of parameters: the blob size upper bound $\sigma_0$ and the regularization parameter $\eta$.

We show renderings of the reconstructed meshes alongside the backprojected solutions, obtained using the Fast Backprojection code provided by Arellano et al.~\shortcite{ArellanoOpEx2017}, and ground truth (Figure~\ref{fig:teaser}(c)).
They show that the quality delivered by our algorithm, in general, outperforms the state-of-the-art method on the synthetic datasets examined in this study.
The meshes produced by our method tend to be more complete, smoother, and overall closer to the true surface.
We also performed more quantitative evaluations.
Figures \ref{fig:deptherror} and \ref{fig:deptherror_plots} show the error of the recovered surface in $z$-direction for three datasets.
We present backprojection results using two different filters: the Laplacian filter, since it is the most popular choice in literature, and the modified Difference of Gaussian filter \cite{Laurenzis:2014a}, which produced the most accurate and low-noise results from a wide range of tested filters.
In general, meshes generated using the backprojection method tend to lie in front of the true surface.
This is due to the way surface geometry is reconstructed from the density volumes obtained by the backprojection algorithm.
Even if the peak of the density distribution lies exactly on the object geometry, extracting an isosurface will displace it by a certain distance.
We were able to reduce this effect for the modified DoG filter by suppressing non-maximum density values along the $z$-axis.
However, using this filter, we were still not able to produce reconstructions that achieved the same high coverage and low error as our method.
It is the combination of a surface-oriented scattering model, in combination with our analysis-by-synthesis scheme, which vitally enables a reconstruction with systematic errors as low as the ones exhibited by our method.

\paragraph{Degradation experiments} To put the robustness of our method to the test, we performed a series of experiments that deliberately deviate from an idealized, noise-free, Lambertian and global-illumination-free light transport model, or reduce the amount of input data used for the reconstruction. In a first series of experiments, we sub-sampled the \texttt{Mannequin} dataset both spatially and temporally, and observed the degradation in reconstructed outcome (\Fig{fig:mannequin:degradations}). In a second series, we added increasing amounts of Poisson noise (\Fig{fig:bunny:noise}). Next, we replaced the diffuse reflectance of the \texttt{BunnyGI} model by a metal BRDF (Blinn model as implemented by \texttt{pbrt}) and decreased the roughness value (\Fig{fig:metalbunny}). Our fast renderer used during reconstruction was set to the same BRDF parameters that were used to generate the input data. Finally, we constructed a strongly concave synthetic scene (\texttt{Bowl}) and used high albedo values in order to test the influence of unaccounted-for global illumination on the reconstructed geometry (\Fig{fig:bowls}).

As expected, in all these examples, the further the data deviates from the ideal case, the more the reconstruction quality decreases. While backprojection tends to be more robust with respect to low-frequency bias (\texttt{Bowl} experiment), our method quite gracefully deals with high-frequency noise by fitting a low-frequent rendering to it. For highly specular materials, the discretization of the surface mesh and the sensing locations on the wall may lead to sampling issues: specular glints that are missed by the forward simulation cannot contribute to the solution.

\paragraph{Experimental datasets} We show reconstructions of {two experimental datasets} obtained using SPAD sensors. 

The first dataset (\texttt{SPADScene}) was measured by Buttafava et al.~\shortcite{Buttafava:2015}, by observing a single location on the wall with a SPAD detector, and scanning a pulsed laser to a rectangular grid of locations. We note that this setup is dual, and hence equivalent for our purpose, to illuminating the single spot and scanning the detector to the grid of different locations. The dataset came included with the Fast Backprojection code provided by Arellano et al.~\shortcite{ArellanoOpEx2017}.
To apply our algorithm on the \texttt{SPADScene} dataset, we first subtracted a lowpass-filtered version (with $\sigma=1000$\,bins) of the signal to reduce noise and background, then downsampled the dataset from its original temporal resolution by a factor of 25.

\begin{figure}[ht]
\includegraphics[height=0.25\linewidth]{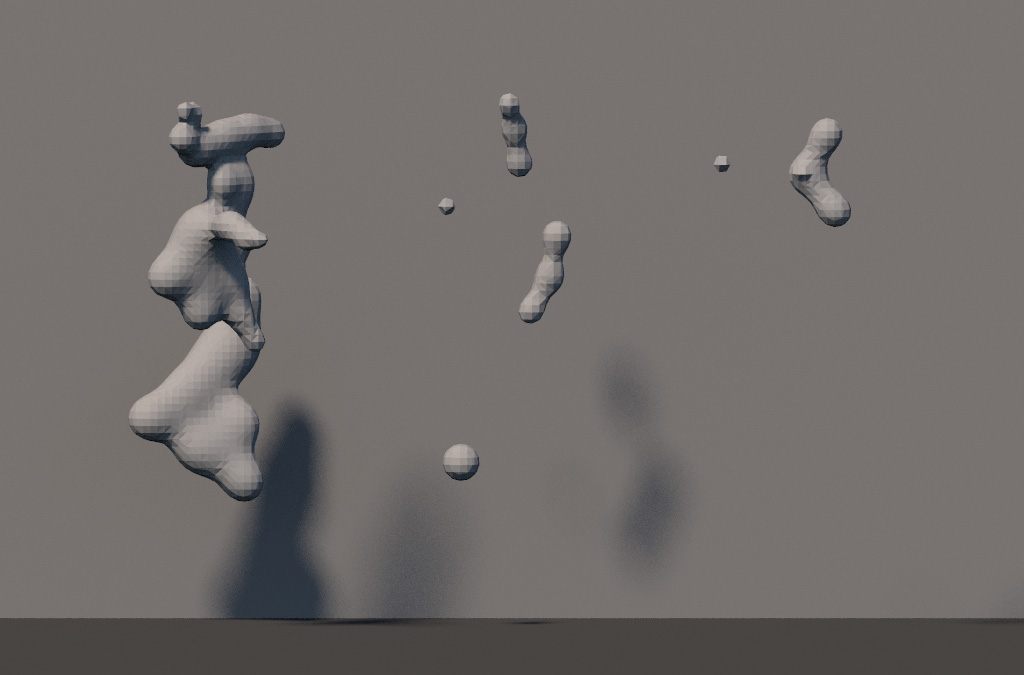}\hfill
\includegraphics[height=0.25\linewidth]{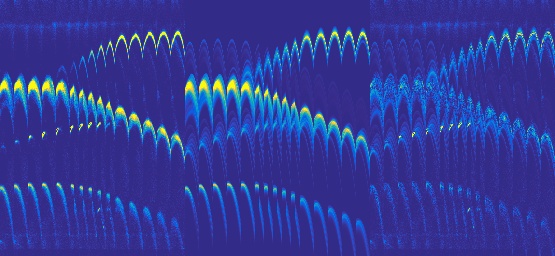}
\caption{Reconstruction of the experimental \texttt{SPADScene} dataset \cite{Buttafava:2015}. Shown is the output mesh and the transient data (from left to right: observation, prediction, residual).}
\label{fig:spadscene}
\end{figure}

\begin{figure}[ht]
\includegraphics[width=0.6\linewidth]{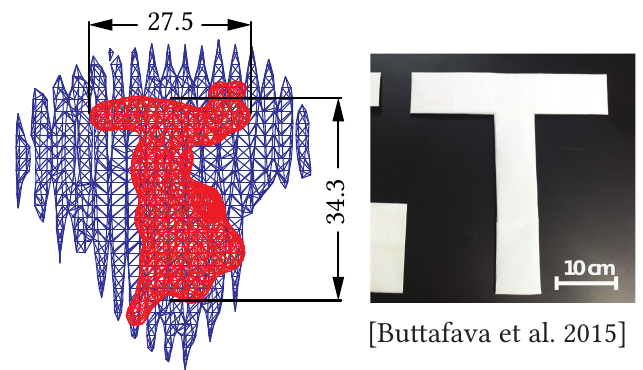}
\caption{The ``T'' object from the experimental \texttt{SPADScene} dataset published by Buttafava et al.~\shortcite{Buttafava:2015}. Shown are reconstructions obtained using backprojection (blue) and the proposed method (red), along with approximate dimensions using the scale provided in the original work (right).}
\label{fig:t}
\end{figure}

Like in the original work, the reconstruction remains vague and precise details are hard to make out (\Fig{fig:spadscene}). The reconstructed blobby objects appear to be in roughly the right places, but their shapes are poorly defined. We note that our method quite clearly carves out the letter ``T'' where backprojection delivers a less clearly defined shape (\Fig{fig:t}).

\begin{figure}[ht]
\centering
\includegraphics[height=0.3\linewidth]{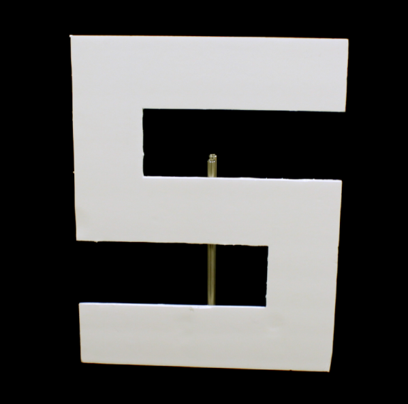}\hfill
\includegraphics[height=0.3\linewidth]{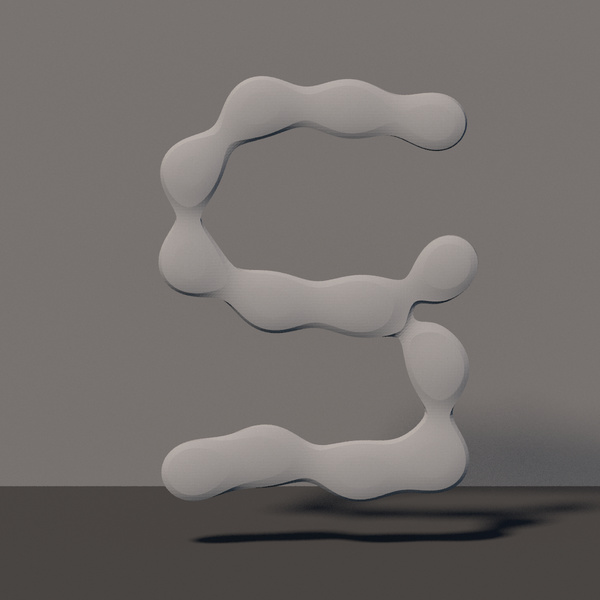}\hfill
\includegraphics[height=0.3\linewidth]{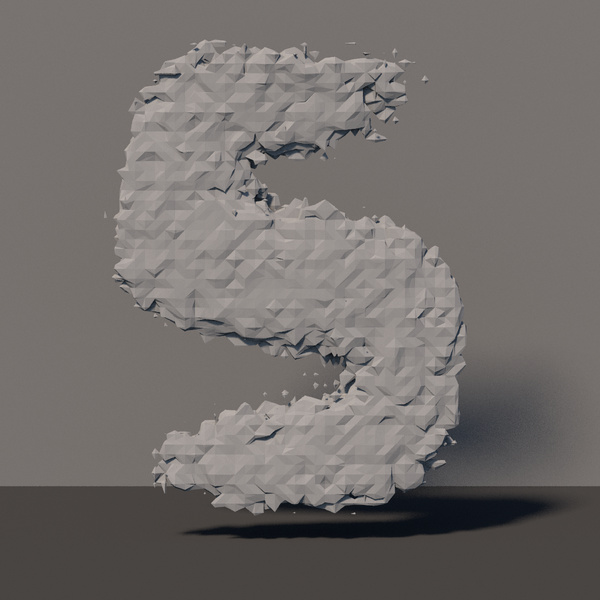} 
\caption{\texttt{OTooleDiffuseS} dataset \cite{otoole2018}. From left to right: photo of diffuse ``S''-shaped cutout
; surface mesh reconstructed using our method; mesh reconstructed using method described in \cite{otoole2018}.}
\label{fig:diffuseS}
\end{figure}

The second dataset (\texttt{OTooleDiffuseS}) is a measurement of a letter ``S'' cut from white cardboard, which O'Toole et al.~measured via a diffuse wall using their confocal setup \cite{otoole2018}. In this setup, illumination and observation share the same optical path and are scanned across the surface. We downsampled the input data by a factor of $4\times 4\times 4$ in the spatial and temporal domains. Although the inclusion of the direct reflection in the data allowed for a better background subtraction and white point correction than in the case of the previous dataset, it becomes clear that there must be more sources of bias. In particular, we identified a temporal blur of roughly 3 time bins. Adding a similar blur to our renderer (a box filter of width 3 bins), made the reconstructed ``S'' shape much more clearly recognizable as such (\Fig{fig:diffuseS}).

 \section{Discussion}
In the proposed approach, we develop computer graphics methodology (a near-physical, extremely efficient rendering scheme) to reconstruct occluded 3D shape from three-bounce indirect reflections. To our knowledge, this marks the first instance of a non-line-of-sight reconstruction algorithm that is consistent with a physical forward model. This solid theoretical foundation leads to results that, under favorable conditions, show higher object coverage and detail than the de-facto state of the art, error backprojection. In extreme situations, like very low spatial / temporal resolutions or high noise levels, we have shown that our method breaks down significantly later than the current state of the art (Figures~\ref{fig:mannequin:degradations} and~\ref{fig:bunny:noise}). Under conditions that are not covered by the forward model (noise, bias/background, global illumination) the results are on par or slightly inferior to existing methods. In terms of runtime, our method typically takes several hours or even days for a reconstruction run (\Tab{tbl:parameters}) and therefore cannot compete with recent optimized versions of error backprojection \cite{ArellanoOpEx2017} or GPU-based deconvolvers \cite{otoole2018a}, which are typically on the order of \unit{10}{s} to \unit{100}{s} and \unit{1}{s} respectively. However, we consider this a soft hindrance that has to be considered together with the fact that the capture of suitable input data, too, is far from being instantaneous. This latter factor is governed by the physics of light and therefore may turn out, in the long run, to impose more severe limitations to the practicality of non-line-of-sight sensing solutions. 

We noted that the reconstruction quality of the SPAD datasets stays behind the quality of the synthetic datasets (whether path-traced or using our own renderer). Our image formation model approximates the physical light transport up to very high accuracy (as shown in \Sec{sec:rendereval}), but does not explicitly model the SPAD sensor response to the incoming light. The SPAD data is biased due to background noise and dark counts, and the temporal impulse response is asymmetric and smeared out due to time jitter and afterpulsing \cite{gulinatti2011, hernandez2017computational}. While these effects could easily be incorporated into our forward model, doing so would require either a careful calibration of the imaging setup (which was not provided with the public datasets) or an estimation of the noise parameters from input data. In this light, we find the presented results very promising for this line of research, and consider the explicit application of measured noise profiles and the modeling of additional imaging setups as future work.

A key feature of our method is that, within the limitations of the forward model (opaque, but not necessarily diffuse, light transport without further interreflections) good solutions can be immediately identified by a low residual error.
However, the non-convex objective and possibly unknown noise and background terms may make it challenging to reach this point. Our optimization scheme, while delivering good results in the provided examples, offers no guarantee of global convergence. As of today, it is unclear which of the two factors will prove more important in practice, the physical correctness of the forward model or the minimizability of the objective derived from it. 

\section{Future work}
We imagine that extended versions of our method could be used to jointly estimate geometry and material. Advanced global optimization heuristics could further improve the convergence behavior and the overall quality of the outcome.
We imagine that hierarchical approaches or hybrid solutions might bring further improvement, for instance by using the (physically inaccurate but global) solution of one reconstruction scheme to warm-start another local optimization run using a more accurate model like ours.

\begin{figure}[t]
\centering\includegraphics[width=0.9\columnwidth]{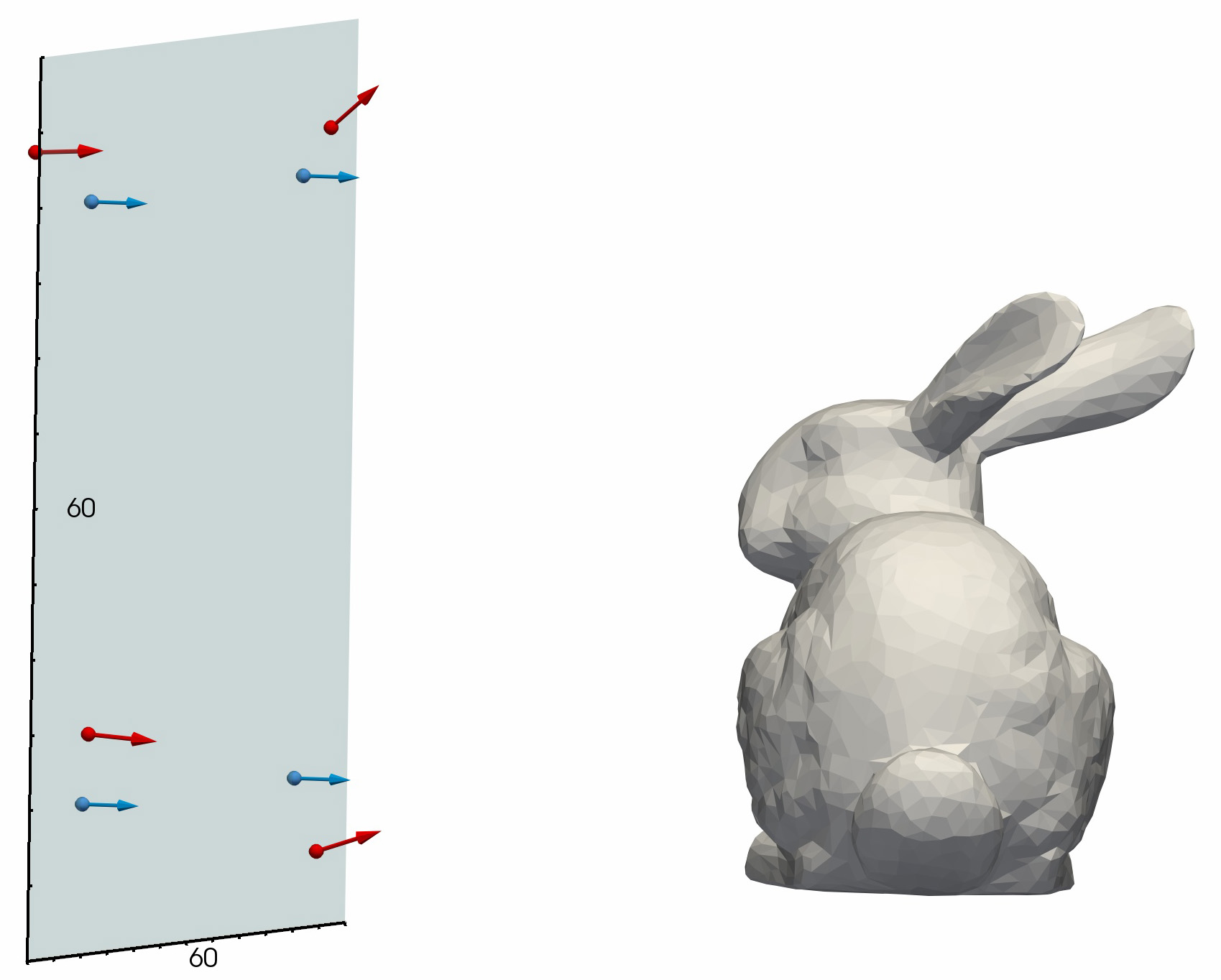}
\caption{An illustration of our preliminary NLOS camera calibration experiment. A transient image of the \texttt{Bunny} scene has been acquired using our transient renderer. The blue arrows denote the ground truth positions and normals of the projected camera pixels. The red arrows show an initial guess before optimization with a positional RMSE of $2.7$ units and an average angular error of \unit{16.6}{\degree}. After optimization using \Eq{eq:objective}, the optimized pixel positions and normals coincide with the ground truth up to floating point precision. Due to the greatly reduced number of variables compared to the geometry reconstruction problem, the optimization concluded in less than one minute.}
\label{fig:camera_calibration}
\end{figure}

The extrinsic and intrinsic calibration of traditional 2D imaging setups is well understood 
\cite{zhang1999}.
However, this problem has not been satisfyingly solved by the NLOS reconstruction community so far.
The current best practice is to manually estimate the positions and normals of the projected camera pixels, potentially leading to a systematic bias in the (typically non-metric) reconstructions.
Our proposed method presents not only an alternative solution to NLOS reconstruction, but also lays out a foundation for solving related problems.
Here, we presented a method for recovering the scene geometry, where the acquisition geometry was assumed to be known.
In future work, we would like to study the \textit{dual problem}, where the scene geometry is known (a calibration target), but the acquisition geometry is unknown.
We conducted initial experiments with our synthetic \texttt{Bunny} dataset and were able to recover the positions and normals of four projected pixels up to a very high precision, regardless of an overly imprecise initial guess, see \Fig{fig:camera_calibration}.
Again, we utilize \Eq{eq:objective} as the objective function, but the parameter vector $\vec{P}$ consists of the positions and normals of the projected pixels.
Challenges will include the generalization to real-world data, the design of an optimum calibration target, and the validation against measured data.
We could also imagine utilizing our forward model to estimate the parameters of a SPAD sensor response model \cite{hernandez2017computational}.

Finally, our renderer is not constrained to use in a costly iterative solver.
Just as well, we can imagine using it to enable new machine learning approaches to the problem.
A suitably trained feedforward neural network, for example, would deliver instant results.
Whereas existing renderers are too slow for generating large amounts of training data, our renderer would be fast enough to obtain millions of datasets in a single day.
Together with a suitable signal degradation model \cite{hernandez2017computational}, we expect that it will be possible to closely approximate the most relevant real-world scenarios.

\begin{table*}[t!]\centering
\footnotesize
\begin{tabular}{l|r|r|r|r|r|r|r|r|r|r|r|r|r}
\multicolumn{10}{c}{}\\[-1mm] {\bf Name} & {Reference} & {Resolution} & {\# Lasers} & $s_\text{geom}$ & $s_\text{camera}$ & {$\eta$} & {$\sigma_0$} & {$t_0$} & {$\delta_t$} & {$\nicefrac{c_n}{c_0}$ [\%]}& T [min] & $n_\text{iters}$ & $n_\text{blobs}$\\
\hline
\texttt{Bunny} & Ours & $16 \times 16 \times 256$ & 4 & $40 \times 40$ & $80 \times 80$ & 1.01 & 1.5 & 80 & 0.4 & 0.32 & 5096 & 660 & 156\\
\texttt{BunnyGI} & pbrt & $16 \times 16 \times 256$ & 4 & $40 \times 40$ & $80 \times 80$ & 1.01 & 1.5 & 80 & 0.4 & 0.59 & 5611 & 181 & 109\\
\texttt{BunnyMetal0.05} & pbrt &  $16 \times 16 \times 256$ & 4 & $40 \times 40$ & $80 \times 80$ & 1.01 & 1.5 & 80 & 0.4 & 2.02 & 3419 & 259 & 101\\
\texttt{BunnyMetal0.01} & pbrt &  $16 \times 16 \times 256$ & 4 & $40 \times 40$ & $80 \times 80$ & 1.01 & 1.5 & 80 & 0.4 & 11.75 & 3005 & 361 & 87\\
\texttt{BowlAlbedo0.3} & pbrt &  $16 \times 16 \times 256$ & 4 & $26 \times 26$ & $80 \times 80$ & 1.001 & 0.4 & 80 & 0.4 & 4.36 & 5579 & 167 & 155\\
\texttt{BowlAlbedo1} & pbrt &  $16 \times 16 \times 256$ & 4 & $26 \times 26$ & $80 \times 80$ & 1.001 & 0.4 & 80 & 0.4 & 31.53 & 4280 & 267 & 197\\
\texttt{Mannequin} & Ours & $16 \times 16 \times 256$ & 4 & $40 \times 494$ & $480 \times 80$ & 1.005 & 1.5 & 90 & 0.4 & 1.46 & 2326 & 505 & 69\\
\texttt{MannequinLowRes} & Ours & $4 \times 4 \times 256$ & 4 & $40 \times 49$ & $80 \times 80$ & 1.005 & 1.5 & 90 & 0.4 &  1.41 & 1251 & 252 & 76\\
\texttt{MannequinMinRes} & Ours & $2 \times 2 \times 256$ & 4 & $40 \times 49$ & $80 \times 80$ & 1.005 & 1.5 & 90 & 0.4 &  4.44 & 931 & 350 & 101\\
\texttt{MannequinLowTemp} & Ours & $16 \times 16 \times 32$ & 4 & $40 \times 49$ & $80 \times 80$ & 1.005 & 1.5 & 90 & 3.2 &  1.21 & 1322 & 166 & 67\\
\texttt{MannequinMinTemp} & Ours & $16 \times 16 \times 8$ & 4 & $40 \times 49$ & $80 \times 80$ & 1.005 & 1.5 & 90 & 12.8 & 7.95 & 420 & 102 & 23\\
\texttt{Mannequin1Laser} & Ours & $16 \times 16 \times 256$ & 1 & $40 \times 49$ & $80 \times 80$ & 1.005 & 1.5 & 90 & 0.4 & 0.59 & 1419 & 243 & 57\\
\texttt{SPADScene} & Measured & $185 \times 1 \times 256$ & 1 & --- & --- & 1.005 & 4.5 & 373 & 0.748 & 20.31 & 1280 & 328 & 43\\
\texttt{OTooleDiffuseS} & Measured & $64 \times 64 \times 2048$ & 1 & --- & --- & 1.01 & 0.015 & 0.756 & 0.0012 & 33.43 & 67 & 13 & 13
\end{tabular}
\caption{Parameters of our reconstructed scenes, where $s_\text{geom}$ is the size of the ground truth object projected onto the diffuse camera wall in world units, $s_\text{camera}$ is the area covered by the camera in world units, $\eta$ is the drop deletion factor in \Alg{alg:subroutines}, $\sigma_0$ is the initial blob standard deviation, $t_0$ is the time stamp of the first time bin, $\delta_t$ is the size of a time bin, and $\nicefrac{c_n}{c_0}$ is the residual cost after optimization (relative to the initial cost). The total reconstruction times $T$ are taken from file timestamps and vary due to manual termination of the reconstruction procedure, execution on different GPU models, overhead through parallel execution of multiple jobs, as well as debugging output. The optimizations terminated after $n_{\text{iters}}$ iterations and consist of $n_\text{blobs}$ Gaussian blobs. Please note that the exact scene geometry is only known for the synthetic experiments.}
\label{tbl:parameters}
\end{table*}
 \balance
\bibliography{main}


\begin{thebibliography}{46}


\ifx \showCODEN    \undefined \def \showCODEN     #1{\unskip}     \fi
\ifx \showDOI      \undefined \def \showDOI       #1{#1}\fi
\ifx \showISBNx    \undefined \def \showISBNx     #1{\unskip}     \fi
\ifx \showISBNxiii \undefined \def \showISBNxiii  #1{\unskip}     \fi
\ifx \showISSN     \undefined \def \showISSN      #1{\unskip}     \fi
\ifx \showLCCN     \undefined \def \showLCCN      #1{\unskip}     \fi
\ifx \shownote     \undefined \def \shownote      #1{#1}          \fi
\ifx \showarticletitle \undefined \def \showarticletitle #1{#1}   \fi
\ifx \showURL      \undefined \def \showURL       {\relax}        \fi
\providecommand\bibfield[2]{#2}
\providecommand\bibinfo[2]{#2}
\providecommand\natexlab[1]{#1}
\providecommand\showeprint[2][]{arXiv:#2}

\bibitem[\protect\citeauthoryear{Abramson}{Abramson}{1978}]%
        {Abramson:78}
\bibfield{author}{\bibinfo{person}{Nils Abramson}.}
  \bibinfo{year}{1978}\natexlab{}.
\newblock \showarticletitle{Light-in-flight recording by holography}.
\newblock \bibinfo{journal}{\emph{Optics Letters}} \bibinfo{volume}{3},
  \bibinfo{number}{4} (\bibinfo{date}{Oct} \bibinfo{year}{1978}),
  \bibinfo{pages}{121--123}.
\newblock
\urldef\tempurl%
\url{https://doi.org/10.1364/OL.3.000121}
\showDOI{\tempurl}


\bibitem[\protect\citeauthoryear{Agarwal, Mierle, and {Others~[sic]}}{Agarwal
  et~al\mbox{.}}{2015}]%
        {ceres-solver}
\bibfield{author}{\bibinfo{person}{Sameer Agarwal}, \bibinfo{person}{Keir
  Mierle}, {and} \bibinfo{person}{{Others~[sic]}}.}
  \bibinfo{year}{2015}\natexlab{}.
\newblock \bibinfo{title}{Ceres Solver}.
\newblock \bibinfo{howpublished}{\url{http://ceres-solver.org}}.
\newblock


\bibitem[\protect\citeauthoryear{Arellano, Gutierrez, and Jarabo}{Arellano
  et~al\mbox{.}}{2017}]%
        {ArellanoOpEx2017}
\bibfield{author}{\bibinfo{person}{Victor Arellano}, \bibinfo{person}{Diego
  Gutierrez}, {and} \bibinfo{person}{Adrian Jarabo}.}
  \bibinfo{year}{2017}\natexlab{}.
\newblock \showarticletitle{Fast back-projection for non-line of sight
  reconstruction}.
\newblock \bibinfo{journal}{\emph{Optics Express}} \bibinfo{volume}{25},
  \bibinfo{number}{10} (\bibinfo{year}{2017}).
\newblock


\bibitem[\protect\citeauthoryear{Boger-Lombard and Katz}{Boger-Lombard and
  Katz}{2018}]%
        {boger2018non}
\bibfield{author}{\bibinfo{person}{Jeremy Boger-Lombard} {and}
  \bibinfo{person}{Ori Katz}.} \bibinfo{year}{2018}\natexlab{}.
\newblock \showarticletitle{Non line-of-sight localization by passive optical
  time-of-flight}.
\newblock \bibinfo{journal}{\emph{arXiv preprint arXiv:1808.01000}}
  (\bibinfo{year}{2018}).
\newblock


\bibitem[\protect\citeauthoryear{Bouman, Ye, Yedidia, Durand, Wornell,
  Torralba, and Freeman}{Bouman et~al\mbox{.}}{2017}]%
        {bouman2017turning}
\bibfield{author}{\bibinfo{person}{Katherine~L Bouman}, \bibinfo{person}{Vickie
  Ye}, \bibinfo{person}{Adam~B Yedidia}, \bibinfo{person}{Fr{\'e}do Durand},
  \bibinfo{person}{Gregory~W Wornell}, \bibinfo{person}{Antonio Torralba},
  {and} \bibinfo{person}{William~T Freeman}.} \bibinfo{year}{2017}\natexlab{}.
\newblock \showarticletitle{Turning Corners into Cameras: Principles and
  Methods}. In \bibinfo{booktitle}{\emph{Proceedings of the IEEE Conference on
  Computer Vision and Pattern Recognition}}. \bibinfo{pages}{2270--2278}.
\newblock


\bibitem[\protect\citeauthoryear{Buttafava, Zeman, Tosi, Eliceiri, and
  Velten}{Buttafava et~al\mbox{.}}{2015}]%
        {Buttafava:2015}
\bibfield{author}{\bibinfo{person}{Mauro Buttafava}, \bibinfo{person}{Jessica
  Zeman}, \bibinfo{person}{Alberto Tosi}, \bibinfo{person}{Kevin Eliceiri},
  {and} \bibinfo{person}{Andreas Velten}.} \bibinfo{year}{2015}\natexlab{}.
\newblock \showarticletitle{Non-line-of-sight imaging using a time-gated single
  photon avalanche diode}.
\newblock \bibinfo{journal}{\emph{Optics Express}} \bibinfo{volume}{23},
  \bibinfo{number}{16} (\bibinfo{year}{2015}), \bibinfo{pages}{20997--21011}.
\newblock


\bibitem[\protect\citeauthoryear{Carr, Beatson, Cherrie, Mitchell, Fright,
  McCallum, and Evans}{Carr et~al\mbox{.}}{2001}]%
        {carr2001reconstruction}
\bibfield{author}{\bibinfo{person}{Jonathan Carr}, \bibinfo{person}{Richard
  Beatson}, \bibinfo{person}{Jon Cherrie}, \bibinfo{person}{Tim Mitchell},
  \bibinfo{person}{W Fright}, \bibinfo{person}{Bruce McCallum}, {and}
  \bibinfo{person}{Tim Evans}.} \bibinfo{year}{2001}\natexlab{}.
\newblock \showarticletitle{Reconstruction and representation of {3D} objects
  with radial basis functions}. In \bibinfo{booktitle}{\emph{Proc.~28th Annual
  Conf.~on Computer Graphics and Interactive Techniques}}. ACM,
  \bibinfo{pages}{67--76}.
\newblock


\bibitem[\protect\citeauthoryear{Debevec}{Debevec}{1998}]%
        {debeveclightprobes}
\bibfield{author}{\bibinfo{person}{Paul Debevec}.}
  \bibinfo{year}{1998}\natexlab{}.
\newblock \bibinfo{title}{Light Probe Image Gallery}.
\newblock
\newblock
\newblock
\shownote{\texttt{http://www.pauldebevec.com/Probes/}.}


\bibitem[\protect\citeauthoryear{Fuchs}{Fuchs}{2010}]%
        {fuchs2010multipath}
\bibfield{author}{\bibinfo{person}{Stefan Fuchs}.}
  \bibinfo{year}{2010}\natexlab{}.
\newblock \showarticletitle{Multipath interference compensation in
  time-of-flight camera images}. In \bibinfo{booktitle}{\emph{Pattern
  Recognition (ICPR), 2010 20th International Conference on}}. IEEE,
  \bibinfo{pages}{3583--3586}.
\newblock


\bibitem[\protect\citeauthoryear{Gariepy, Krstajic, Henderson, Li, Thomson,
  Buller, Heshmat, Raskar, Leach, and Faccio}{Gariepy et~al\mbox{.}}{2015}]%
        {gariepy2015single}
\bibfield{author}{\bibinfo{person}{G Gariepy}, \bibinfo{person}{N Krstajic},
  \bibinfo{person}{R Henderson}, \bibinfo{person}{C Li}, \bibinfo{person}{RR
  Thomson}, \bibinfo{person}{GS Buller}, \bibinfo{person}{B Heshmat},
  \bibinfo{person}{R Raskar}, \bibinfo{person}{J Leach}, {and}
  \bibinfo{person}{D Faccio}.} \bibinfo{year}{2015}\natexlab{}.
\newblock \showarticletitle{Single-photon sensitive light-in-flight imaging}.
\newblock \bibinfo{journal}{\emph{Nature Communications}}  \bibinfo{volume}{6}
  (\bibinfo{year}{2015}).
\newblock


\bibitem[\protect\citeauthoryear{Gariepy, Tonolini, Henderson, Leach, and
  Faccio}{Gariepy et~al\mbox{.}}{2016}]%
        {Gariepy:2016}
\bibfield{author}{\bibinfo{person}{Genevieve Gariepy},
  \bibinfo{person}{Francesco Tonolini}, \bibinfo{person}{Robert Henderson},
  \bibinfo{person}{Jonathan Leach}, {and} \bibinfo{person}{Daniele Faccio}.}
  \bibinfo{year}{2016}\natexlab{}.
\newblock \showarticletitle{Detection and tracking of moving objects hidden
  from view}.
\newblock \bibinfo{journal}{\emph{Nature Photonics}} \bibinfo{volume}{10},
  \bibinfo{number}{1} (\bibinfo{year}{2016}).
\newblock


\bibitem[\protect\citeauthoryear{Gkioulekas, Zhao, Bala, Zickler, and
  Levin}{Gkioulekas et~al\mbox{.}}{2013}]%
        {gkioulekas2013inverse}
\bibfield{author}{\bibinfo{person}{Ioannis Gkioulekas}, \bibinfo{person}{Shuang
  Zhao}, \bibinfo{person}{Kavita Bala}, \bibinfo{person}{Todd Zickler}, {and}
  \bibinfo{person}{Anat Levin}.} \bibinfo{year}{2013}\natexlab{}.
\newblock \showarticletitle{Inverse volume rendering with material
  dictionaries}.
\newblock \bibinfo{journal}{\emph{ACM Transactions on Graphics (TOG)}}
  \bibinfo{volume}{32}, \bibinfo{number}{6} (\bibinfo{year}{2013}),
  \bibinfo{pages}{162}.
\newblock


\bibitem[\protect\citeauthoryear{Gulinatti, Rech, Assanelli, Ghioni, and
  Cova}{Gulinatti et~al\mbox{.}}{2011}]%
        {gulinatti2011}
\bibfield{author}{\bibinfo{person}{A. Gulinatti}, \bibinfo{person}{I. Rech},
  \bibinfo{person}{M. Assanelli}, \bibinfo{person}{M. Ghioni}, {and}
  \bibinfo{person}{S. Cova}.} \bibinfo{year}{2011}\natexlab{}.
\newblock \showarticletitle{A physically based model for evaluating the photon
  detection efficiency and the temporal response of {{SPAD}} detectors}.
\newblock \bibinfo{journal}{\emph{Journal of Modern Optics}}
  \bibinfo{volume}{58}, \bibinfo{number}{3-4} (\bibinfo{year}{2011}),
  \bibinfo{pages}{210--224}.
\newblock
\urldef\tempurl%
\url{https://doi.org/10.1080/09500340.2010.536590}
\showDOI{\tempurl}


\bibitem[\protect\citeauthoryear{Heide, Hullin, Gregson, and Heidrich}{Heide
  et~al\mbox{.}}{2013}]%
        {Heide:2013:LBT}
\bibfield{author}{\bibinfo{person}{Felix Heide}, \bibinfo{person}{Matthias~B.
  Hullin}, \bibinfo{person}{James Gregson}, {and} \bibinfo{person}{Wolfgang
  Heidrich}.} \bibinfo{year}{2013}\natexlab{}.
\newblock \showarticletitle{Low-budget Transient Imaging using Photonic Mixer
  Devices}.
\newblock \bibinfo{journal}{\emph{ACM Transactions on Graphics (Proc.
  SIGGRAPH)}} \bibinfo{volume}{32}, \bibinfo{number}{4} (\bibinfo{year}{2013}),
  \bibinfo{pages}{45:1--45:10}.
\newblock


\bibitem[\protect\citeauthoryear{Heide, O'Toole, Zhang, Lindell, Diamond, and
  Wetzstein}{Heide et~al\mbox{.}}{2017}]%
        {heideNLOS2017}
\bibfield{author}{\bibinfo{person}{Felix Heide}, \bibinfo{person}{Matthew
  O'Toole}, \bibinfo{person}{Kai Zhang}, \bibinfo{person}{David~B. Lindell},
  \bibinfo{person}{Steven Diamond}, {and} \bibinfo{person}{Gordon Wetzstein}.}
  \bibinfo{year}{2017}\natexlab{}.
\newblock \showarticletitle{Robust Non-line-of-sight Imaging with Single Photon
  Detectors}.
\newblock \bibinfo{journal}{\emph{CoRR}}  \bibinfo{volume}{abs/1711.07134}
  (\bibinfo{year}{2017}).
\newblock
\showeprint[arxiv]{1711.07134}
\urldef\tempurl%
\url{http://arxiv.org/abs/1711.07134}
\showURL{%
\tempurl}


\bibitem[\protect\citeauthoryear{Heide, Xiao, Heidrich, and Hullin}{Heide
  et~al\mbox{.}}{2014}]%
        {Heide:2014}
\bibfield{author}{\bibinfo{person}{Felix Heide}, \bibinfo{person}{Lei Xiao},
  \bibinfo{person}{Wolfgang Heidrich}, {and} \bibinfo{person}{Matthias~B.
  Hullin}.} \bibinfo{year}{2014}\natexlab{}.
\newblock \showarticletitle{Diffuse Mirrors: {3D} Reconstruction from Diffuse
  Indirect Illumination Using Inexpensive Time-of-Flight Sensors}.
\newblock \bibinfo{journal}{\emph{IEEE Conf. on Computer Vision and Pattern
  Recognition (CVPR)}} (\bibinfo{year}{2014}).
\newblock


\bibitem[\protect\citeauthoryear{Hernandez, Gutierrez, and Jarabo}{Hernandez
  et~al\mbox{.}}{2017}]%
        {hernandez2017computational}
\bibfield{author}{\bibinfo{person}{Quercus Hernandez}, \bibinfo{person}{Diego
  Gutierrez}, {and} \bibinfo{person}{Adrian Jarabo}.}
  \bibinfo{year}{2017}\natexlab{}.
\newblock \showarticletitle{A Computational Model of a Single-Photon Avalanche
  Diode Sensor for Transient Imaging}.
\newblock \bibinfo{journal}{\emph{arXiv preprint arXiv:1703.02635}}
  (\bibinfo{year}{2017}).
\newblock


\bibitem[\protect\citeauthoryear{Jakob}{Jakob}{2010}]%
        {Mitsuba}
\bibfield{author}{\bibinfo{person}{Wenzel Jakob}.}
  \bibinfo{year}{2010}\natexlab{}.
\newblock \bibinfo{title}{Mitsuba renderer}.
\newblock
\newblock
\newblock
\shownote{http://www.mitsuba-renderer.org.}


\bibitem[\protect\citeauthoryear{Jarabo, Marco, Mu{\~n}oz, Buisan, Jarosz, and
  Gutierrez}{Jarabo et~al\mbox{.}}{2014}]%
        {jarabo2014framework}
\bibfield{author}{\bibinfo{person}{Adrian Jarabo}, \bibinfo{person}{Julio
  Marco}, \bibinfo{person}{Adolfo Mu{\~n}oz}, \bibinfo{person}{Raul Buisan},
  \bibinfo{person}{Wojciech Jarosz}, {and} \bibinfo{person}{Diego Gutierrez}.}
  \bibinfo{year}{2014}\natexlab{}.
\newblock \showarticletitle{A framework for transient rendering}.
\newblock \bibinfo{journal}{\emph{ACM Transactions on Graphics (TOG)}}
  \bibinfo{volume}{33}, \bibinfo{number}{6} (\bibinfo{year}{2014}),
  \bibinfo{pages}{177}.
\newblock


\bibitem[\protect\citeauthoryear{Jarabo, Masia, Marco, and Gutierrez}{Jarabo
  et~al\mbox{.}}{2017}]%
        {jarabo2017recent}
\bibfield{author}{\bibinfo{person}{Adrian Jarabo}, \bibinfo{person}{Belen
  Masia}, \bibinfo{person}{Julio Marco}, {and} \bibinfo{person}{Diego
  Gutierrez}.} \bibinfo{year}{2017}\natexlab{}.
\newblock \showarticletitle{Recent advances in transient imaging: A computer
  graphics and vision perspective}.
\newblock \bibinfo{journal}{\emph{Visual Informatics}} \bibinfo{volume}{1},
  \bibinfo{number}{1} (\bibinfo{year}{2017}), \bibinfo{pages}{65--79}.
\newblock


\bibitem[\protect\citeauthoryear{Kadambi, Zhao, Shi, and Raskar}{Kadambi
  et~al\mbox{.}}{2016}]%
        {Kadambi:2016:OIT:2882845.2836164}
\bibfield{author}{\bibinfo{person}{Achuta Kadambi}, \bibinfo{person}{Hang
  Zhao}, \bibinfo{person}{Boxin Shi}, {and} \bibinfo{person}{Ramesh Raskar}.}
  \bibinfo{year}{2016}\natexlab{}.
\newblock \showarticletitle{Occluded Imaging with Time-of-Flight Sensors}.
\newblock \bibinfo{journal}{\emph{ACM Transactions on Graphics}}
  \bibinfo{volume}{35}, \bibinfo{number}{2}, Article \bibinfo{articleno}{15}
  (\bibinfo{date}{March} \bibinfo{year}{2016}), \bibinfo{numpages}{12}~pages.
\newblock
\showISSN{0730-0301}
\urldef\tempurl%
\url{https://doi.org/10.1145/2836164}
\showDOI{\tempurl}


\bibitem[\protect\citeauthoryear{Katz, Heidmann, Fink, and Gigan}{Katz
  et~al\mbox{.}}{2014}]%
        {Katz:2014}
\bibfield{author}{\bibinfo{person}{Ori Katz}, \bibinfo{person}{Pierre
  Heidmann}, \bibinfo{person}{Mathias Fink}, {and} \bibinfo{person}{Sylvain
  Gigan}.} \bibinfo{year}{2014}\natexlab{}.
\newblock \showarticletitle{Non-invasive single-shot imaging through scattering
  layers and around corners via speckle correlations}.
\newblock \bibinfo{journal}{\emph{Nature Photonics}} \bibinfo{volume}{8},
  \bibinfo{number}{10} (\bibinfo{year}{2014}), \bibinfo{pages}{784--790}.
\newblock


\bibitem[\protect\citeauthoryear{Kirmani, Hutchison, Davis, and Raskar}{Kirmani
  et~al\mbox{.}}{2009}]%
        {Kirmani:2009}
\bibfield{author}{\bibinfo{person}{A. Kirmani}, \bibinfo{person}{T. Hutchison},
  \bibinfo{person}{J. Davis}, {and} \bibinfo{person}{R. Raskar}.}
  \bibinfo{year}{2009}\natexlab{}.
\newblock \showarticletitle{Looking around the corner using transient imaging}.
  In \bibinfo{booktitle}{\emph{Proc. ICCV}}. \bibinfo{pages}{159--166}.
\newblock


\bibitem[\protect\citeauthoryear{Klein, Peters, Mart{\'\i}n, Laurenzis, and
  Hullin}{Klein et~al\mbox{.}}{2016}]%
        {klein2016tracking}
\bibfield{author}{\bibinfo{person}{Jonathan Klein}, \bibinfo{person}{Christoph
  Peters}, \bibinfo{person}{Jaime Mart{\'\i}n}, \bibinfo{person}{Martin
  Laurenzis}, {and} \bibinfo{person}{Matthias~B. Hullin}.}
  \bibinfo{year}{2016}\natexlab{}.
\newblock \showarticletitle{Tracking objects outside the line of sight using
  {2D} intensity images}.
\newblock \bibinfo{journal}{\emph{Scientific Reports}} \bibinfo{volume}{6},
  \bibinfo{number}{32491} (\bibinfo{year}{2016}).
\newblock


\bibitem[\protect\citeauthoryear{{La Manna}, Kine, Breitbach, Jackson, Sultan,
  and Velten}{{La Manna} et~al\mbox{.}}{2018}]%
        {lamanna}
\bibfield{author}{\bibinfo{person}{M. {La Manna}}, \bibinfo{person}{F. Kine},
  \bibinfo{person}{E. Breitbach}, \bibinfo{person}{J. Jackson},
  \bibinfo{person}{T. Sultan}, {and} \bibinfo{person}{A. Velten}.}
  \bibinfo{year}{2018}\natexlab{}.
\newblock \showarticletitle{Error Backprojection Algorithms for
  Non-Line-of-Sight Imaging}.
\newblock \bibinfo{journal}{\emph{IEEE Transactions on Pattern Analysis and
  Machine Intelligence}} (\bibinfo{year}{2018}), \bibinfo{pages}{1--1}.
\newblock
\showISSN{0162-8828}
\urldef\tempurl%
\url{https://doi.org/10.1109/TPAMI.2018.2843363}
\showDOI{\tempurl}


\bibitem[\protect\citeauthoryear{Laurenzis and Velten}{Laurenzis and
  Velten}{2014a}]%
        {Laurenzis:2014a}
\bibfield{author}{\bibinfo{person}{Martin Laurenzis} {and}
  \bibinfo{person}{Andreas Velten}.} \bibinfo{year}{2014}\natexlab{a}.
\newblock \showarticletitle{Feature selection and back-projection algorithms
  for nonline-of-sight laser–gated viewing}.
\newblock \bibinfo{journal}{\emph{Journal of Electronic Imaging}}
  \bibinfo{volume}{23}, \bibinfo{number}{6} (\bibinfo{year}{2014}),
  \bibinfo{pages}{1 -- 6}.
\newblock
\urldef\tempurl%
\url{https://doi.org/10.1117/1.JEI.23.6.063003}
\showDOI{\tempurl}


\bibitem[\protect\citeauthoryear{Laurenzis and Velten}{Laurenzis and
  Velten}{2014b}]%
        {Laurenzis:2014}
\bibfield{author}{\bibinfo{person}{Martin Laurenzis} {and}
  \bibinfo{person}{Andreas Velten}.} \bibinfo{year}{2014}\natexlab{b}.
\newblock \showarticletitle{Nonline-of-sight laser gated viewing of scattered
  photons}.
\newblock \bibinfo{journal}{\emph{Optical Engineering}} \bibinfo{volume}{53},
  \bibinfo{number}{2} (\bibinfo{year}{2014}), \bibinfo{pages}{023102--023102}.
\newblock


\bibitem[\protect\citeauthoryear{Levenberg}{Levenberg}{1944}]%
        {levenberg}
\bibfield{author}{\bibinfo{person}{Kenneth Levenberg}.}
  \bibinfo{year}{1944}\natexlab{}.
\newblock \showarticletitle{A Method for the Solution of Certain Non-linear
  Problems in Least Squares}.
\newblock \bibinfo{journal}{\emph{Quart. Appl. Math.}} \bibinfo{volume}{2},
  \bibinfo{number}{2} (\bibinfo{year}{1944}), \bibinfo{pages}{164--168}.
\newblock


\bibitem[\protect\citeauthoryear{Lorensen and Cline}{Lorensen and
  Cline}{1987}]%
        {Lorensen:1987:MCH:37401.37422}
\bibfield{author}{\bibinfo{person}{William Lorensen} {and}
  \bibinfo{person}{Harvey Cline}.} \bibinfo{year}{1987}\natexlab{}.
\newblock \showarticletitle{Marching {Cubes}: A High Resolution {3D} Surface
  Construction Algorithm}. In \bibinfo{booktitle}{\emph{Proc.~14th Annual
  Conf.~on Computer Graphics and Interactive Techniques}}
  \emph{(\bibinfo{series}{SIGGRAPH '87})}. \bibinfo{publisher}{ACM},
  \bibinfo{pages}{163--169}.
\newblock
\showISBNx{0-89791-227-6}
\urldef\tempurl%
\url{https://doi.org/10.1145/37401.37422}
\showDOI{\tempurl}


\bibitem[\protect\citeauthoryear{Marco, Jarosz, Gutierrez, and Jarabo}{Marco
  et~al\mbox{.}}{2017}]%
        {marco17transient}
\bibfield{author}{\bibinfo{person}{Julio Marco}, \bibinfo{person}{Wojciech
  Jarosz}, \bibinfo{person}{Diego Gutierrez}, {and} \bibinfo{person}{Adrian
  Jarabo}.} \bibinfo{year}{2017}\natexlab{}.
\newblock \showarticletitle{Transient Photon Beams}. In
  \bibinfo{booktitle}{\emph{Spanish Computer Graphics Conference (CEIG)}}.
  \bibinfo{publisher}{The Eurographics Association}.
\newblock
\showISBNx{978-3-03868-046-8}
\urldef\tempurl%
\url{https://doi.org/10.2312/ceig.20171216}
\showDOI{\tempurl}


\bibitem[\protect\citeauthoryear{Marquardt}{Marquardt}{1963}]%
        {marquardt}
\bibfield{author}{\bibinfo{person}{Donald~W. Marquardt}.}
  \bibinfo{year}{1963}\natexlab{}.
\newblock \showarticletitle{An algorithm for Least-Squares Estimation of
  Nonlinear Parameters}.
\newblock \bibinfo{journal}{\emph{SIAM J. Appl. Math.}} \bibinfo{volume}{11},
  \bibinfo{number}{2} (\bibinfo{year}{1963}), \bibinfo{pages}{431--441}.
\newblock
\urldef\tempurl%
\url{https://doi.org/10.1137/0111030}
\showDOI{\tempurl}


\bibitem[\protect\citeauthoryear{Naik, Zhao, Velten, Raskar, and Bala}{Naik
  et~al\mbox{.}}{2011}]%
        {Naik:2011}
\bibfield{author}{\bibinfo{person}{N. Naik}, \bibinfo{person}{S. Zhao},
  \bibinfo{person}{A. Velten}, \bibinfo{person}{R. Raskar}, {and}
  \bibinfo{person}{K. Bala}.} \bibinfo{year}{2011}\natexlab{}.
\newblock \showarticletitle{Single view reflectance capture using multiplexed
  scattering and time-of-flight imaging}.
\newblock \bibinfo{journal}{\emph{ACM Trans. Graph.}} \bibinfo{volume}{30},
  \bibinfo{number}{6} (\bibinfo{year}{2011}), \bibinfo{pages}{171}.
\newblock


\bibitem[\protect\citeauthoryear{O'Toole, Lindell, and Wetzstein}{O'Toole
  et~al\mbox{.}}{2018a}]%
        {otoole2018}
\bibfield{author}{\bibinfo{person}{Matthew O'Toole}, \bibinfo{person}{David~B.
  Lindell}, {and} \bibinfo{person}{Gordon Wetzstein}.}
  \bibinfo{year}{2018}\natexlab{a}.
\newblock \showarticletitle{Confocal non-line-of-sight imaging based on the
  light-cone transform}.
\newblock \bibinfo{journal}{\emph{Nature}} \bibinfo{volume}{555},
  \bibinfo{number}{25489} (\bibinfo{year}{2018}), \bibinfo{pages}{338--341}.
\newblock


\bibitem[\protect\citeauthoryear{O'Toole, Lindell, and Wetzstein}{O'Toole
  et~al\mbox{.}}{2018b}]%
        {otoole2018a}
\bibfield{author}{\bibinfo{person}{Matthew O'Toole}, \bibinfo{person}{David~B.
  Lindell}, {and} \bibinfo{person}{Gordon Wetzstein}.}
  \bibinfo{year}{2018}\natexlab{b}.
\newblock \showarticletitle{Real-time Non-line-of-sight Imaging}. In
  \bibinfo{booktitle}{\emph{ACM SIGGRAPH 2018 Emerging Technologies}}
  \emph{(\bibinfo{series}{SIGGRAPH '18})}. \bibinfo{publisher}{ACM},
  \bibinfo{address}{New York, NY, USA}, Article \bibinfo{articleno}{14},
  \bibinfo{numpages}{2}~pages.
\newblock
\showISBNx{978-1-4503-5810-1}
\urldef\tempurl%
\url{https://doi.org/10.1145/3214907.3214920}
\showDOI{\tempurl}


\bibitem[\protect\citeauthoryear{Pediredla, Buttafava, Tosi, Cossairt, and
  Veeraraghavan}{Pediredla et~al\mbox{.}}{2017}]%
        {pediredla2017reconstructing}
\bibfield{author}{\bibinfo{person}{Adithya~Kumar Pediredla},
  \bibinfo{person}{Mauro Buttafava}, \bibinfo{person}{Alberto Tosi},
  \bibinfo{person}{Oliver Cossairt}, {and} \bibinfo{person}{Ashok
  Veeraraghavan}.} \bibinfo{year}{2017}\natexlab{}.
\newblock \showarticletitle{Reconstructing rooms using photon echoes: A plane
  based model and reconstruction algorithm for looking around the corner}. In
  \bibinfo{booktitle}{\emph{Computational Photography (ICCP), 2017 IEEE
  International Conference on}}. IEEE, \bibinfo{pages}{1--12}.
\newblock


\bibitem[\protect\citeauthoryear{Pharr and Humphreys}{Pharr and
  Humphreys}{2010}]%
        {Pharr:2010:PBR:1854996}
\bibfield{author}{\bibinfo{person}{Matt Pharr} {and} \bibinfo{person}{Greg
  Humphreys}.} \bibinfo{year}{2010}\natexlab{}.
\newblock \bibinfo{booktitle}{\emph{Physically Based Rendering, Second Edition:
  From Theory To Implementation} (\bibinfo{edition}{2nd} ed.)}.
\newblock \bibinfo{publisher}{Morgan Kaufmann Publishers Inc.},
  \bibinfo{address}{San Francisco, CA, USA}.
\newblock
\showISBNx{0123750792, 9780123750792}


\bibitem[\protect\citeauthoryear{Schroeder, Martin, and Lorensen}{Schroeder
  et~al\mbox{.}}{2006}]%
        {VTK4}
\bibfield{author}{\bibinfo{person}{Will Schroeder}, \bibinfo{person}{Ken
  Martin}, {and} \bibinfo{person}{Bill Lorensen}.}
  \bibinfo{year}{2006}\natexlab{}.
\newblock \bibinfo{booktitle}{\emph{{The Visualization Toolkit--An
  Object-Oriented Approach To 3D Graphics}} (\bibinfo{edition}{fourth} ed.)}.
\newblock \bibinfo{publisher}{Kitware, Inc.}
\newblock


\bibitem[\protect\citeauthoryear{Slaney and Chou}{Slaney and Chou}{2014}]%
        {toftracer}
\bibfield{author}{\bibinfo{person}{Malcolm Slaney} {and}
  \bibinfo{person}{Philip~A. Chou}.} \bibinfo{year}{2014}\natexlab{}.
\newblock \bibinfo{booktitle}{\emph{Time of Flight Tracer}}.
\newblock \bibinfo{type}{{T}echnical {R}eport}. \bibinfo{institution}{Microsoft
  Research}.
\newblock
\urldef\tempurl%
\url{https://www.microsoft.com/en-us/research/publication/time-of-flight-tracer/}
\showURL{%
\tempurl}


\bibitem[\protect\citeauthoryear{Smith, Skorupski, and Davis}{Smith
  et~al\mbox{.}}{2008}]%
        {smith2008transient}
\bibfield{author}{\bibinfo{person}{Adam Smith}, \bibinfo{person}{James
  Skorupski}, {and} \bibinfo{person}{James Davis}.}
  \bibinfo{year}{2008}\natexlab{}.
\newblock \bibinfo{booktitle}{\emph{Transient Rendering}}.
\newblock \bibinfo{type}{{T}echnical {R}eport} UCSC-SOE-08-26.
  \bibinfo{institution}{School of Engineering, University of California, Santa
  Cruz}.
\newblock


\bibitem[\protect\citeauthoryear{Thrampoulidis, Shulkind, Xu, Freeman, Shapiro,
  Torralba, Wong, and Wornell}{Thrampoulidis et~al\mbox{.}}{2017}]%
        {thrampoulidis2017exploiting}
\bibfield{author}{\bibinfo{person}{Christos Thrampoulidis},
  \bibinfo{person}{Gal Shulkind}, \bibinfo{person}{Feihu Xu},
  \bibinfo{person}{William~T Freeman}, \bibinfo{person}{Jeffrey~H Shapiro},
  \bibinfo{person}{Antonio Torralba}, \bibinfo{person}{Franco~NC Wong}, {and}
  \bibinfo{person}{Gregory~W Wornell}.} \bibinfo{year}{2017}\natexlab{}.
\newblock \showarticletitle{Exploiting Occlusion in Non-Line-of-Sight Active
  Imaging}.
\newblock \bibinfo{journal}{\emph{arXiv preprint arXiv:1711.06297}}
  (\bibinfo{year}{2017}).
\newblock


\bibitem[\protect\citeauthoryear{Velten, Raskar, and Bawendi}{Velten
  et~al\mbox{.}}{2011}]%
        {Velten:2011}
\bibfield{author}{\bibinfo{person}{Andreas Velten}, \bibinfo{person}{Ramesh
  Raskar}, {and} \bibinfo{person}{Moungi Bawendi}.}
  \bibinfo{year}{2011}\natexlab{}.
\newblock \showarticletitle{Picosecond Camera for Time-of-Flight Imaging}, In
  \bibinfo{booktitle}{Imaging and Applied Optics}.
\newblock \bibinfo{journal}{\emph{Imaging and Applied Optics}},
  \bibinfo{pages}{IMB4}.
\newblock
\urldef\tempurl%
\url{https://doi.org/10.1364/ISA.2011.IMB4}
\showDOI{\tempurl}


\bibitem[\protect\citeauthoryear{Velten, Willwacher, Gupta, Veeraraghavan,
  Bawendi, and Raskar}{Velten et~al\mbox{.}}{2012}]%
        {Velten:2012:Recovering}
\bibfield{author}{\bibinfo{person}{A. Velten}, \bibinfo{person}{T. Willwacher},
  \bibinfo{person}{O. Gupta}, \bibinfo{person}{A. Veeraraghavan},
  \bibinfo{person}{M.G. Bawendi}, {and} \bibinfo{person}{R. Raskar}.}
  \bibinfo{year}{2012}\natexlab{}.
\newblock \showarticletitle{Recovering three-dimensional shape around a corner
  using ultrafast time-of-flight imaging}.
\newblock \bibinfo{journal}{\emph{Nature Communications}}  \bibinfo{volume}{3}
  (\bibinfo{year}{2012}), \bibinfo{pages}{745}.
\newblock


\bibitem[\protect\citeauthoryear{Velten, Wu, Jarabo, Masia, Barsi, Joshi,
  Lawson, Bawendi, Gutierrez, and Raskar}{Velten et~al\mbox{.}}{2013}]%
        {Velten:2013:FCV:2461912.2461928}
\bibfield{author}{\bibinfo{person}{Andreas Velten}, \bibinfo{person}{Di Wu},
  \bibinfo{person}{Adrian Jarabo}, \bibinfo{person}{Belen Masia},
  \bibinfo{person}{Christopher Barsi}, \bibinfo{person}{Chinmaya Joshi},
  \bibinfo{person}{Everett Lawson}, \bibinfo{person}{Moungi Bawendi},
  \bibinfo{person}{Diego Gutierrez}, {and} \bibinfo{person}{Ramesh Raskar}.}
  \bibinfo{year}{2013}\natexlab{}.
\newblock \showarticletitle{Femto-photography: Capturing and Visualizing the
  Propagation of Light}.
\newblock \bibinfo{journal}{\emph{ACM Trans. Graph.}} \bibinfo{volume}{32},
  \bibinfo{number}{4}, Article \bibinfo{articleno}{44} (\bibinfo{date}{July}
  \bibinfo{year}{2013}), \bibinfo{numpages}{8}~pages.
\newblock
\showISSN{0730-0301}
\urldef\tempurl%
\url{https://doi.org/10.1145/2461912.2461928}
\showDOI{\tempurl}


\bibitem[\protect\citeauthoryear{Wu, Velten, O'Toole, Masia, Agrawal, Dai, and
  Raskar}{Wu et~al\mbox{.}}{2014}]%
        {Wu2014}
\bibfield{author}{\bibinfo{person}{Di Wu}, \bibinfo{person}{Andreas Velten},
  \bibinfo{person}{Matthew O'Toole}, \bibinfo{person}{Belen Masia},
  \bibinfo{person}{Amit Agrawal}, \bibinfo{person}{Qionghai Dai}, {and}
  \bibinfo{person}{Ramesh Raskar}.} \bibinfo{year}{2014}\natexlab{}.
\newblock \showarticletitle{Decomposing Global Light Transport Using Time of
  Flight Imaging}.
\newblock \bibinfo{journal}{\emph{International Journal of Computer Vision}}
  \bibinfo{volume}{107}, \bibinfo{number}{2} (\bibinfo{date}{01 Apr}
  \bibinfo{year}{2014}), \bibinfo{pages}{123--138}.
\newblock
\showISSN{1573-1405}
\urldef\tempurl%
\url{https://doi.org/10.1007/s11263-013-0668-2}
\showDOI{\tempurl}


\bibitem[\protect\citeauthoryear{Wu, Wetzstein, Barsi, Willwacher, O'Toole,
  Naik, Dai, Kutulakos, and Raskar}{Wu et~al\mbox{.}}{2012}]%
        {wu2012frequency}
\bibfield{author}{\bibinfo{person}{Di Wu}, \bibinfo{person}{Gordon Wetzstein},
  \bibinfo{person}{Christopher Barsi}, \bibinfo{person}{Thomas Willwacher},
  \bibinfo{person}{Matthew O'Toole}, \bibinfo{person}{Nikhil Naik},
  \bibinfo{person}{Qionghai Dai}, \bibinfo{person}{Kyros Kutulakos}, {and}
  \bibinfo{person}{Ramesh Raskar}.} \bibinfo{year}{2012}\natexlab{}.
\newblock \showarticletitle{Frequency analysis of transient light transport
  with applications in bare sensor imaging}.
\newblock \bibinfo{journal}{\emph{Computer Vision--ECCV 2012}}
  (\bibinfo{year}{2012}), \bibinfo{pages}{542--555}.
\newblock


\bibitem[\protect\citeauthoryear{Zhang}{Zhang}{2000}]%
        {zhang1999}
\bibfield{author}{\bibinfo{person}{Zengyou Zhang}.}
  \bibinfo{year}{2000}\natexlab{}.
\newblock \showarticletitle{A Flexible New Technique for Camera Calibration}.
\newblock \bibinfo{journal}{\emph{IEEE Transactions on Pattern Analysis and
  Machine Intelligence}}  \bibinfo{volume}{22} (\bibinfo{date}{December}
  \bibinfo{year}{2000}), \bibinfo{pages}{1330--1334}.
\newblock
\urldef\tempurl%
\url{https://www.microsoft.com/en-us/research/publication/a-flexible-new-technique-for-camera-calibration/}
\showURL{%
\tempurl}
\newblock
\shownote{MSR-TR-98-71, Updated March 25, 1999.}


\end{thebibliography}
\appendix

\end{document}